\documentclass[amsmath,
amssymb,
aps,
prb,
preprint,
floatfix]{revtex4-2}

\usepackage{graphicx}
\usepackage{dcolumn}
\usepackage{bm}
\usepackage{xfrac}
\usepackage[hidelinks]{hyperref}

\DeclareGraphicsExtensions{.png}

\begin{document}

\title{Landau Level Phases in Bilayer Graphene under Pressure at Charge Neutrality}

\author{Brett R. Green}
\email{brg5241@psu.edu}
\affiliation{Department of Physics, The Pennsylvania State University}
\author{Jorge O. Sofo}
\email{sofo@psu.edu}
\homepage{http://sites.psu.edu/sofo}
\affiliation{%
Department of Physics, Department of Materials Science and Engineering, and Materials Research Institute\\
The Pennsylvania State University}%

\date{\today}% It is always \today, today,

\begin{abstract}

Bilayer graphene in a magnetic field hosts a variety of ordered phases built from eight Landau levels close in energy to the neutrality point. These levels are characterized by orbital $n=0,1$, valley $\xi=+,-$ and spin $\sigma=\;\uparrow,\downarrow$; their relative energies depend strongly on the Coulomb interaction, magnetic field, and interlayer bias. We treat interactions at the Hartree-Fock level, including the effects of metallic gates, layer separation, spatial extent of the p\textsubscript{z} orbitals, all Slonczewski-Weiss-McClure tight-binding parameters, and pressure. We obtain the ground state as function of the applied magnetic field, bias, and pressure. The gates, layer separation and extent of the p\textsubscript{z} orbitals weaken the Coulomb interaction at different length scales; these effects distort the phase diagram but do not change its topology. However, previously-predicted continuous transitions become discontinuous when all tight-binding parameters are included nonperturbatively. We find that pressure increases the importance of the noninteracting scale with respect to the Coulomb energy, which drives phase transitions to occur at lower fields. This brings two orbitally polarized states not yet predicted or observed into the experimentally accessible region of the phase diagram, in addition to previously-identified valley-, spin-, and partially orbitally polarized states.

\end{abstract}

\maketitle

%\tableofcontents

\section{Introduction}

Electrons in a magnetic field occupy highly degenerate states known as Landau levels (LLs). In multilayer 2D materials, a perpendicular electric field can change the relative position in energy of electronic states near the Fermi level, offering an exciting platform for the exploration of quantum order in condensed matter systems. Bilayer graphene (BLG) is no exception, and it has been shown experimentally \cite{hunt_direct_2017,li_effective_2018,li_metallic_2019,chuang_landau_2019,weitz_broken-symmetry_2010} to produce different macroscopic states, such as a fully spin-polarized state, a fully valley-polarized state, and others to be described below. The appearance of these states as a function of applied fields generates a phase diagram, which is a target of research in this area and provides a map for the study of these phases. Of course, the experimental identification of each ground state is challenging, and this work needs to be complemented by theoretical understanding of the system.

Recent experiments \cite{hunt_direct_2017,li_effective_2018,chuang_landau_2019} on undoped (filling factor $\nu=0$) BLG found a single sharp transition as the electric field was increased while the magnetic field was low, but at higher magnetic fields, the phase boundary splits into two. These transitions were identified by peaks in the sample's two-terminal conductivity. One low-field state is a fully spin-polarized or ferromagnetic state evolving from a canted antiferromagnetic state, identified by edge state conductivity measurements \cite{kharitonov_canted_2012,maher_evidence_2013,li_metallic_2019}. The other two have been characterized by layer polarization measurements \cite{hunt_direct_2017}, which support the identification of the low-magnetic-field, high-electric-field state as a fully valley-polarized state, and of the intermediate state as one with mixed polarization in both spin and valley. The intermediate state is also the first to be observed with polarization in the orbital index $n$, an additional low-energy degree of freedom in BLG deriving from its unique LL spectrum: $E_0\approx0$, $E_1\approx0$, $E_{\pm n}\approx\pm\hbar\omega\sqrt{n(n-1)}$ for $n\ge 2$ \cite{mccann_electronic_2013}.

The $\nu=0$ phase diagram has proven to be highly sensitive to experimental perturbations, such as screening by an atomically-thin dielectric \cite{chuang_landau_2019} or changes in device geometry and size \cite{li_metallic_2019}, underlining the possibilities for quantum state engineering and the importance of a careful treatment of interactions. In this work, we add a new method of manipulating states: pressure. We show that pressure can be used to control the orbital degree of freedom, and that this is achieved by changing the energy scale of the noninteracting dynamics relative to the interaction energy scale.

Regarding the treatment of interactions, two approaches have been used in previous work: one based on the bare Coulomb potential \cite{cote_orbital_2010,cote_biased_2011,lambert_quantum_2013,lambert_ferro-aimants_2013,knothe_phase_2016}, and the other using only short-range interactions which may break symmetries of the bare Coulomb potential, an approach introduced by Kharitonov \cite{kharitonov_phase_2012,kharitonov_canted_2012,murthy_spin-valley_2017}. Additionally, Hunt \textit{et al.} \cite{hunt_direct_2017} treat the direct Coulomb interaction with a random phase approximation including metallic gates in the bare propagator in addition to symmetry-breaking parameters. On one hand, the former approach has no free parameters but has not yet reproduced the experimentally-observed intermediate phase; on the other, the latter approach has succeeded in reproducing the intermediate phase but requires undetermined parameters whose physical origins are not transparent. So that we can understand the underlying physics while exploring the effects of pressure, we take the parameter-free approach.

Previous use of this approach has included the effects of layer separation in Refs.~\cite{cote_orbital_2010,cote_biased_2011,lambert_quantum_2013,lambert_ferro-aimants_2013,knothe_phase_2016} and screening by metallic gates in Ref.~\cite{hunt_direct_2017} when treating the interaction. We unify these by deriving a propagator which includes both effects, and also address the out-of-plane spatial extent of the p\textsubscript{z} orbitals with layer-resolved 3D LL wavefunctions, which had previously been taken as 2D in each layer. These wavefunctions are derived by exact diagonalization of a four-band tight-binding Hamiltonian including all Slonczewski-Weiss-McClure tight-binding (TB) parameters, which we show are key in determining the nature (discontinuous, or continuously interpolating between ground states) of phase transitions. In particular, our model reproduces experimental findings of a single sharp spin- to valley-polarized transition at low fields, which contrasts with the continuous transition mixing the states found in previous parameter-free studies \cite{cote_orbital_2010,cote_biased_2011,lambert_quantum_2013,lambert_ferro-aimants_2013,knothe_phase_2016}.

Under pressure, we find two orbitally polarized states not yet predicted or observed. These states appear because pressure increases the energy gap between orbitals so that it overcomes the interaction energy scale, which had stabilized the spin- and valley-polarized states observed at low magnetic fields. Hence, pressure effectively tunes the strength of interactions relative to the noninteracting energy scale. Pressure can also be treated as a theoretical proxy in our results for other effects that influence the noninteracting energy scales.

The paper is outlined as follows. In Sec.~\ref{H_ni}, we solve the TB model in a magnetic field to find LL energies and wavefunctions. We then address interactions at the Hartree-Fock level in Sec.~\ref{H_HF}, and describe our approach to the interacting problem. Solving the interacting problem as a function of magnetic field and bias yields phase diagrams which we present in Sec.~\ref{results}. We also characterize the possible ground states in this section, and discuss how the effects we include in treatment affect our results. We summarize our work and findings, and suggest next steps, in Sec.~\ref{conclusion}.

\section{Methods}

\subsection{Noninteracting Hamiltonian} \label{H_ni}

\begin{figure}
    \centering
    \includegraphics[width=6cm]{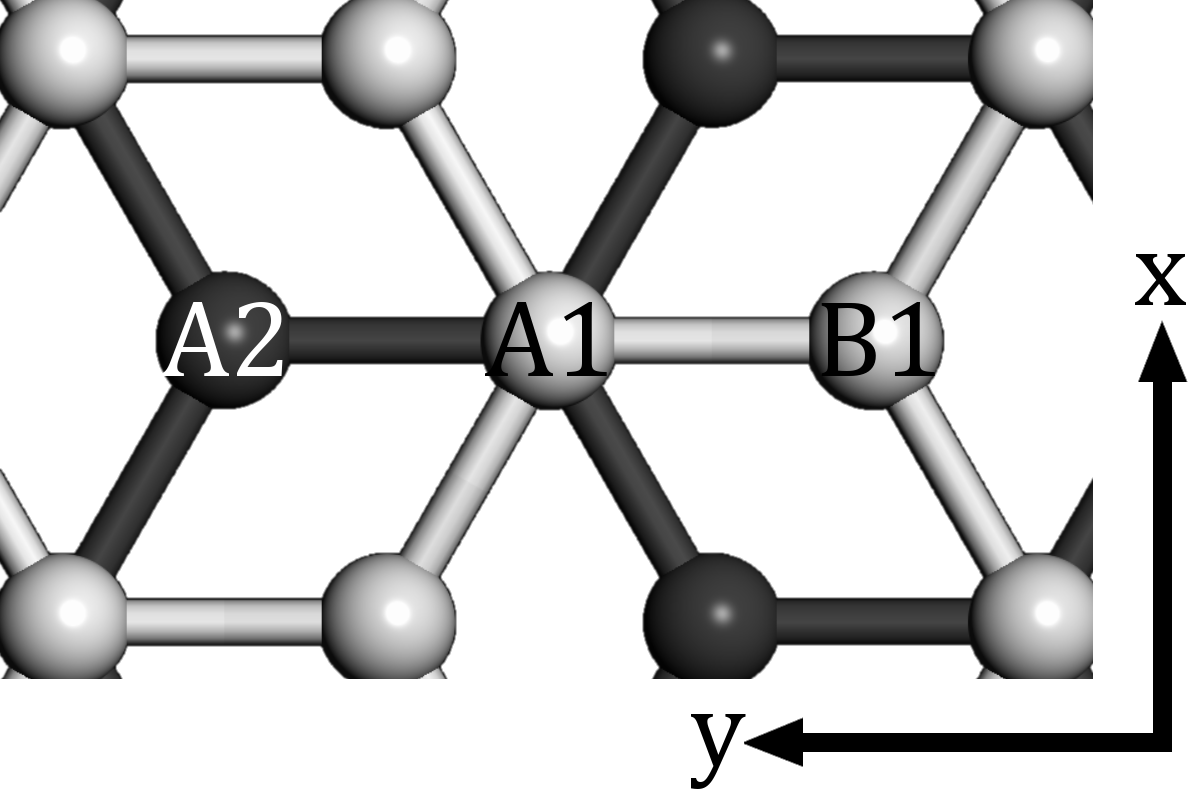}
    \caption{The BLG unit cell has a four-atom basis with inequivalent A and B sites in each layer. They are shown here with upper layer sites are denoted as $A1$, $B1$ and lower layer sites as $A2$, $B2$. The $A1$ sites and $B2$ sites are stacked.}
    \label{fig:BLGunitcell}
\end{figure}

We begin with the spin-free TB Bloch Hamiltonian
\begin{equation}
H_{\bm{k}}=\left[ \begin{array}{cccc}{2 \varepsilon+\frac{\Delta}{2}} & {t \phi} & {t_{4} \phi^{*}} & {t_{\perp}} \\ {t \phi^{*}} & {\frac{\Delta}{2}} & {t_{3} \phi} & {t_{4} \phi^{*}} \\ {t_{4} \phi} & {t_{3} \phi^{*}} & {-\frac{\Delta}{2}} & {t \phi} \\ {t_{\perp}} & {t_{4} \phi} & {t \phi^{*}} & {2 \varepsilon-\frac{\Delta}{2}}\end{array}\right] \end{equation}
written in the basis $\left\{ |A1,\bm{k}\rangle, |B1,\bm{k}\rangle, |A2,\bm{k}\rangle, |B2,\bm{k}\rangle \right\}$. Here $|T\bm{k}\rangle=\frac{1}{\sqrt{N}} \sum_{\bm{R}} e^{i \bm{k} \cdot \bm{R}}|T\bm{R}\rangle$ is the Fourier transform of the p\textsubscript{z} orbitals $|T\bm{R}\rangle$ on the lattice site $T=T_{2D}T_z$ with sublattice $T_{2D}=A,B$ and layer $T_z=1,2$, located in the unit cell at $\bm{\tau_T}=\bm{\tau^{2D}_T}+\tau^z_T\bm{\hat{z}}$ with $\bm{T^{2D}_T}$ in the hexagonal lattice and layer $\tau^z_T=(-1)^{T_z+1}d/2$. $\bm{R}$ gives the location of the unit cell, and $N$ gives the number of unit cells in the sample. $t$, $t_\perp$, $t_3$ and $t_4$ are the hopping parameters, $\epsilon$ gives the site energy for stacked $A1$ and $B2$ atoms, and $\Delta$ is an interlayer bias induced by a perpendicular electric field. The TB parameters vary with pressure and are given by Munoz et al. \cite[Table II]{munoz_bilayer_2016} We expand $\phi=\phi\left(\bm{k}\right)$ to linear order in $\bm{q}=\bm{k}-\bm{K}_\xi$ about valley $\xi$:
\begin{equation}\begin{aligned}
\phi\left(\bm{k}\right)=e^{i a k_{y}}\left(1+2 e^{-i \frac{3 a}{2} k_{y}} \cos \left(\frac{a \sqrt{3}}{2} k_{x}\right)\right) \, , \\
\phi\left(\bm{K}_{\xi}+\bm{q}\right) \approx-\xi \frac{3 a}{2}\left(q_{x}-\xi i q_{y}\right)=-\xi \frac{3 a}{2} q_{-\xi} \, ,
\end{aligned}\end{equation}
where $q_{ \pm}=q_{x} \pm i q_{y}$. The lattice sites and coordinate system are depicted in Fig.~\ref{fig:BLGunitcell}.

We represent the magnetic field $\bm{B}=B\bm{\hat{z}}$ by a vector potential in the Landau gauge given by $\bm{A}=Bx\bm{\hat{y}}$. This will enter the Hamiltonian through a Peierls substitution, $\bm{k} \to \bm{k}+\frac{e}{\hbar} \bm{A}$, which is analogous to the replacement of momentum with canonical momentum, $\bm{p} \to \bm{p}+e \bm{A}$. The result is $q_{ \pm} \to q_{x} \pm i \left( q_{y} + \frac{e}{\hbar}Bx \right) = \kappa_\pm$.

With this substitution, the Hamiltonian may be written in terms of harmonic oscillator raising and lowering operators. Denoting the harmonic oscillator wavefunctions by $Q_j (x)$ and working on the prototypical Landau level wavefunction,
\begin{equation} \label{LLenvelope}
h_{j X}(\bm{R}) = \frac{1}{\sqrt{L_{y}}} e^{i \frac{X}{l_{B}^{2}}R_y} Q_j (R_x-X) \, ,
\end{equation}
we verify the commutation relation $[\kappa_-,\kappa_+]h_{j X}(\bm{R})=\frac{2}{l_B^2}h_{j X}(\bm{R})$, where $l_B=\sqrt{\frac{\hbar}{eB}}$ is the magnetic length. Hence, $\kappa_{ \pm}$ satisfies $\kappa_{+}=\frac{\sqrt{2}}{l_{B}} a^{+}, \kappa_{-}=\frac{\sqrt{2}}{l_{B}} a$. In particular, if we define the basis states
\begin{equation}
\left|T j X\right\rangle = \sum_{\bm{R}} h_{jX}(\bm{R}) \left|T\bm{R}\right\rangle \, ,
\end{equation}
where $h_{jX}(\bm{R})$ is an envelope on the p\textsubscript{z} orbitals $\left|T\bm{R}\right\rangle$, then $a^+ |TjX\rangle = \sqrt{j+1} |T(j+1)X\rangle$ and $a |TjX\rangle = \sqrt{j} |T(j-1)X\rangle$.

Letting $C_{\mu}=\frac{3 a}{\sqrt{2} l_{B}} t_{\mu}$, then, we have for example in valley $\xi=+$ the LL Hamiltonian
\begin{equation} \label{equation_H_LL}
H_{+}=\left[ \begin{array}{cccc}{2 \varepsilon+\frac{\Delta}{2}} & {-C a} & {-C_{4} a^{+}} & {t_{\perp}} \\ {-C a^{+}} & {\frac{\Delta}{2}} & {-C_{3} a} & {-C_{4} a^{+}} \\ {-C_{4} a} & {-C_{3} a^{+}} & {-\frac{\Delta}{2}} & {-C a} \\ {t_{\perp}} & {-C_{4} a} & {-C a^{+}} & {2 \varepsilon-\frac{\Delta}{2}}\end{array}\right] .
\end{equation}
The Hamiltonian $H_-$ for the other valley, $\xi=-$, is obtained by replacing $a$ and $a^+$ with $-a^+$ and $-a$, respectively, in the same basis. The full noninteracting Hamiltonian (both spatial and spin parts) is then
\begin{equation}
\hat{H}_{ni}=\left(\frac{1}{2}\left(1+\lambda^v_z\right)\hat{H}_+ + \frac{1}{2}\left(1-\lambda^v_z\right)\hat{H}_- \right) \otimes -\mu_{B} B \sigma_{z}
\end{equation}
where $\lambda^v_z$ is a Pauli matrix acting on the valley space $\{+,-\}$.

To diagonalize $H_\xi$, which contains operators as represented in Eq.~(\ref{equation_H_LL}), we express it as a matrix of scalars by taking matrix elements in a truncated basis of oscillator states
\begin{equation}
\left\langle TjX \right| H_\xi \left| T j^\prime X \right\rangle \, , \quad
\begin{array}{c}
T,T^\prime=A1,B1,A2,B2 \, , \\ j,j^\prime=0,1,2,...j_{max} \, .
\end{array}
\end{equation}
The coefficients of the wavefunctions for the states near the neutrality point decrease as $j$ increases. Therefore, we take $j_{max} = 15$, for which the greatest coefficient after $j>12$ in the expansion of the LLL eigenstates was below $0.01$ for all magnetic fields and pressures we consider. There are two eigenstates near zero energy, which we index by the orbital quantum number $n$. The LLL eigenstates are then
\begin{equation} \label{LLLbasisstates}
|n \xi \sigma X\rangle= \sum_{T j} c_{n \xi}^{T j} |T j X\rangle \times |\sigma\rangle
\end{equation}
when spin is inlcuded. The eight combinations of three binary indices $n, \xi, \sigma$ give the eight nearly-degenerate low-energy (LLLs). Each LLL is highly degenerate because its energy does not depend on the guiding center $X$.

At zero bias, there is a useful symmetry between the valleys. It arises from the relation between the Hamiltonians $H_+$ and $H_-$, which can also be described as $H_-$ being the transpose of $H_+$ with ladder operators $a,a^+$. As a result, their eigenvectors are related by the signed permutation
\begin{equation}
c_{n-}^{T j}=(-1)^{j} c_{n+}^{\pi_{T} j} \, , \quad \left[\begin{array}{l}{T} \\ {\pi_{T}}\end{array}\right]=\left[\begin{array}{llll}{A 1} & {B 1} & {A 2} & {B 2} \\ {B 2} & {A 2} & {B 1} & {A 1}\end{array}\right] ,
\end{equation}
so that the valley $-$ states have the same spatial distribution of valley $+$ states but in the opposite layer and lattice sites. (This symmetry identifies $\xi=+$ with the upper layer and $\xi=-$ with the lower layer, which known as the valley-layer correspondence.) Furthermore, their energies are degenerate and may be labeled $E_n$ independently of valley. Over the range of high magnetic fields that we are interested in, bias has negligible effect on coefficients, so the symmetry may be treated as exact and bias can be addressed as a perturbation to the energy. Defining the layer polarization of the LLL $n$ by
\begin{equation}
\Pi_{n}=\left(\sum_{j=0}^{\infty} \sum_{T_{2D}}\left|c_{n+}^{T_{2D} 1 j}\right|^{2}\right)-\left(\sum_{j=0}^{\infty} \sum_{T_{2D}}\left|c_{n+}^{T_{2D} 2 j}\right|^{2}\right)
\end{equation}
and using the symmetry between the valleys, the full noninteracting energy is
\begin{equation} \label{noninteractingenergy}
E_{n \xi \sigma}=E_{n}-\xi \Pi_{n} \frac{\Delta}{2}-\sigma \mu_{B} B \, .
\end{equation}

Energies $E_{n\xi\sigma}$ versus bias, orbital gap $E_1-E_0$, layer polarization $\Pi_n$ versus magnetic field and pressure, and eigenvector coefficients $c_{n\xi}^{Tj}$ are illustrated in Sec.~S1 of the Supplemental Material. Note that when we refer to orbital gap, we mean the splitting caused strictly by noninteracting orbital dynamics, not the energy gap between two LLLs of different orbital, which in general also depends on bias, magnetic field, and interactions.

\begin{figure}
    \centering
    \includegraphics[width=10cm]{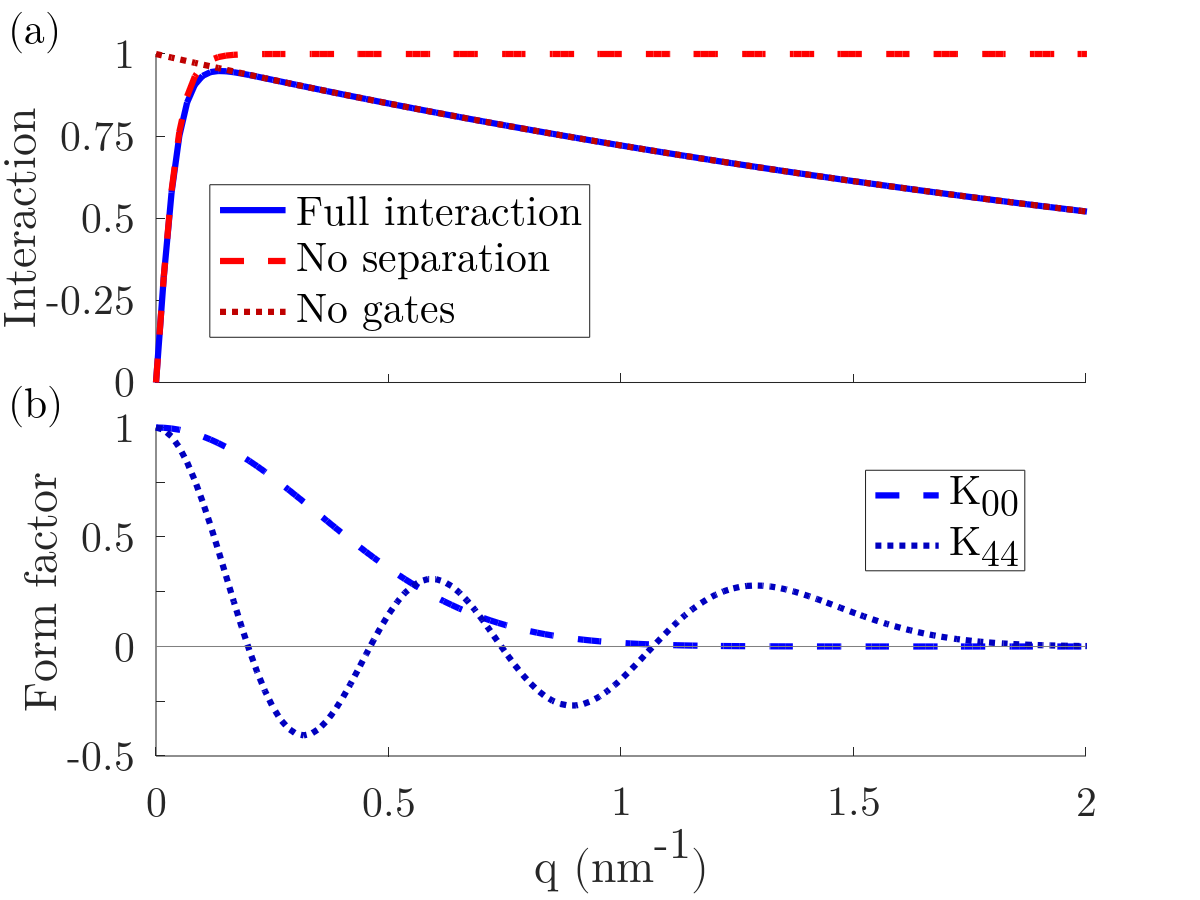}
    \caption{(a) The interaction strength given by Eq.~(\ref{Vfit}) versus wavevector is plotted here, in units of $\frac{1}{N_\Phi} \alpha \frac{1}{ql_B}$. We use the interlayer case, $T_{z} \ne T_{z}^{\prime}$, for demonstration. It can be seen that gating weakens the long-range (small $q$) interactions and layer separation weakens the short-range (large $q$). The dashed curve corresponds to the interaction neglecting separation, $d^{T_{z} T_{z}^\prime}_{eff}=0$, and the dotted curve corresponds to the absence of gates, $D\to\infty$. (b) The amplitude of the elementary form factors, as defined in Eq.~(\ref{elem_ff}), are plotted versus wavevector on the same scale. As the form factors are integrated against the interaction in the exchange integral, Eq.~(\ref{equation_fockint}), in this figure we can see that both length scales are relevant in the reciprocal-space support of the wavefunctions.}
    \label{fig:VvsFFcomparison}
\end{figure}

\begin{widetext}\subsection{Coulomb interaction} \label{H_HF}

The Coulomb interaction,
\begin{equation}
\hat{V} = \frac{1}{2} \int d^2 \bm{r} \int d z \int d^2 \bm{r^{\prime}} \int d z^\prime \psi^{+}(\bm{r},z) \psi^{+}\left(\bm{r^{\prime}},z^\prime\right) V\left(\bm{r}-\bm{r^{\prime}},z,z^\prime\right) \psi\left(\bm{r^{\prime}},z^\prime\right) \psi(\bm{r},z) \, ,
\end{equation}
is treated in the Hartree-Fock (HF) approximation, similarly to previous works \cite{barlas_intra-landau-level_2008,cote_orbital_2010,cote_biased_2011,kharitonov_phase_2012,kharitonov_canted_2012,kharitonov_edge_2012,lambert_quantum_2013,lambert_ferro-aimants_2013,knothe_phase_2016,murthy_spin-valley_2017}. Throughout this work, we use $\bm{r}$ for the 2D in-plane position vector, and retain $z$-dependence to address the effects of layer separation, gating, and the spatial extent of the p\textsubscript{z} orbitals.

Expanding the Coulomb interaction as its Fourier transform in the in-plane direction as ${V(\bm{r},z,z^\prime)=\sum_{\bm{q}} e^{i \bm{q} \cdot \bm{r}} V\left(q,z,z^\prime\right)}$, and expanding the field operators in the LLL basis,
\begin{equation}
\psi (\bm{r},z) = \sum_{n \xi \sigma X} \phi_{n \xi \sigma X}(\bm{r},z) c_{n \xi \sigma X}
= \sum_{n \xi \sigma X} \left(
\sum_{T \bm{R}} c_{n \xi X}^{T}(\bm{R})\langle \bm{r},z|T\bm{R}\rangle \times|\sigma\rangle
\right)
c_{n \xi \sigma X} \, ,
\end{equation}
we have
\begin{equation}\begin{aligned} \label{Vexpandedfieldops} \hat{V} = \frac{1}{2} \sum_{\substack{n_j \xi_j \sigma_j X_j \\ j=1,2,3,4}} \sum_{\bm{q}} & \Bigl(\int d z \int d z^{\prime} V\left(q, z, z^{\prime}\right) \\ \times & \left( \int d^{2} \bm{r} e^{i \bm{q} \cdot \bm{r}} \phi_{n_{1} \xi_{1} \sigma_{1} X_1}^{*}\left(\bm{r}, z\right) \phi_{n_{4} \xi_{4} \sigma_{4} X_4}\left(\bm{r}, z\right)\right) \\ \times & \Bigl(\int d^{2} \bm{r^{\prime}} e^{-i \bm{q} \cdot \bm{r^{\prime}}} \phi_{n_{2} \xi_{2} \sigma_{2} X_2}^{*}\left(\bm{r^{\prime}}, z^{\prime}\right) \phi_{n_{3} \xi_{3} \sigma_{3} X_3}\left(\bm{r^{\prime}}, z^{\prime}\right)\Bigr) \Bigr) \\ \times & c^+_{n_1 \xi_1 \sigma_1 X_1}c^+_{n_2 \xi_2 \sigma_2 X_2}c_{n_3 \xi_3 \sigma_3 X_3}c_{n_4 \xi_4 \sigma_4 X_4} \, . \end{aligned}\end{equation}
%\begin{equation}\begin{aligned} \label{Vexpandedfieldops} \hat{V} = \frac{1}{2} & \sum_{\substack{n_j \xi_j \sigma_j X_j \\ j=1,2,3,4}} \sum_{\bm{q}} c^+_{n_1 \xi_1 \sigma_1 X_1}c^+_{n_2 \xi_2 \sigma_2 X_2}c_{n_3 \xi_3 \sigma_3 X_3}c_{n_4 \xi_4 \sigma_4 X_4} \Bigg( \int d z \int d z^{\prime} V\left(q, z, z^{\prime}\right) \\ \times & \left( \int d^{2} \bm{r} e^{i \bm{q} \cdot \bm{r}} \phi_{n_{1} \xi_{1} \sigma_{1} X_1}^{*}\left(\bm{r}, z\right) \phi_{n_{4} \xi_{4} \sigma_{4} X_4}\left(\bm{r}, z\right)\right) \left(\int d^{2} \bm{r^{\prime}} e^{-i \bm{q} \cdot \bm{r^{\prime}}} \phi_{n_{2} \xi_{2} \sigma_{2} X_2}^{*}\left(\bm{r^{\prime}}, z^{\prime}\right) \phi_{n_{3} \xi_{3} \sigma_{3} X_3}\left(\bm{r^{\prime}}, z^{\prime}\right)\right) \Bigg) \, . \end{aligned}\end{equation}

To incorporate both layer separation and the screening effect of metallic double gates used in recent experiments \cite{hunt_direct_2017,li_effective_2018,li_metallic_2019}, we use a propagator of the Coulomb interaction corresponding to equipotential walls at $\pm D$. The Fourier transform of this propagator is
\begin{equation} \label{Vintegrand}
V\left(q, z, z^{\prime}\right) =
\frac{2\pi}{A} \frac{e^{2}}{4 \pi \epsilon_{0}} \frac{1}{q} \frac{\cosh q\left(2 D-\left|z^{\prime}-z\right|\right)-\cosh q\left(z+z^{\prime}\right)}{\sinh 2 q D}
\end{equation}\end{widetext}
where $D=20~\text{nm}$ \cite{hunt_direct_2017,li_effective_2018}, $\epsilon_r=6.9$ \cite{laturia_dielectric_2018} and $\alpha=\frac{e^{2}}{4 \pi \epsilon_r \epsilon_{0} l_{B} }$. The effective dielectric constant has been taken to be the dielectric constant of hexagonal boron nitride. The normalization and energy scale may be rewritten as $\frac{2\pi}{A} \frac{e^{2}}{4 \pi \epsilon_{0}} \frac{1}{q} = \frac{1}{N_\Phi} \alpha \frac{1}{q l_{B}}$, where $N_\Phi$ is the number of flux quanta penetrating the bilayer and hence the degeneracy of the system. We include both gating and layer separation because both affect wavevector scales relevant the LLs, as illustrated in Fig.~\ref{fig:VvsFFcomparison}. The pressure-dependent layer separation is given by Munoz et al. \cite[Table I]{munoz_bilayer_2016}

Note that if one neglects the layer separation $d$ when compared to $D$, i.e. takes $D+d \approx D$, we have
\begin{equation} \label{V_d<<D_propagator}
V\left(q,+\frac{d}{2},-\frac{d}{2}\right) \approx \frac{1}{N_\Phi} \alpha \frac{1}{q l_{B}} \tanh (q D) e^{-q d} \, .
\end{equation}
Taking $D\to\infty$ yields the propagator of Refs.~\cite{cote_orbital_2010,cote_biased_2011,lambert_quantum_2013,lambert_ferro-aimants_2013,knothe_phase_2016}, while taking $d=0$ yields the propagator of Ref.~\cite{hunt_direct_2017}.

The tight-binding orbitals contribute a $z$-direction density, $P\left(z\right)$, which is integrated out to obtain the layer-resolved Coulomb interaction,
\begin{equation}\begin{aligned} \label{Vlayerprojected} V_{T_{z} T_{z}^{\prime}}(q) & = \int dz \int dz^{\prime} V\left(q, z, z^{\prime}\right) \\ & \quad \, \, \times P\!\left(z+(\textrm{--} 1)^{T_z}\frac{d}{2}\right) P\!\left(z+(\textrm{--} 1)^{T_z^\prime}\frac{d}{2}\right) .
\end{aligned}\end{equation}
We find that this integral can be well approximated by
\begin{equation} \label{Vfit}
V_{T_{z} T_{z}^{\prime}}(q) = \frac{1}{N_\Phi} \alpha \frac{1}{ql_B}\tanh{(qD)}e^{-q d^{T_{z} T_{z}^{\prime}}_{eff}}
\end{equation}
which has the form of Eq.~(\ref{V_d<<D_propagator}) but uses an effective layer separation $d^{T_{z} T_{z}^{\prime}}_{eff}$ in place of the physical layer separation $d$. This expression is a fit to exact evaluations of Eq.~(\ref{Vlayerprojected}). A complete derivation of these expressions may be found in Sec.~S2 of the Supplemental Material, together with Fig.~S3 which illustrates the validity of the fit given by Eq~(\ref{Vfit}). In the limit $P(z)\to\delta (z)$, the effective interlayer separation becomes the actual layer separation so that the effective intralayer separation vanishes, ${d^{T_{z} T_{z}^{\prime}}_{eff}=d\left(1-\delta_{T_{z}T_{z}^\prime}\right)}$, and we have ${V_{11}(q) \to V\left(q,\frac{d}{2},\frac{d}{2}\right)}$ and ${V_{12}(q) \to V\left(q,\frac{d}{2},-\frac{d}{2}\right)}$. By symmetry, ${V_{11}(q)=V_{22}(q)}$ and ${V_{12}(q)=V_{21}(q)}$.

Returning now to Eq.~(\ref{Vexpandedfieldops}), it remains to calculate the Fourier transforms of the wavefunction overlaps, or form factors. These are evaluated as
\begin{align} \label{fullformfac}
& \int d^{2} \bm{r} e^{i \bm{q} \cdot \bm{r}} \phi_{n_{1} \xi_1 \sigma X_1}^{*}(\bm{r}, z) \phi_{n_{4} \xi_4 \sigma X_4}(\bm{r}, z) \\
& \, \, = \delta_{X_{4},X_{1} \! - \! q_{y} l_{B}^{2}} e^{i \frac{q_{x}}{2} \! \left(X_{1} \! + \! X_{4}\right)} \sum_{T_{z}} P \! \left(z+(\textrm{--}1)^{T_z}\frac{d}{2}\right) J_{\substack{n_1 \xi_1 \\ n_4 \xi_4}}^{T_z} \! (\bm{q}) \nonumber
\end{align}
with the layer-projected form factors (writing $c_{n \xi}^{T_{2 D} T_z j}$ in place of $c_{n \xi}^{T j}$)
\begin{equation}
J_{\substack{n_1 \xi_1 \\ n_4 \xi_4}}^{T_z}(\bm{q})=\sum_{j_{1} j_{4}} K_{j_{1} j_{4}}(\bm{q}) \sum_{T_{2 D}} {c_{n_{1} \xi_1}^{T_{2 D} T_{z} j_{1}}}^* c_{n_{4} \xi_4}^{T_{2 D} T_{z} j_{4}}
\end{equation}
and elementary form factors
\begin{equation} \label{elem_ff}
K_{j_{1} j_{4}}(\bm{q}) = \int d x e^{i q_{x} x} Q_{j_{1}} \! \left(x-\frac{q_{y} l_{B}^{2}}{2}\right) Q_{j_{4}} \! \left(x+\frac{q_{y} l_{B}^{2}}{2}\right) .
\end{equation}

Each of these expressions is derived in detail in Sec.~S2 as well, and a general expression for the elementary form factors follows in Sec.~S4, in the Supplemental Material. We will approximate $J_{\substack{n_{1} \xi_{1} \\ n_{4} \xi_{4}}}^{T_{z}}(\bm{q}) = J_{\substack{n_{1} \xi_{1} \\ n_{4} \xi_{1}}}^{T_{z}}(\bm{q}) \delta_{\xi_1 \xi_4}$ in the following because the $\xi=+$ and $\xi=-$ LLLs have very little overlap. We have broken the full form factors into the layer-projected form factors $J_{\substack{n_{1} \xi_{1} \\ n_{4} \xi_{4}}}^{T_{z}}(\bm{q})$ because each LL wavefunction has support on both layers. This splitting between layers is important because it delocalizes charge and weakens interactions.

The Coulomb interaction is now expressed as
\begin{equation}\begin{aligned} \hat{V} \! = \! \sfrac{1}{2} & \!\!\!\!\! \sum_{\substack{n_j X_j \\ j=1,2,3,4}} \!\! \sum_{\substack{\xi \xi^\prime \sigma \sigma^{\prime}}} \!\! \sum_{\bm{q}} \! \sum_{T_z T_{z^\prime}} c^+_{n_1\xi\sigma\!X_1} \! c^+_{n_2\xi^\prime \sigma^\prime\!X_2} \! c_{n_3\xi^\prime \sigma^\prime\!X_3} c_{n_4\xi \sigma\!X_4} \\ & \times V_{T_{z} T_{z}^{\prime}}(\bm{q}) \left( \delta_{X_{4},X_{1}-q_{y} l_{B}^{2}} e^{i\frac{q_{x}}{2}\left(X_{1}+X_{4}\right)} J_{\substack{n_{1} \xi \\ n_{4} \xi}}^{T_{z}}(\bm{q})\right) \\ & \quad \quad \times \left( \delta_{X_{3},X_{2}-q_{y} l_{B}^{2}} e^{i\frac{q_{x}}{2}\left(X_{2}+X_{3}\right)} J_{\substack{n_2 \xi^\prime \\ n_3 \xi^\prime}}^{T_{z}}(-\bm{q})\right) . \end{aligned}\end{equation}
%**Not compatible with vertical figures - needs widetext**
%\begin{equation}\begin{aligned} \hat{V} = \frac{1}{2} \sum_{\substack{n_j X_j \\ j=1,2,3,4}} \sum_{\substack{\xi \xi^\prime \\ \sigma \sigma^{\prime}}} \sum_{\bm{q}} \sum_{T_z T_{z^\prime}} & V_{T_{z} T_{z}^{\prime}}(\bm{q}) \left( \delta_{X_{4},X_{1}-q_{y} l_{B}^{2}} e^{i\frac{q_{x}}{2}\left(X_{1}+X_{4}\right)} J_{\substack{n_{1} \xi \\ n_{4} \xi}}^{T_{z}}(\bm{q})\right) \left( \delta_{X_{3},X_{2}-q_{y} l_{B}^{2}} e^{i\frac{q_{x}}{2}\left(X_{2}+X_{3}\right)} J_{\substack{n_2 \xi^\prime \\ n_3 \xi^\prime}}^{T_{z}}(-\bm{q})\right) \\ & \times c^+_{n_1 \xi \sigma X_1}c^+_{n_2 \xi^\prime \sigma^\prime X_2}c_{n_3 \xi^\prime \sigma^\prime X_3}c_{n_4 \xi \sigma X_4} \, . \end{aligned}\end{equation}

In the Hartree-Fock approximation, we replace
\begin{equation}\begin{aligned} & \frac{1}{2} c^+_{n_1 \xi_1 \sigma_1 X_1}c^+_{n_2 \xi_2 \sigma_2 X_2}c_{n_3 \xi_3 \sigma_3 X_3}c_{n_4 \xi_4 \sigma_4 X_4} \\ & \to \langle c^+_{n_1 \xi_1 \sigma_1 X_1}c_{n_4 \xi_4 \sigma_4 X_4} \rangle c^+_{n_2 \xi_2 \sigma_2 X_2}c_{n_3 \xi_3 \sigma_3 X_3} \\ & \quad - \langle c^+_{n_1 \xi_1 \sigma_1 X_1}c_{n_3 \xi_3 \sigma_3 X_3} \rangle c^+_{n_2 \xi_2 \sigma_2 X_2}c_{n_4 \xi_4 \sigma_4 X_4} \\ & \hat{V} \to \hat{V}_D - \hat{V}_X \end{aligned}\end{equation}
%\begin{equation}\begin{aligned} & \frac{1}{2} \, c^+_{n_1 \xi_1 \sigma_1 X_1}c^+_{n_2 \xi_2 \sigma_2 X_2}c_{n_3 \xi_3 \sigma_3 X_3}c_{n_4 \xi_4 \sigma_4 X_4} \\ & \quad \quad \rightarrow \langle c^+_{n_1 \xi_1 \sigma_1 X_1}c_{n_4 \xi_4 \sigma_4 X_4} \rangle c^+_{n_2 \xi_2 \sigma_2 X_2}c_{n_3 \xi_3 \sigma_3 X_3} - \langle c^+_{n_1 \xi_1 \sigma_1 X_1}c_{n_3 \xi_3 \sigma_3 X_3} \rangle c^+_{n_2 \xi_2 \sigma_2 X_2}c_{n_4 \xi_4 \sigma_4 X_4} \, , \\ & \quad \quad \hat{V} \rightarrow \hat{V}_D - \hat{V}_X \end{aligned}\end{equation}\end{widetext}
where $\hat{V}_D$ is the direct term and $\hat{V}_X$ is the exchange term. We then define the density operators
\begin{equation}
\rho_{n n^{\prime}}^{\xi \xi^{\prime} \sigma \sigma^{\prime}} \! (\bm{q})=\frac{1}{N_\Phi} \! \sum_{X X^{\prime}} \delta_{X^{\prime} \! , X \! - \! q_{y} l_B^2} e^{i \frac{q_{x}}{2} \! \left(X \! + \! X^\prime\right)} c_{n \xi \sigma X}^{+} c_{n^{\prime} \xi^{\prime} \sigma^{\prime} X^{\prime}}
\end{equation}
which give a natural basis for studying the system and interactions. In terms of the density operators, the direct term is written as
\begin{equation}
\hat{V}_D = N_\Phi \! \sum_{\bm{q}} \!\! \sum_{\substack{n_1 n_2 n_3 n_4 \\ \xi \xi^\prime \sigma \sigma^\prime}} \!\!\!\!\!\! H^{\xi \xi^\prime}_{n_1 n_2 n_3 n_4}(\bm{q}) \! \left< \rho_{n_{1} n_{2}}^{\xi \xi \sigma \sigma} \! (\bm{q}) \right> \! \rho_{n_{3} n_{4}}^{\xi^{\prime} \xi^{\prime} \sigma^{\prime} \sigma^{\prime}} \! (-\bm{q}) \, ,
\end{equation}
\begin{equation}
H^{\xi \xi^\prime}_{n_1 n_2 n_3 n_4}(\bm{q}) = N_\Phi \sum_{T_z T_{z^\prime}} V_{T_{z}T_{z}^{\prime}}(\bm{q})
J_{\substack{n_{1} \xi \\ n_{2} \xi}}^{T_{z}}(\bm{q})
J_{\substack{n_3 \xi^\prime \\ n_4 \xi^\prime}}^{T_z^\prime}(-\bm{q}) \, ,
\end{equation}
and the exchange term is written as
\begin{equation}
\hat{V}_X = N_\Phi \! \sum_{\bm{q}} \!\! \sum_{\substack{n_1 n_2 n_3 n_4 \\ \xi \xi^\prime \sigma \sigma^\prime}} \!\!\!\!\!\! X_{n_{1} n_{4} n_{3} n_{2}}^{\xi \xi^{\prime}} \! (\bm{q}) \! \left<\rho_{n_{1} n_{2}}^{\xi \xi^\prime \sigma \sigma^\prime} \! (\bm{q})\right> \! \rho_{n_{3} n_{4}}^{\xi^{\prime} \xi \sigma^{\prime} \sigma} \! (-\bm{q}) \, ,
\end{equation}
\begin{equation} \label{equation_fockint}
X_{n_{1} n_{4} n_{3} n_{2}}^{\xi \xi^{\prime}}(\bm{q}) = \sum_{T_z T_{z^\prime}} \int \frac{d^2\bm{p} \, l_B^2}{2\pi} H^{\xi \xi^\prime}_{n_1 n_4 n_3 n_2}(\bm{q}) e^{i \bm{q} l_{B} \times \bm{p} l_{B}} \, .
\end{equation}
Since the exchange integral has the symmetries $X_{k l m n}^{\xi \xi^{\prime}}(\bm{q})=X_{k l m n}^{\xi^{\prime} \xi}(\bm{q})$ and $X_{k l m n}^{++}(\bm{q})=X_{k l m n}^{--}(\bm{q})$, we can write all exchange integrals in terms of the two $X_{k l m n}^{++}(\bm{q})$ and $X_{k l m n}^{+-}(\bm{q})$. Further information on the properties and calculation of the exchange integrals is given in Sec.~S4 in the Supplemental Material.

We will focus only on spatially-uniform solutions and find the lowest energy state in this subspace. This can be later compared with possible states that break translational symmetry. In other words, we assume $\left<\rho_{n_{1} n_{2}}^{\xi \xi \sigma \sigma}(\bm{q})\right>=0$ if $\bm{q}\ne0$. (After making this assumption this we will generally drop the argument $(0)$, e.g. write $X_{n_{1} n_{4} n_{3} n_{2}}$ instead of $X_{n_{1} n_{4} n_{3} n_{2}}(0)$.)

The direct term in the Coulomb interaction is a Coulomb blockade that penalizes layer polarization. In the present case of uniform states, it takes the form of a capacitive correction, as noted in previous work \cite{cote_orbital_2010,lambert_quantum_2013,lambert_ferro-aimants_2013,knothe_phase_2016,hunt_direct_2017}. We find that, up to a constant for fixed total filling $\nu$,
\begin{equation} \label{VD_formula}
\hat{V}_{D} = - N_\Phi \Delta V \sum_{n \xi \sigma}\left(\nu_{2} \frac{1+\xi \Pi_{n}}{2}+\nu_{1} \frac{1-\xi \Pi_{n}}{2}\right) \rho_{n n}^{\xi \xi \sigma \sigma} \, ,
\end{equation}
where we have defined the upper and lower layer occupations by
\begin{equation}
\nu_{T_z} = \frac{1}{2}\Big(\tilde{\nu} - (-1)^{T_z} \sum_{n \xi \sigma}\left\langle\rho_{n n}^{\xi \xi \sigma \sigma}\right\rangle \xi \Pi_{n}\Big) ,
\end{equation}
with $\tilde{\nu}=\nu+4$ being the number of filled LLLs, and
\begin{equation}\begin{aligned} \label{equation_capintegral}
\Delta V = \frac{\alpha}{l_{B}} \int d z \int d z^{\prime} & \Bigg( \left|z^{\prime}-z+d\right| - \left|z^{\prime}-z\right| \\ & \quad - \frac{d(2z+d)}{2 D}\Bigg) P(z) P\left(z^{\prime}\right) \, .
\end{aligned}\end{equation}

Hereafter we will frequently refer to this simply as the Coulomb blockade. In the limits $P(z)\to\delta (z)$, $D\to\infty$ this reproduces the result of previous work, $\Delta V = \alpha \frac{d}{l_B}$ \cite{cote_orbital_2010,cote_biased_2011,lambert_quantum_2013,lambert_ferro-aimants_2013,knothe_phase_2016,hunt_direct_2017}. Hence, $\Delta V$ can also be written in terms of an effective layer separation $d^{CB}_{eff}$ defined by ${\Delta V = \alpha \frac{d^{CB}_{eff}}{l_B}}$, and we find that the extend of the p\textsubscript{z} orbitals weakens the Coulomb blockade: ${d^{CB}_{eff}<d}$. This is shown in Fig.~S3, and the derivation of these equations from the direct term is given in Sec.~S3, in the Supplemental Material.

As pointed out by Shizuya \cite{shizuya_structure_2012}, exchange interactions with the "Dirac sea" of occupied LLs lower the energy of the $n=1$ orbitals relative to $n=0$:
\begin{equation}
\hat{V}_{DS} = \frac{1}{2} N_\Phi \left(X_{1111}^{++}-X_{0000}^{++}\right)\frac{1}{2}\left(1-\lambda_z\right) \, ,
\end{equation}
where $\lambda_z^o$ is a Pauli matrix acting on the orbital space $\{0,1\}$. This exactly compensates for the difference in exchange energy for fully occupied $n=0$ LLLs compared to $n=1$. Ref.~\cite{shizuya_structure_2012} also indicates that the direct interaction with the Dirac sea screens the bias. Because rescaling bias exclusively affects the valley gap, it does not change the balance between any energy scales in a way that would change which ground states appear as a function of magnetic field and bias. Hence, we do not address the direct DS interaction, though it could be relevant for quantitative results in future studies. Adding this "Lamb-like shift" $\Delta_{Lamb}=\frac{1}{2}\left(X_{1111}^{++}-X_{0000}^{++}\right)$ to the noninteracting Hamiltonian, we have
\begin{equation}
\hat{H}_{ni+DS}=N_\Phi\sum_{n\xi\sigma}\left(E_{n\xi\sigma} + \Delta_{Lamb}\delta_{1n}\right) \rho^{\xi\xi\sigma\sigma}_{nn} \, .
\end{equation}

The full HF Hamiltonian is then
\begin{equation} \label{HF_Hamiltonian}
\hat{H}_{HF}=\hat{H}_{ni+DS}+\hat{V}_D-\hat{V}_X \, 
\end{equation}

The Hamiltonian matrix element $\left(H_{HF}\right)_{\left(n\xi\sigma\right),\left(n^\prime\xi^\prime\sigma^\prime\right)}$ is the coefficient of the density operator $\rho_{nn^\prime}^{\xi\xi^\prime\sigma\sigma^\prime}$. Because the Hamiltonian for a spatially-uniform system is block diagonal in $X$, with $8$-dimensional blocks indexed by $n\xi\sigma$, the HF problem is reduced to an $8 \times 8$. If the filling factor is $\nu$, then $\tilde{\nu}=\nu+4$ LLLs are filled, so the many-body eigenstate is 
\begin{equation}
|\Psi\rangle=\prod_{X} \left( \prod_{j=1}^{\tilde{\nu}} \left(\sum_{n \xi \sigma} A_{n \xi \sigma}^{j} c_{n \xi \sigma X}^{+}\right) \right) |\varnothing\rangle
\end{equation}
where $A_{n \xi \sigma}^{j}$ are the coefficients of the $j$th eigenvector of the matrix $\left(H_{HF}\right)_{\left(n\xi\sigma\right),\left(n^\prime\xi^\prime\sigma^\prime\right)}$, ordered by energy with the lowest first. The density matrix elements are given by
\begin{equation}
\left\langle\rho_{n n^{\prime}}^{\xi \xi^{\prime} \sigma \sigma^{\prime}}\right\rangle=\sum_{j=1}^{\tilde{v}}\left(A_{n \xi \sigma}^{j}\right)^{*} A_{n^{\prime} \xi^{\prime} \sigma^{\prime}}^{j} \, .
\end{equation}
In the self-consistent approach to solving the HF problem, these density matrix elements are then used to generate a new HF Hamiltonian, and the cycle is iterated until a self-consistent solution has been found. When the solution is found, we refer to it as an LLSD (Landau level Slater determinant) or LLC (Landau level coherent) state if it is given by a diagonal or non-diagonal density matrix, respectively. LLC states can be thought of as the result of LLSD states mixing via coherent superpositions.

It is very useful to calculate the average energy per particle as well. If there are $N_e$ electrons in the LLLs, then since $\tilde{\nu} = \frac{N_e}{N_\Phi}$, up to a constant we have
\begin{widetext}\begin{equation}\begin{aligned} \label{systemenergy} \frac{E_{HF}}{N_e} = & \frac{1}{\tilde{\nu}} \Bigg( \sum_{n\xi\sigma}\left(E_{n\xi\sigma} + \frac{1}{2}\left(X_{1111}^{++}-X_{0000}^{++}\right)\delta_{1n}\right) \langle\rho^{\xi\xi\sigma\sigma}_{nn}\rangle \\ & - \Delta V \nu_1 \nu_2  - \frac{1}{2}\sum_{\substack{n_1 n_2 n_3 n_4 \\ \xi \xi^\prime \sigma \sigma^\prime}} X_{n_{1} n_{4} n_{3} n_{2}}^{\xi \xi^{\prime}} \left<\rho_{n_{1} n_{2}}^{\xi \xi^\prime \sigma \sigma^\prime}\right> \left<\rho_{n_{3} n_{4}}^{\xi^{\prime} \xi \sigma^{\prime} \sigma}\right>\Bigg) \, . \end{aligned}\end{equation}\end{widetext}
%\begin{widetext}\begin{equation}\begin{aligned} \label{systemenergy} \frac{E_{HF}}{N_e} = \frac{1}{\tilde{\nu}} \Bigg( \sum_{n\xi\sigma}\left(E_{n\xi\sigma} \! + \sfrac{1}{2}\left(X_{1111}^{++}\!-\!X_{0000}^{++}\right)\delta_{1n}\right) \! \langle\rho^{\xi\xi\sigma\sigma}_{nn}\rangle - \Delta V \nu_1 \nu_2  - \sfrac{1}{2} \!\!\!\!\! \sum_{\substack{n_1 n_2 n_3 n_4 \\ \xi \xi^\prime \sigma \sigma^\prime}} \!\!\!\!\! X_{n_{1} n_{4} n_{3} n_{2}}^{\xi \xi^{\prime}} \! \left<\rho_{n_{1} n_{2}}^{\xi \xi^\prime \sigma \sigma^\prime}\right> \! \left<\rho_{n_{3} n_{4}}^{\xi^{\prime} \xi \sigma^{\prime} \sigma}\right>\Bigg) \, . \end{aligned}\end{equation}\end{widetext}
This is the energy that the correct many-body solution will minimize. $E_{n\xi\sigma}$ is the noninteracting energy given by Eq.~(\ref{noninteractingenergy}), $\frac{1}{2}\left(X_{1111}^{++}-X_{0000}^{++}\right)$ is the Lamb-like shift \cite{shizuya_structure_2012}, $\Delta V$ is the Coulomb blockade given by Eq.~(\ref{equation_capintegral}), and $X_{n_{1} n_{4} n_{3} n_{2}}^{\xi \xi^{\prime}}$ are the exchange matrix elements appearing in Eq.~(\ref{equation_fockint}). By comparing the energies of LLSD states and mixing them into LLC states near their crossings, we can also minimize energy as a function of the parameter or parameters that describe the LLC state's superposition. This method allows us to find the ground state analytically, and is the approach we use in this work.

\section{Results} \label{results}

\subsection{Phase diagrams} \label{PDs}

\begin{figure*}
    \centering
    \includegraphics[width=16.1cm]{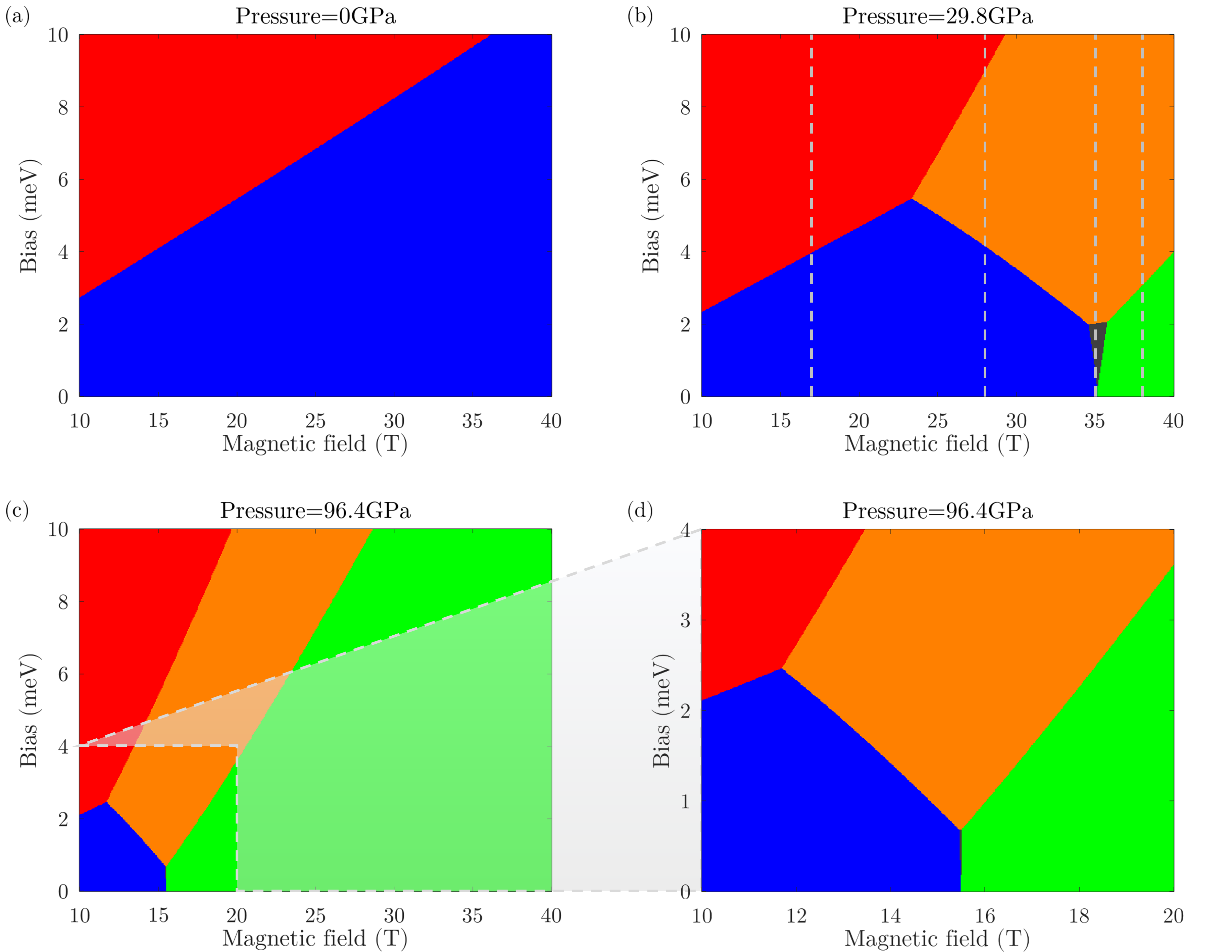}
    
    \vspace{0.5cm}
    
    \includegraphics[width=12cm]{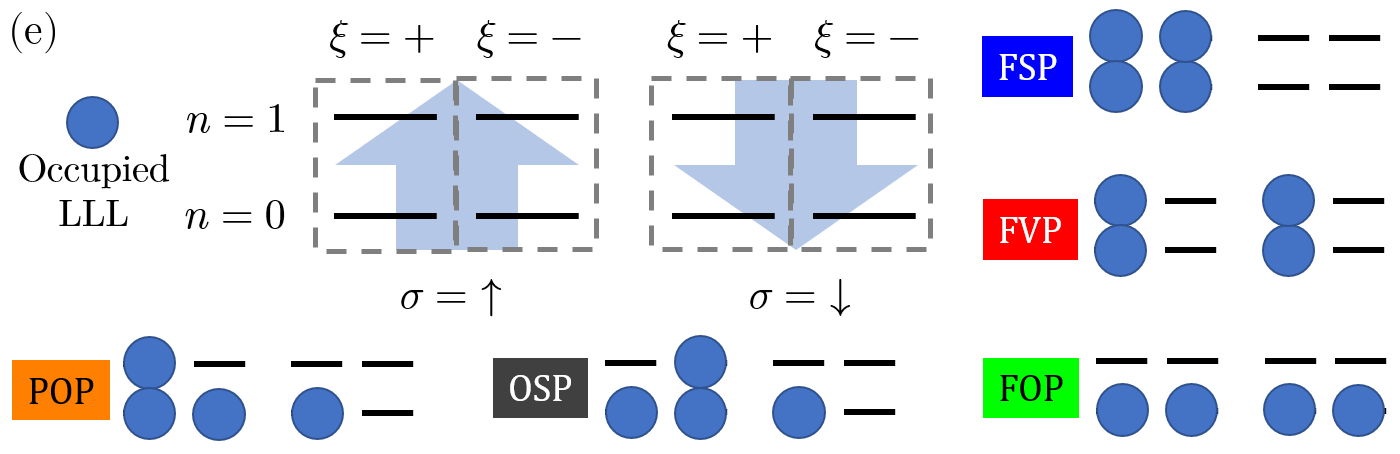}
    \caption{Phase diagrams for (a) zero, (b) intermediate, and (c) high pressures are given; (d) magnifies (c). Five LLSD and no LLC states appear. Notice that applied pressure literally compresses the phase diagram so that all transitions occur at progressively lower fields, as explained in the text, but that the overall topology remains unchanged. The dashed lines on $\text{P}=29.8~\text{GPa}$ correspond to the traces in Fig.~\ref{fig:energycrossings}. (e) A schematic of the dot-diagram depiction of states devised by Lambert and C\^{o}t\'{e} \cite{lambert_quantum_2013,lambert_ferro-aimants_2013}, and the dot-diagram representation of the different states appearing in our phase diagrams.}
    \label{fig:phaseanddotdiagrams}
\end{figure*}

Using the HF calculations presented above, we obtain the ground state for different values of magnetic field, bias and pressure. For fixed pressure, we draw this as a phase diagram whose different regions represent characteristic ground states as a function of magnetic field and bias. The diagrams evolve continuously with pressure, and we give results for the representative cases of zero pressure, an intermediate pressure of $29.8~\text{GPa}$, and a high pressure of $96.4~\text{GPa}$. Since pressure changes the scale of the bias versus magnetic field phase diagrams but does not change their topology, in the following discussion we will use the intermediate pressure case at $29.8~\text{GPa}$ to illustrate.

\begin{figure*}
    \centering
    \includegraphics[width=8cm]{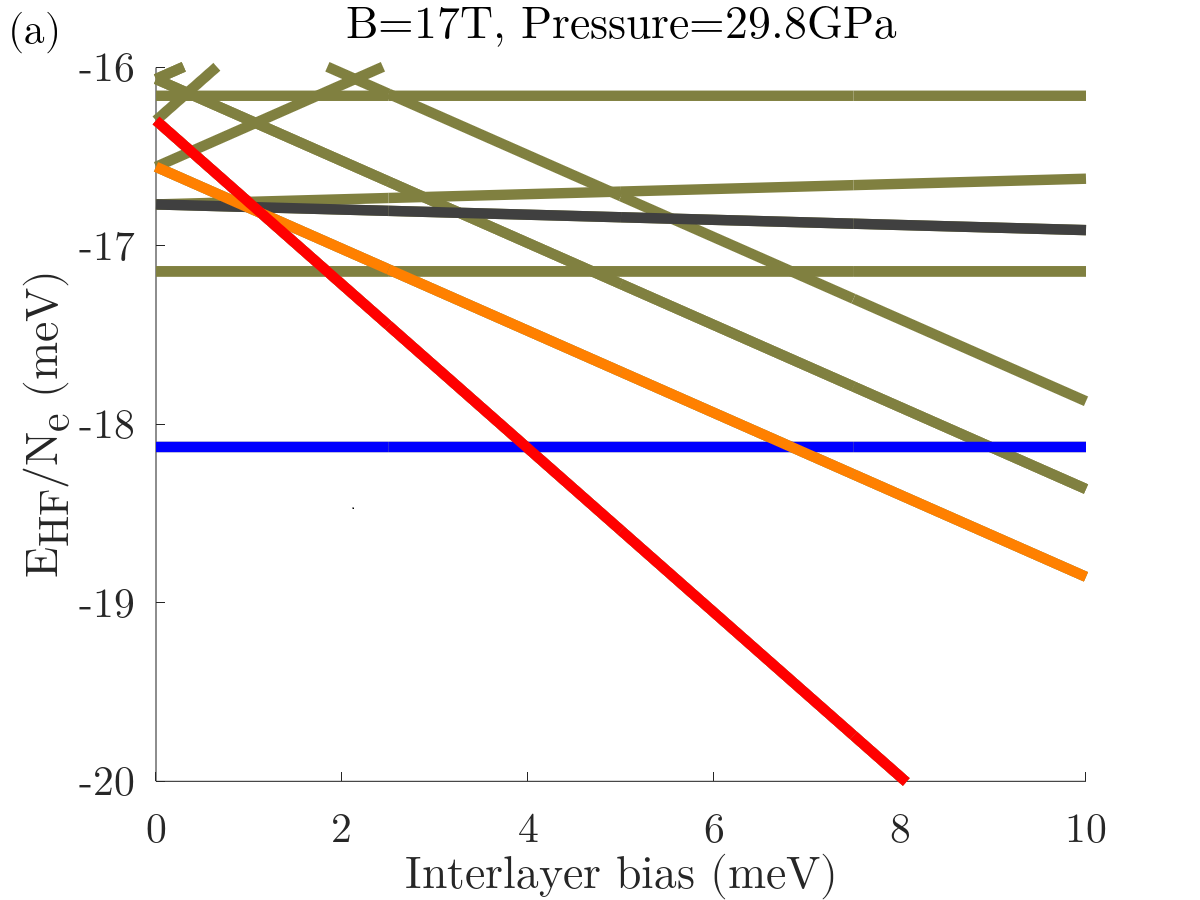}
    \includegraphics[width=8cm]{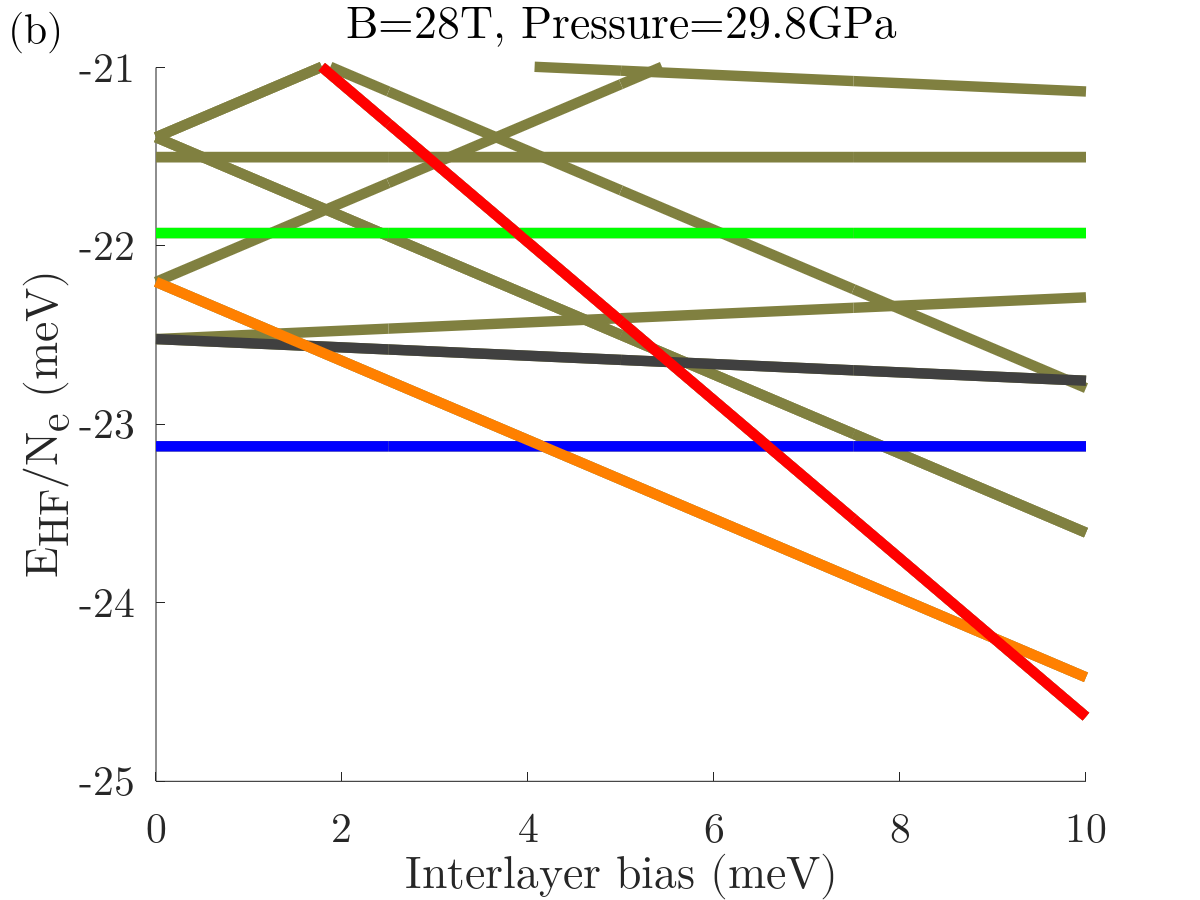}
    
    \vspace{0.5cm}
    
    \includegraphics[width=8cm]{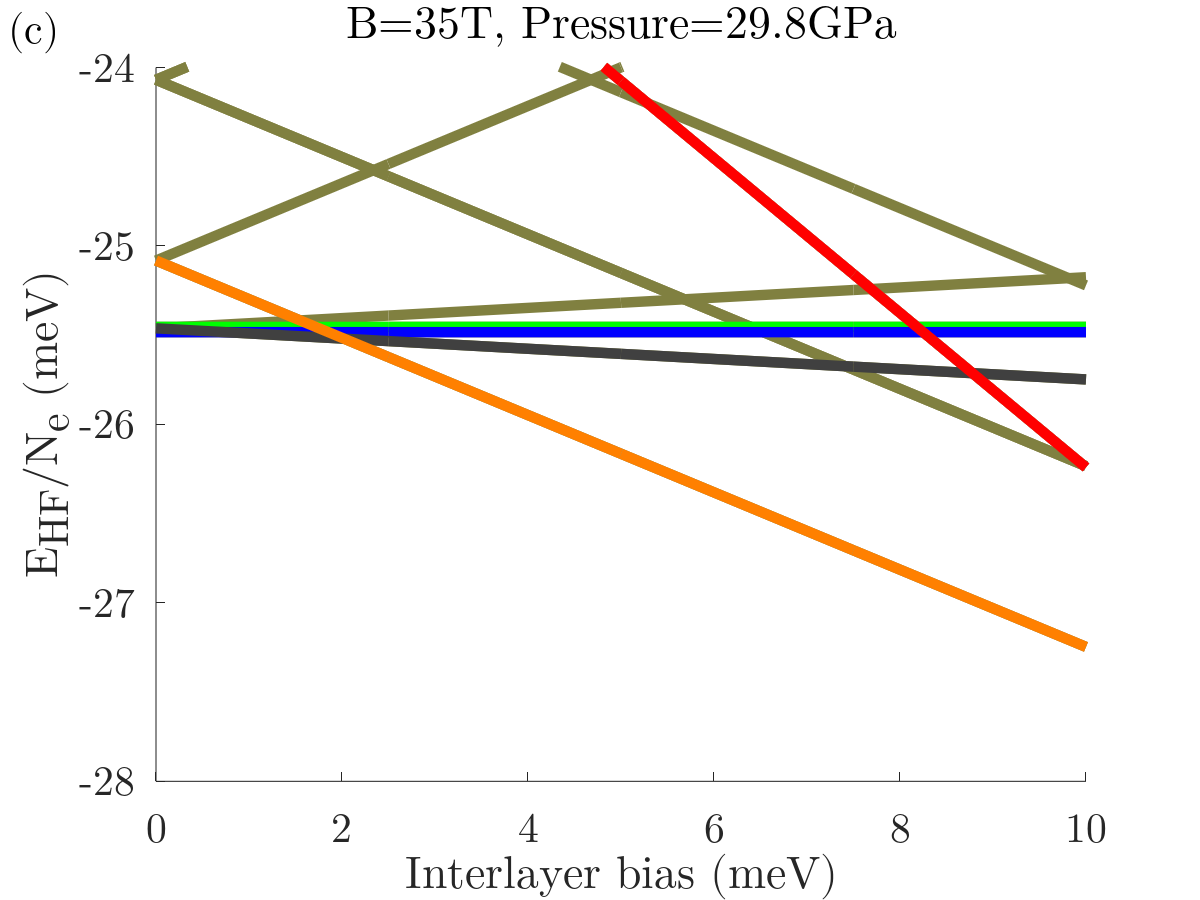}
    \includegraphics[width=8cm]{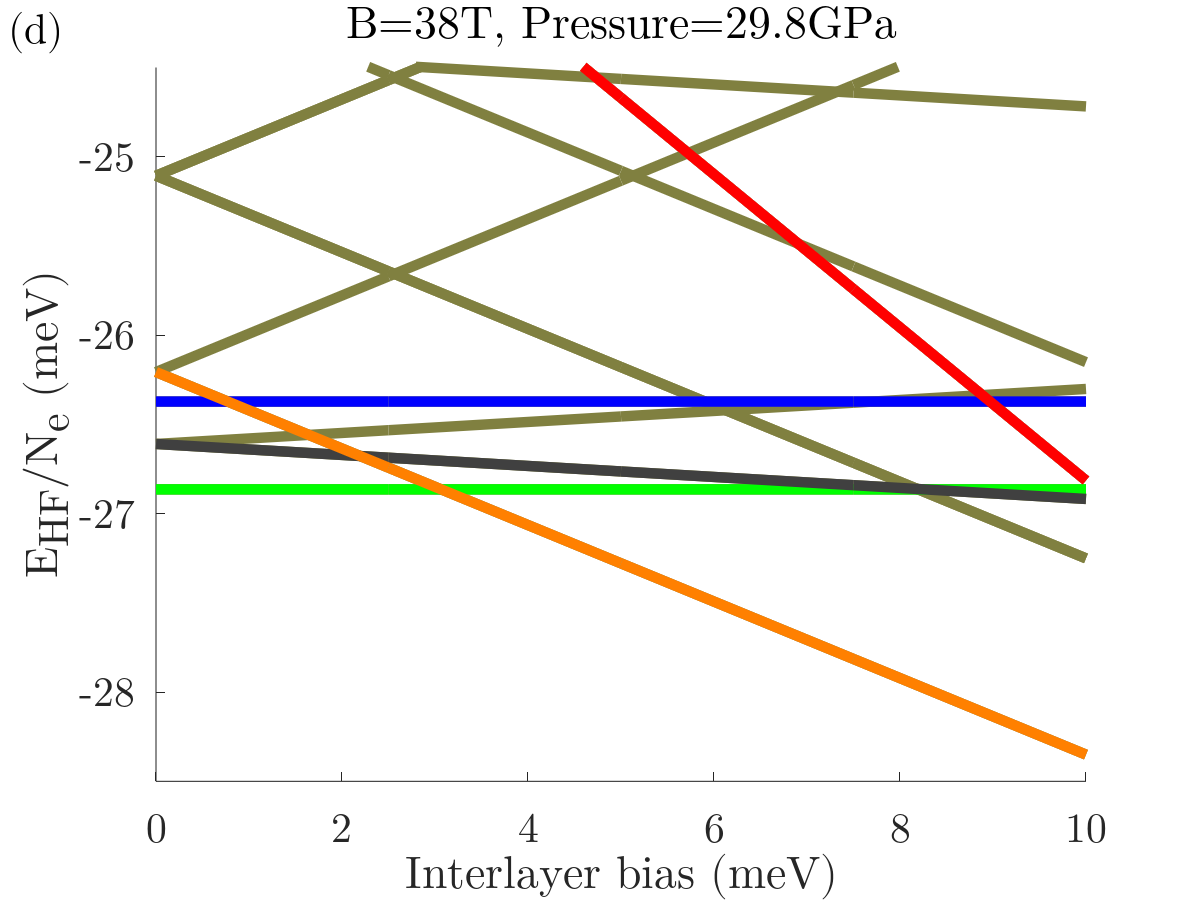}
    
    \vspace{0.5cm}
    
    \caption{The state energies are plotted here as a function of bias for the representative magnetic field linecuts in Fig.~\ref{fig:phaseanddotdiagrams}(b). Pressure is fixed at $\text{P}=29.8~\text{GPa}$ because it does not change the topology of the phase diagram. LLSD states use the same color scheme as in Fig.~\ref{fig:phaseanddotdiagrams}, plus olive green for the numerous excited states.}
    \label{fig:energycrossings}
\end{figure*}

For low magnetic field and bias, the ground state is the fully spin-polarized (FSP) state, which is layer unpolarized and is drawn in blue in Fig.~\ref{fig:phaseanddotdiagrams}. Further information on this state, and all others, is given in Sec.~\ref{StateConfigsAndDescriptions}, and they are represented pictorially in Fig.~\ref{fig:phaseanddotdiagrams}(e). As the bias is increased while the magnetic field is kept low, the FSP state is replaced by the fully valley-polarized (FVP) state, which is fully layer-polarized and drawn in red. This phase transition occurs when the bias is strong enough to overcome the Coulomb blockade energy. The situation described here can be seen by following the first linecut at $B=17~\text{T}$ in Fig.~\ref{fig:phaseanddotdiagrams}(b). To give a more complete picture of the evolution of the ground state with bias, these states' energies and those of excited LLSD states are plotted along in Fig.~\ref{fig:energycrossings}(a) along the same linecut.

This transition demonstrates a general trend: as bias increases and overcomes the Coulomb blockade, states with lower layer polarization are replaced by states with higher layer polarization, Indeed, this pattern persists throughout the phase diagram for all magnetic fields. Consider increasing the magnetic field to $B=28~\text{T}$, shown by the second linecut in Fig.~\ref{fig:phaseanddotdiagrams}(b) and by Fig.~\ref{fig:energycrossings}(b). At this field the transition between the FSP and FVP states no longer occurs directly but has an intermediate state, which is known as the partially orbitally polarized (POP) state and is drawn in orange. This state is partially polarized in all three degrees of freedom; in particular, it is partially valley- and hence partially layer-polarized. In this way the trend continues: as bias overcomes the Coulomb blockade, each successive ground state has greater layer polarization.

The appearance of the POP state is driven by a transition in energy scale dominance that occurs as the magnetic field is increased, much like the transition between bias and the Coulomb blockade energy which happened as the bias was increased, This transition occurs as the orbital gap $E_1-E_0$, which scales as $B$, overcomes the exchange energy, which scales as $\sqrt{B}$. This transition induces the general trend of spin and valley polarization being traded for orbital polarization as the magnetic field increases. This trend follows from the more precise rule that exchange favors states with same-spin, same-valley pairs of LLs occupied, which can be read from Fig.~\ref{fig:phaseanddotdiagrams}(e) as pairs of vertically aligned dots. These pairs are favorable because only states of the same spin and valley have nonzero overlap, so that their energy is reduced by exchange.

This trend explains the appearance of two new orbitally polarized states as the magnetic field is increased further. At $B=35~\text{T}$, shown by the third linecut in Fig.~\ref{fig:phaseanddotdiagrams}(b) and by Fig.~\ref{fig:energycrossings}(c), the low-bias ground state is the orbitally and spin-polarized (OSP) state, drawn in charcoal. It is partially orbitally and spin-polarized, but valley-unpolarized and hence minimally layer-polarized, so that it replaces the FSP state as the bottom rung of the ladder of increasingly layer-polarized states. At $B=38~\text{T}$, shown by the fourth linecut in Fig.~\ref{fig:phaseanddotdiagrams}(b) and by Fig.~\ref{fig:energycrossings}(d), the fully orbitally polarized (FOP) state, drawn in green, likewise replaces the OSP state as the minimally layer-polarized ground state at low bias. The FOP state is fully orbitally polarized and has no valley or spin polarization, so it is layer-unpolarized.

Now that we understand the energy scales driving the phase transitions in Fig.~\ref{fig:phaseanddotdiagrams}(b), it is simple to understand the changes in the phase diagram with pressure. As pressure increases, the orbital gap increases more steeply with magnetic field (see Fig.~S1 in the Supplemental Material) so that the transitions to orbitally polarized states occur at lower magnetic fields. Likewise, pressure decreases layer separation and thus weakens the Coulomb blockade so that transitions to layer-polarized states also occur at slightly lower bias. Hence, pressure literally compresses the phase diagram into a smaller region in the space of magnetic field and bias. In Fig.~\ref{fig:phaseanddotdiagrams}(a), the orbitally polarized states do not appear simply because the orbital gap does not grow quickly enough at zero pressure for these states to appear at an experimentally reasonable magnetic field.

These five LLSD states are all possible states that may appear in our model even at arbitrary magnetic field, bias and pressure. This is clear because increasing pressure beyond $96.4~\text{GPa}$ or the magnetic field beyond $40~\text{T}$ will simply further stabilize the FOP state, and increasing bias beyond $10~\text{mev}$ further stabilizes the FVP state, and the partially orbitally and layer-polarized POP state will always intermediate between them. It is interesting that no LLC states manifest as ground states in our results, because one would generally expect the interaction to mix LLSD states when they are close in energy - namely, at the phase boundaries in Fig.~\ref{fig:phaseanddotdiagrams} or the energy crossings in Fig.~\ref{fig:energycrossings}. This finding contrasts with previous results \cite{lambert_quantum_2013,lambert_ferro-aimants_2013,knothe_phase_2016,murthy_spin-valley_2017}, and we explain this discrepancy in Sec.~\ref{NoEntangledStates}.

\subsection{State configurations and descriptions} \label{StateConfigsAndDescriptions}

Of the five LLSD states we find in the phase diagram, three (the FSP, FVP and FOP states) are fully polarized in one degree of freedom while unpolarized in the other, and two (the POP and OSP states) have mixed partial polarization. We give their wavefunctions and brief characterizations below.

\subsubsection{Fully spin-polarized (FSP) state}

The FSP state is polarized only in spin and is written as
\begin{equation}
\left|\Psi_{FSP}\right\rangle = \prod_{X}\left(c_{0+\uparrow X}^{+} c_{0-\uparrow X}^{+} c_{1+\uparrow X}^{+} c_{1-\uparrow X}^{+}\right)|\varnothing\rangle \, .
\end{equation}
This state has no layer polarization and two same-spin, same-valley pairs. It is maximally favored by the Zeeman splitting, Coulomb blockade, and exchange interaction, so that it appears at low magnetic field and bias. Many previous studies \cite{kharitonov_canted_2012,lambert_quantum_2013,lambert_ferro-aimants_2013,knothe_phase_2016,hunt_direct_2017,murthy_spin-valley_2017} have also found this state.

\subsubsection{Fully valley-polarized (FVP) state}
The FVP state is polarized only in valley and is written as
\begin{equation}
\left|\Psi_{FVP}\right\rangle = \prod_{X}\left(c_{0+\uparrow X}^{+} c_{0+\downarrow X}^{+} c_{1+\uparrow X}^{+} c_{1+\downarrow X}^{+}\right)|\varnothing\rangle \, .
\end{equation}
This state has maximal layer polarization and two same-spin, same-valley pairs. It is maximally favored by the bias and exchange interaction, so that it is found at high bias and low magnetic field. Many previous studies \cite{kharitonov_canted_2012,lambert_quantum_2013,lambert_ferro-aimants_2013,knothe_phase_2016,hunt_direct_2017,murthy_spin-valley_2017} have also found this state.

\subsubsection{Fully orbitally polarized (FOP) state}

The FOP state is polarized only in orbital and is written as
\begin{equation}
\left|\Psi_{FOP}\right\rangle = \prod_{X}\left(c_{0+\uparrow X}^{+} c_{0+\downarrow X}^{+} c_{0-\uparrow X}^{+} c_{0-\downarrow X}^{+}\right)|\varnothing\rangle \, .
\end{equation}
This state has no layer polarization and no same-spin, same-valley pairs. It is maximally favored by the Coulomb blockade and orbital gap, so that it appears at low bias and high magnetic field. This state has not appeared in any previous studies because it requires a large orbital gap to manifest.

\subsubsection{Partially orbitally polarized (POP) state}

The POP state is partially polarized in all three indices, with 3-to-1 ratios of $n=0$ to $1$, $\xi=+$ to $-$, and $\sigma=\;\uparrow$ to $\downarrow$, and is written as
\begin{equation}
\left|\Psi_{POP}\right\rangle=\prod_{X}\left(c_{0+\uparrow X}^{+} c_{0+\downarrow X}^{+} c_{0-\uparrow X}^{+} c_{1+\uparrow X}^{+}\right)|\varnothing\rangle \, .
\end{equation}
This state has partial layer polarization and one same-spin, same-valley pair. It is partially favored by the bias, Zeeman splitting, Coulomb blockade, exchange interaction and orbital gap, so that it appears at intermediate bias and magnetic field. It is has been predicted and observed before \cite{murthy_spin-valley_2017,hunt_direct_2017,li_effective_2018}. 

\subsubsection{Orbitally and spin-polarized (OSP) state}

The OSP state is partially polarized in orbital and spin, but is unpolarized in valley, and is written as
\begin{equation}
\left|\Psi_{OSP}\right\rangle = \prod_{X}\left(c_{0+\uparrow X}^{+} c_{0+\downarrow X}^{+} c_{0-\uparrow X}^{+} c_{1-\uparrow X}^{+}\right)|\varnothing\rangle \, .
\end{equation}
This state has very small layer polarization and one same-spin, same-valley pair. (Layer polarization is nonzero due to unequal polarizations of the orbitals, $\Pi_0\ne\Pi_1$.) It is partially favored by the Zeeman splitting, exchange interaction and orbital gap, and maximally favored by the Coulomb blockade, so that it appears at low bias and intermediate magnetic field. It has neither been predicted nor observed in previous studies.

\subsection{Absence of LLC states} \label{NoEntangledStates}

The five states we observe are all LLSD states, despite the presence of interactions which in general mix the noninteracting eigenstates into LLC states. To explain the absence of LLC states, we focus on a particular example which has appeared in previous work \cite{lambert_quantum_2013,lambert_ferro-aimants_2013,knothe_phase_2016,murthy_spin-valley_2017}, the FSP-FVP state. This state continuously interpolates between the eponymous LLSD states with two spin-valley superpositions, and can be parametrized by two angles $\theta_0,\theta_1$ as
\begin{widetext}\begin{equation}
\left|\Psi_{FSP-FVP}\right\rangle=\prod_{X}\left(c_{0+\uparrow X}^{+} \! \left(\cos \theta_{0} c_{0-\uparrow X}^{+} \! + \! \sin \theta_{0} c_{0+\downarrow X}^{+}\right) \! c_{1+\uparrow X}^{+} \! \left(\cos \theta_{1} c_{1-\uparrow X}^{+} \! + \! \sin \theta_{1} c_{1+\downarrow X}^{+} \right)\right)|\varnothing\rangle \, .
\end{equation}
For this state not to appear at the phase boundary between the FSP and FVP states, it must be energetically unfavorable. We can verify this analytically by calculating the concavity of its energy, given in Eq.~(S32) in the Supplemental Material, with respect to the superposition parameters. To simplify this, we describe the superposition using a single parameter with the common \cite{knothe_phase_2016,murthy_spin-valley_2017} approximation $\theta_0=\theta_1\equiv\theta$. This approximation, that the transitions occur in tandem, is good because exchange couples the $n=0$ occupation to the $n=1$ occupation. (Only same-spin, same-valley pairs lower the energy of the state by exchange, so it is beneficial for the $n=0$ and $n=1$ superpositions to transfer from $-\uparrow$ to $+\downarrow$ together.) In this case, the concavity of the energy with respect to the $-\uparrow$ LLL occupation is
\begin{equation}\begin{aligned} \label{FSPFVP2ndderiv} \left(\frac{d}{d\cos^{2} \theta}\right)^{2}\nu\frac{E_{HF}}{N_e} & = 2\Delta V\left(\Pi_{0}+\Pi_{1}\right)^{2} \\ & \quad + 2\alpha\left( \left(X_{0000}^{+-}+2 X_{0110}^{+-}+X_{1111}^{+-}\right) - \left(X_{0000}^{++}+2 X_{0110}^{++}+X_{1111}^{++}\right) \right) \end{aligned}\end{equation}\end{widetext}
%\begin{equation} \label{FSPFVP2ndderiv} \left(\frac{d}{d\cos^{2} \theta}\right)^{2}\nu\frac{E_{HF}}{N_e} = 2\Delta V\left(\Pi_{0}+\Pi_{1}\right)^{2} + 2\alpha\left( \left(X_{0000}^{+-}+2 X_{0110}^{+-}+X_{1111}^{+-}\right) - \left(X_{0000}^{++}+2 X_{0110}^{++}+X_{1111}^{++}\right) \right) \end{equation}\end{widetext}

If this expression is negative, then the superposition is unfavorable and the energy is minimized at endpoints $\cos^{2} \theta=1$ or $0$, i.e., the FSP or FVP LLSD states. We find that it is negative for all magnetic fields and pressures in our model.

There are two contributions to the concavity in Eq.~(\ref{FSPFVP2ndderiv}): the Coulomb blockade ($\Delta V$) term, which is always $\ge 0$, and the exchange ($X_{k l m n}^{\xi \xi^{\prime}}$) term, which is always $\le 0$. Each exchange integral $X_{k l m n}^{\xi \xi^{\prime}}$ is positive, so the exchange term actually has a positive intervalley $+X_{k l m n}^{+-}$ and negative intravalley $-X_{k l m n}^{++}$ component. Recalling the valley-layer correspondence, however, the intervalley integrals are always smaller because the layer separation $d$ weakens interlayer interactions.

From this we see that the FSP-FVP LLC state will be unfavorable if the Coulomb blockade is too weak, or if the disparity between the intravalley and intervalley exchange integrals is too large. In our model, the spatial extent of the p\textsubscript{z} orbitals weakens the Coulomb blockade, and the layer-resolved form factors derived from exact diagonalization increase the intravalley-intervalley disparity. In contrast, if the extent of the p\textsubscript{z} orbitals is neglected and the valley-layer correspondence is assumed to be exact, then the FSP-FVP state appears as in previous work using similar interaction propagators \cite{lambert_quantum_2013,lambert_ferro-aimants_2013,knothe_phase_2016}.

We further compare the effects of the spatial extent of the p\textsubscript{z} orbitals, layer separation, gating, and layer-resolved form factors in Sec.~\ref{modeleffects} to explain their impacts on the model. We find that the layer-resolved exact diagonalization form factors are principally responsible for the absence of superpositions. Determining whether superpositions are favorable using energy concavity extends similarly to other pairs of LLSD states, and we use this method to confirm that no other LLC states appear in our model. We give the concavities of the relevant superpositions in Sec.~S5 and discuss the physical interpretation of each term in Sec.~S6 in the Supplemental Material.

\subsection{Effects of 3D p\textsubscript{z} orbitals, layer separation, gating, form factors, and heterostructures} \label{modeleffects}

\begin{figure*}
    \centering
    \includegraphics[width=4cm]{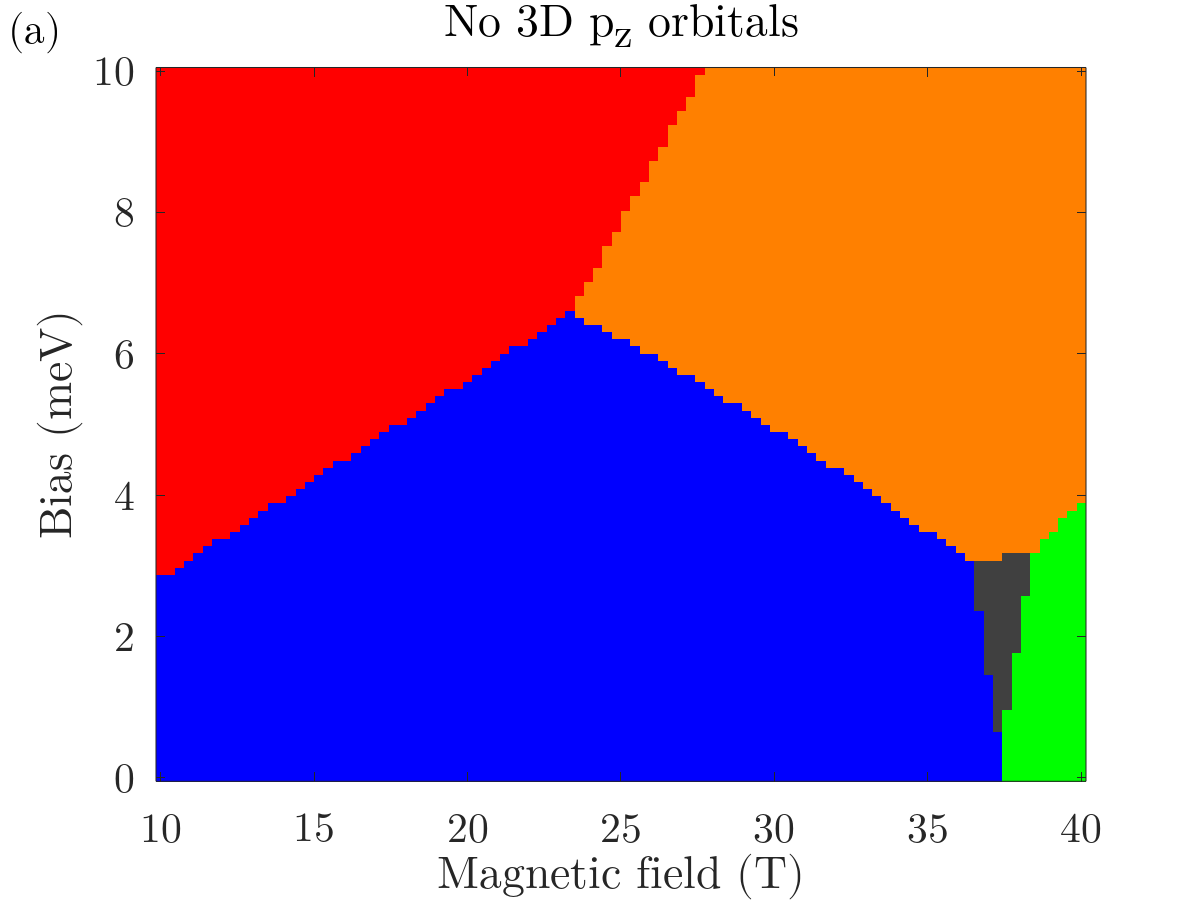}
    \includegraphics[width=4cm]{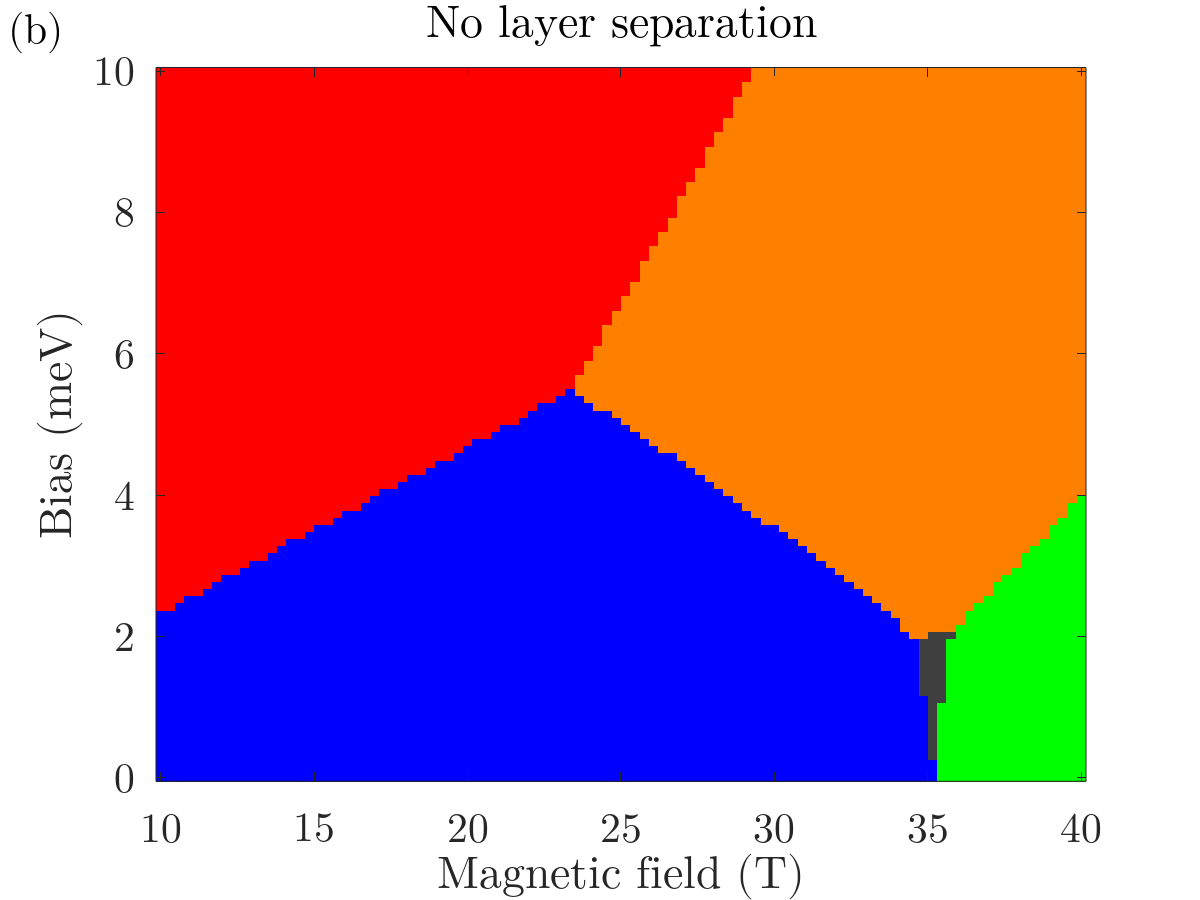}
    \includegraphics[width=4cm]{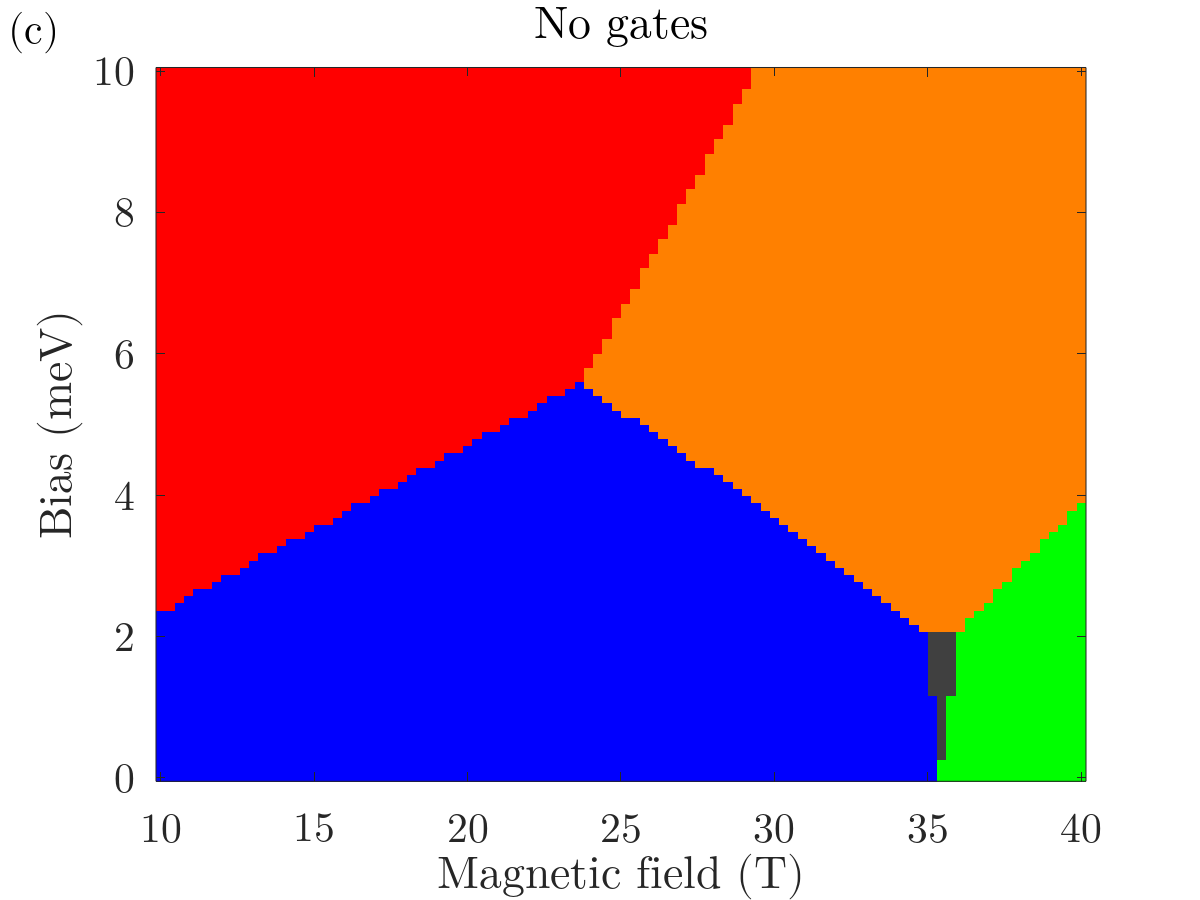}
    \includegraphics[width=4cm]{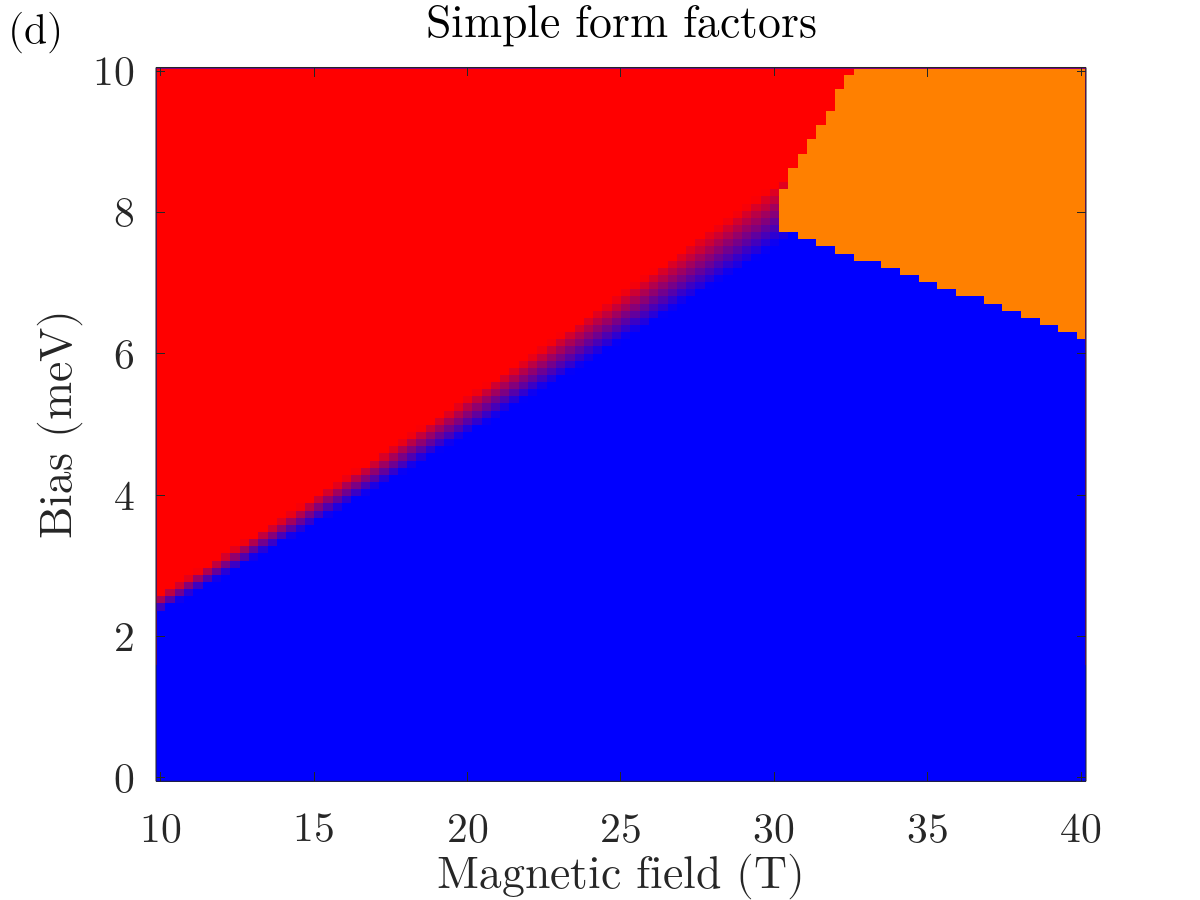}
    
    \vspace{0.5cm}
    
    \includegraphics[width=4cm]{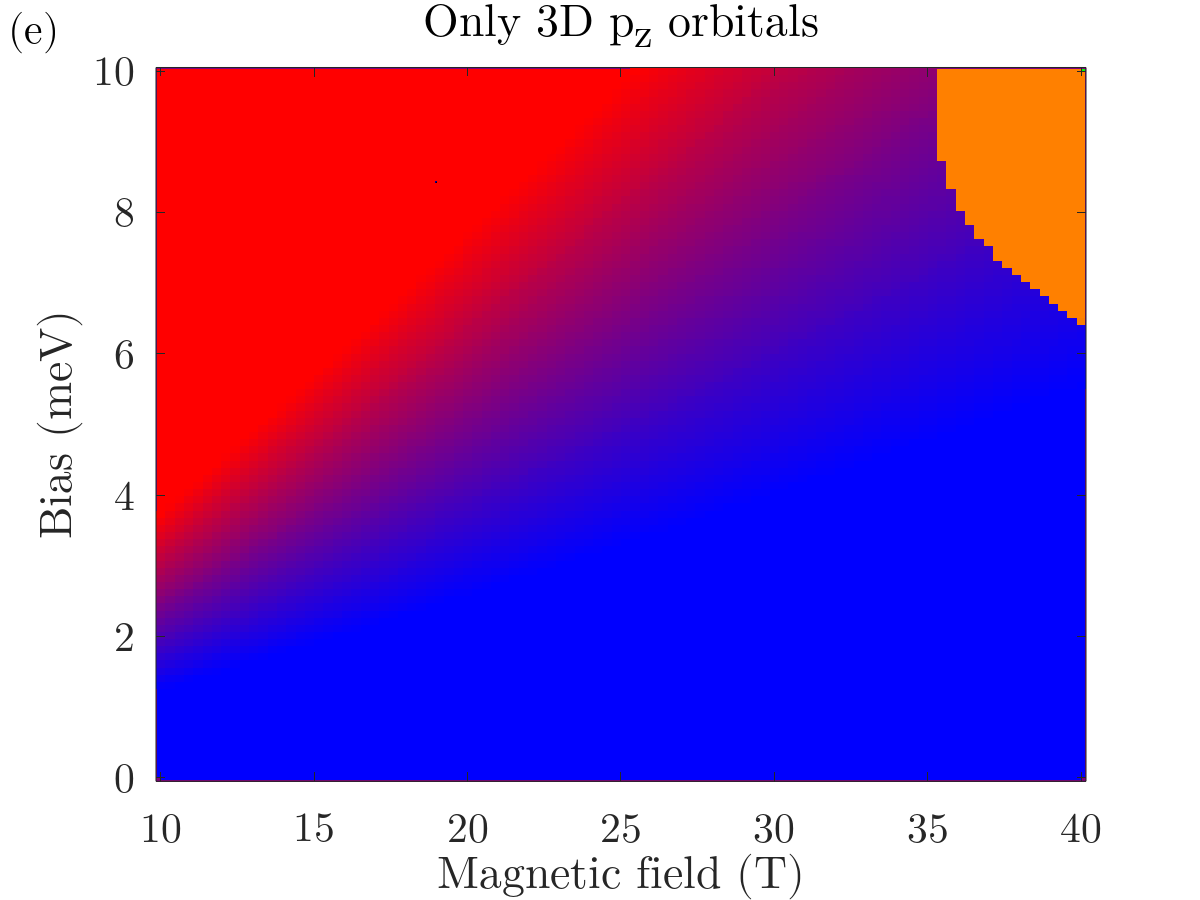}
    \includegraphics[width=4cm]{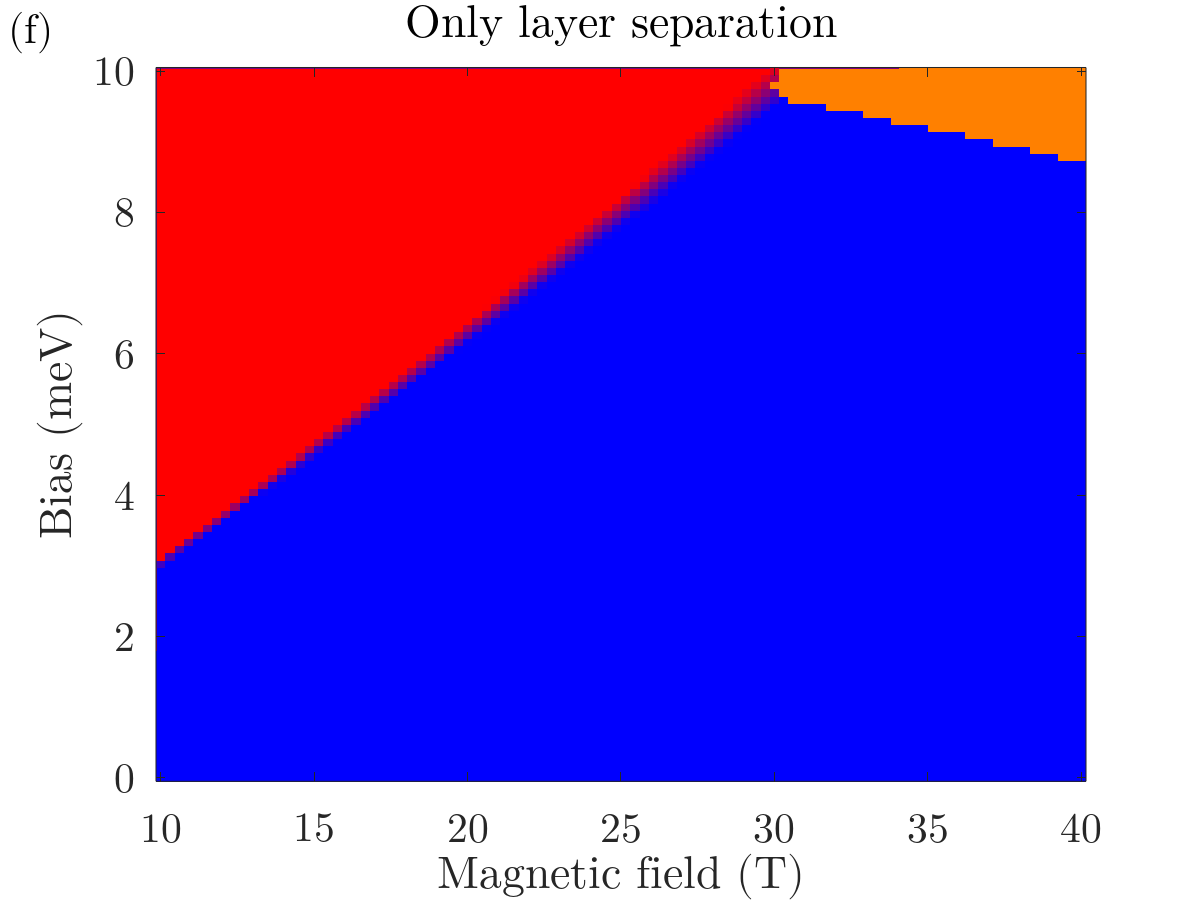}
    \includegraphics[width=4cm]{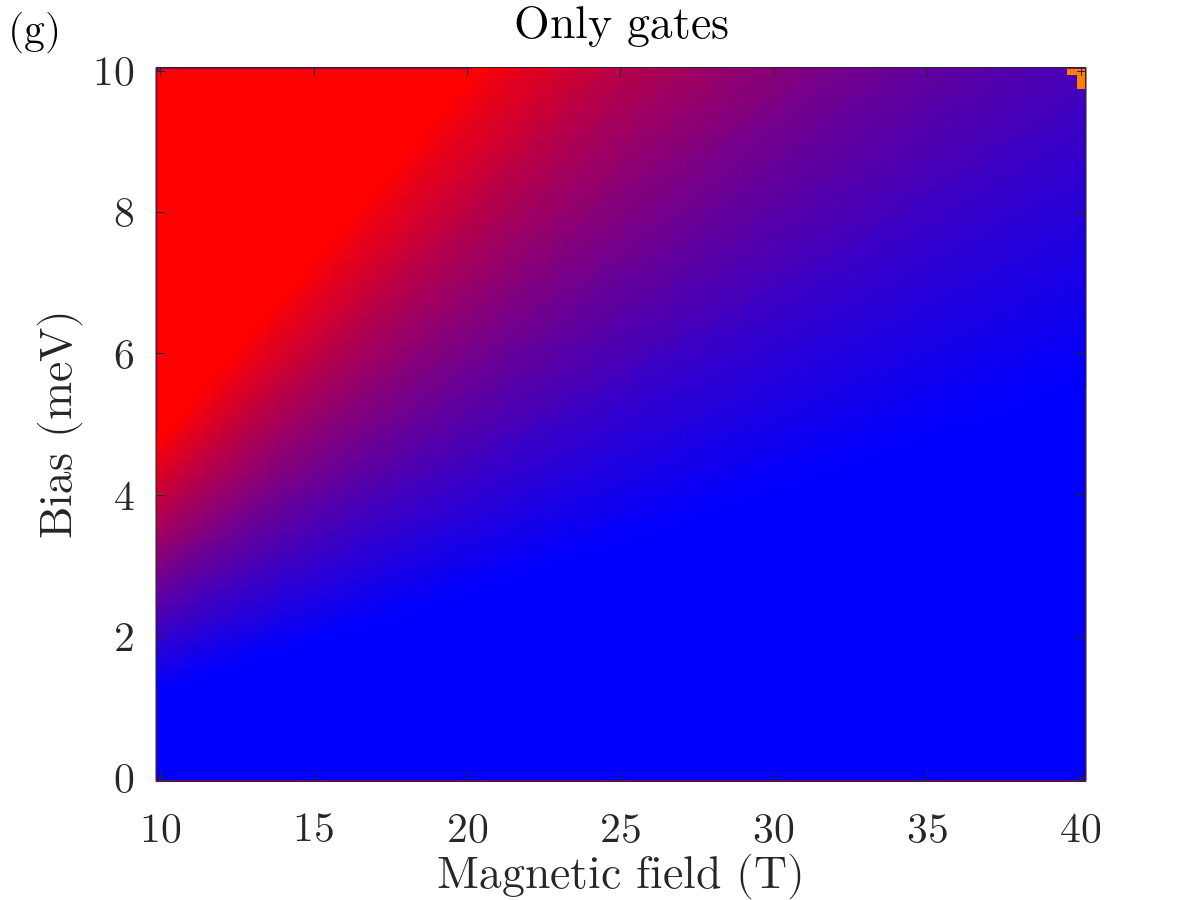}
    \includegraphics[width=4cm]{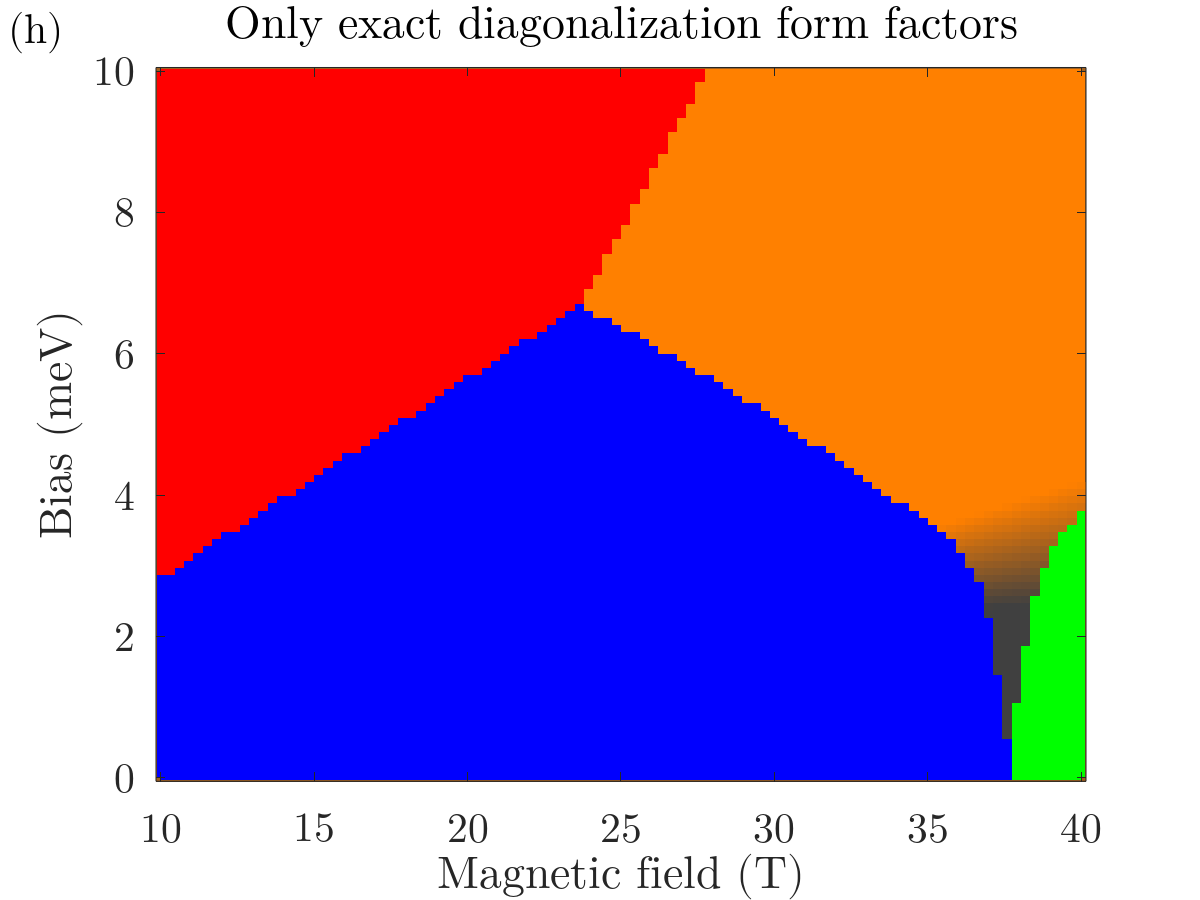}
    
    \caption{These phase diagrams show what our model would predict if we turned effects off one by one (upper row, a-d) or on one by one (lower row, e-f). By "turning off" the spatial extent of the p\textsubscript{z} orbitals, layer separation, gates, and exact diagonalization form factors, we respectively mean taking ${P(z)\to\delta\left(z\right)}$, ${d=0}$, ${D\to\infty}$, or ${c_{n+}^{Tj}=\delta_{T,B1}\delta_{j,n}}$ in Coulomb interaction calculations. If $d=0$ then $\Delta V = 0$, so when turning off layer separation, we only neglect it in exchange calculations, as in Ref.~\cite{hunt_direct_2017}. ${c_{n+}^{Tj}=\delta_{T,B1}\delta_{j,n}}$ is only used in the Coulomb interaction calculations; we always use exact diagonalization results for the polarizations $\Pi_n$ in the noninteracting energies $E_{n\xi\sigma}$, as in Refs.~\cite{cote_orbital_2010,cote_biased_2011,lambert_quantum_2013,lambert_ferro-aimants_2013,knothe_phase_2016}. Only two LLC states appear, the FSP-FVP and OSP-POP states.}
    \label{fig:modelcomparisonPDs}
\end{figure*}

Our model includes the spatial extent of the p\textsubscript{z} orbitals, layer separation, metallic gates, and layer-resolved form factors found by exact diagonalization. Since previous models have included some of these effects while neglecting others, it is worthwhile to explore their respective impacts on the phase diagram. To this end, in Fig.~\ref{fig:modelcomparisonPDs} we plot phase diagrams in which we have either neglected only one of these effects each, or included only one each, and we compare these to our main result in Fig.~\ref{fig:phaseanddotdiagrams}(b).

The spatial extent of the p\textsubscript{z} orbitals in general weakens the Coulomb interaction, as it spreads the electron density out vertically. We can see this effect by comparing Fig.~\ref{fig:phaseanddotdiagrams}(b), where it is included, to \ref{fig:modelcomparisonPDs}(a), where it is neglected. When it is neglected, the FSP state extends to higher magnetic field because the p\textsubscript{z} orbitals' extent weakens exchange (which favors the FSP state over orbitally polarized states) and to higher bias because the p\textsubscript{z} orbitals' extent weakens the Coulomb blockade (which favors the FSP state over the FVP state). Indeed, the effective layer separations plotted in Fig.~S3 in the Supplemental Material also shows that the p\textsubscript{z} orbitals' extent weakens both the Coulomb blockade and exchange interactions.

With regard to layer separation, when $d=0$, the Coulomb blockade integral ${\Delta V=0}$ regardless of other effects. When neglecting layer separation, therefore, we reuse the Coulomb blockade for $d \ne 0$ as in Ref.~\cite{hunt_direct_2017}, and only neglect the layer separation in exchange integral calculations, which greatly decreases the intervalley exchange integrals. These are the off-diagonal matrix elements in the Hamiltonian that mix LLLs of different valleys, producing avoided crossings that we see these as LLC states. Therefore, neglecting $d$ narrows the FSP-FVP LLC state in Fig.~\ref{fig:modelcomparisonPDs}(f). In Fig.~\ref{fig:modelcomparisonPDs}(b), there is no change in comparison to Fig.~\ref{fig:phaseanddotdiagrams}(b) because the LLC state does not appear.

At the large distance $D=20~\text{nm}$ in our model, the gates have only a minimal effect on the phase diagram. They slightly screen both the Coulomb blockade and the exchange interaction. Without the gates, the FSP and FVP states in Fig.~\ref{fig:modelcomparisonPDs}(c) take up a slightly larger region of phase space than with the gates in Fig.~\ref{fig:phaseanddotdiagrams}(b).

The layer-resolved exact diagonalization form factors, which physically describe the spatial distribution of the LL wavefunctions (cf. Eq.~(\ref{LLLbasisstates}) and the coefficients plotted in Fig.~S2 the Supplemental Material) split between the two layers, have the most substantial impacts. They not only weaken interactions more than any other effect but also render superpositions unfavorable. Weakening the Coulomb blockade brings phase transitions to lower bias and weakening exchange interaction brings phase transitions to lower magnetic fields, so that the phase diagram is scaled down. This is seen when comparing Fig.~\ref{fig:modelcomparisonPDs}(d) to the other upper and Fig.~\ref{fig:modelcomparisonPDs}(h) to the other lower row figures. The suppression of superpositions is evinced by the facts that \ref{fig:modelcomparisonPDs}(d) is the only the upper row figure to feature the FSP-FVP LLC state, and that \ref{fig:modelcomparisonPDs}(h) is the only the lower row figure which does \textit{not} feature the aforementioned LLC state.

Fig.~\ref{fig:modelcomparisonPDs}(h) is also notably the only diagram to feature the OSP-POP state, a superposition between the OSP and POP states. It has constant partial orbital and spin polarization and continuously evolving partial valley polarization, and is given by
%\begin{equation}
%\left|\Psi_{O S P-P O P}\right\rangle=\prod_{X}\left(c_{0+\uparrow X}^{+} c_{0+\downarrow X}^{+} c_{0-\uparrow X}^{+}\left(\cos \theta c_{1+\uparrow X}^{+}+\sin \theta c_{1-\uparrow X}^{+}\right)\right)|\varnothing\rangle \, .
%\end{equation}
\begin{align}
& \left|\Psi_{OSP-POP}\right\rangle = \\ & \,\,\, \prod_{X}\left(c_{0+\uparrow X}^{+} c_{0+\downarrow X}^{+} c_{0-\uparrow X}^{+}\left(\cos \theta c_{1+\uparrow X}^{+} \! + \! \sin \theta c_{1-\uparrow X}^{+}\right)\right) \! |\varnothing\rangle . \nonumber
\end{align}
Further information on this state is in Sec.~S5 in the Supplemental Material.

We have examined here only a representative subset of the possible combinations of included and neglected parameters. Our model is also compatible with previous models by changing the parameters described above, plus a few constants. For example, we have reproduced the onset and end of the FSP-FVP state given by Ref.~\cite{lambert_quantum_2013} by removing gates and the spatial extent of the p\textsubscript{z} orbitals, using simplified form factors, neglecting the Lamb-like shift, and using the TB parameters and dielectric constant given therein; and we have reproduced the LLL energy levels of Ref.~\cite{barlas_intra-landau-level_2008} by using the same approximations and additionally setting the orbital gap to $0$.

A modification of the Coulomb interaction we have not addressed in our model is that of screening in a heterostructure. Recently, the experiment of Chuang et al. \cite{chuang_landau_2019} on stacked BLG and $\text{WSe}_2$ mono- or bilayers showed that $\text{WSe}_2$ brings the appearance of the POP state to lower magnetic fields, and noted that thin dielectric layers primarily screen short-range interactions. Though this preferentially weakens exchange for the $n=1$ orbitals, which have more relatively more high-$q$ weight as seen in Fig.~\ref{fig:VvsFFcomparison}, this change is counteracted by the Lamb-like shift. This suggests that it is simply weakening exchange which drives the change, regardless of length scale. Weakening exchange disfavors the FSP and FVP states, so that the POP state appears at a lower magnetic field.

\section{Conclusion} \label{conclusion}

We have produced Landau level phase diagrams of charge-neutral ($\nu=0$) BLG as a function of magnetic field, bias, and pressure. We found noninteracting eigenstates and energies using a four-band tight-binding model with hoppings between each pair of lattice sites. Projecting into the eight LLLs near the Fermi level and treating the Coulomb interaction through the Hartree-Fock approximation, we studied how gate screening, layer separation, the spatial extent of the p\textsubscript{z} orbitals, and layer-resolved form factors found by exact diagonalization impact the interaction and phase diagrams. All parameters were determined by \textit{ab initio} calculations \cite{munoz_bilayer_2016,clementi_atomic_1963} or independent experimental measurements \cite{laturia_dielectric_2018}.

Five LLSD states (FSP, FVP, POP, OSP, and FOP) manifest as ground states. Two of these (OSP and FOP) previously have been neither theoretically predicted to appear nor observed experimentally. The appearance of the orbitally-polarized states (POP, OSP and FOP) is driven by noninteracting dynamics overtaking the Coulomb interaction as the dominant energy scale, and this transition is controlled by pressure and the magnetic field. The absence of LLC states in our results, in comparison to similar theoretical work using parameter-free long-range Coulomb propagators \cite{cote_orbital_2010,cote_biased_2011,lambert_quantum_2013,lambert_ferro-aimants_2013,knothe_phase_2016}, is unique to our model. We isolated the use of exact diagonalization form factors which respect the inequivalence between valley and layer as the source of this change. This emphasizes that, due to the small energy scales involved in this system, even parameters or effects which appear small may in fact be significant.

We chose to focus on $\nu=0$, but our model may readily be applied for other filling factors. Likewise, we focused on ground state phase diagrams, but our model can also be used to calculate excited state energies and single-particle energy gaps to explain transport or cyclotron resonance experiments, as in Ref.~\cite{lambert_quantum_2013} or \cite{barlas_intra-landau-level_2008} respectively, for example. These are natural follow-up topics for us to explore in future work.

Currently, only zero-pressure experimental comparisons are available. Our results agree with experimental indications that the boundary between the FSP and FVP states does not host an LLC state \cite{hunt_direct_2017,li_effective_2018}, which had previously been a source of disagreement in parameter-free models. However, we have not been able to reproduce the experimental appearance \cite{hunt_direct_2017,li_effective_2018} of the POP state at $B=12~\text{T}$ at zero pressure; to date, this has only been reproduced in phenomenological models by fitting the orbital gap \cite{murthy_spin-valley_2017} or screening and symmetry-breaking interaction parameters \cite{hunt_direct_2017} to experimental results. Thus, we have found a physical cause for the discontinuous transition from FSP to FVP, but the cause of the POP state's appearance at moderate magnetic field and zero pressure remains unknown. Though the POP state only appears in our results at elevated pressure, we agree with the identification of POP as the intermediate state. (One experiment by Li \textit{et al.} \cite{li_metallic_2019} using a different device geometry and tilted magnetic field found a metallic state between the FSP and FVP states, but we have not addressed these characteristics in our model.)

Comparing Refs.~\cite{lambert_quantum_2013,knothe_phase_2016} and this work, which did not find the POP state around $B=12~\text{T}$ at zero pressure, with Refs.~\cite{hunt_direct_2017,murthy_spin-valley_2017}, which did, suggests some missing ingredients. For example, LL mixing provides screening \cite{nandkishore_dynamical_2010,gorbar_broken_2012,gorbar_magneto-optical_2012,lambert_quantum_2013,hunt_direct_2017} and, together with the electron-phonon interaction, induces symmetry-breaking interactions \cite{kharitonov_phase_2012,kharitonov_canted_2012,murthy_spin-valley_2017} which may stabilize the POP state. These symmetry-breaking interactions stabilize a canted antiferromagnetic state \cite{kharitonov_canted_2012,kharitonov_edge_2012,murthy_spin-valley_2017}, which does not appear in our model but is supported by experimental evidence \cite{maher_evidence_2013,li_metallic_2019}.

Even small or weak effects may be important due to the small energy scale of the the LLLs. This is demonstrated by the above comparison of published results, by our comparison the effects of model details (particularly the nature of the form factors), and by a comparison of the orbital gap plotted in Fig.~S1 in the Supplemental Material to the phase diagrams in Fig.~\ref{fig:phaseanddotdiagrams}, which shows the significant impact of increasing orbital splitting by only a few meV/T. We are working to understand the aforementioned effects in a physically transparent way. Such understanding will be necessary to answer the many remaining open questions for research in this field and to explore its continually expanding possibilities, both theoretical and experimental,

\begin{acknowledgments}

This material is based upon work supported by the National Science Foundation Graduate Research Fellowship Program under Grant No. DGE1255832. Any opinions, findings, and conclusions or recommendations expressed in this material are those of the author and do not necessarily reflect the views of the National Science Foundation.

\end{acknowledgments}

\bibliographystyle{apsrev4-2}

\clearpage

\newcommand{\beginsupplement}{
	\setcounter{section}{0}
	\renewcommand{\thesection}{S\arabic{section}}
	\setcounter{equation}{0}
	\renewcommand{\theequation}{S\arabic{equation}}
	\setcounter{table}{0}
	\renewcommand{\thetable}{S\arabic{table}}
	\setcounter{figure}{0}
	\renewcommand{\thefigure}{S\arabic{figure}}
	\newcounter{SIsect}
	\renewcommand{\theSIsect}{S\arabic{section}}
	\newcounter{SIeq}
	\renewcommand{\theSIeq}{S\arabic{equation}}
	\newcounter{SIfig}
	\renewcommand{\theSIfig}{S\arabic{SIfig}}}
\newcommand{\supplementtitle}[2]{\noindent{\textbf{ \large{#1} \\ \normalsize{#2}}}}

\beginsupplement

\begin{widetext}
\supplementtitle{Supplemental Material}{Landau Level Phases in Bilayer Graphene under Pressure at Charge Neutrality}

\section{Noninteracting results} \refstepcounter{SIsect}\label{appendix_spresults}

In Fig.~\ref{fig:spresults2} we give the eigenvector coefficients $c_{n\xi}^{Tj}$, used in the expansion given by LLL wavefunctions, up to sign. The pattern of nonvanishing coefficients is 3-periodic; in particular,
\begin{equation}\begin{aligned} \refstepcounter{SIeq}\label{nonzerocoeffs_appendix}
& c_{0+}^{Tj} \ne 0 \text{ for } (T,j)\in\left\{(A1,3m+2),(B1,3m),(A2,3m+1),(B2,3m+2) : m\ge0\right\} \, , \\
& c_{1+}^{Tj} \ne 0 \text{ for } (T,j)\in\left\{(A1,3m),(B1,3m+1),(A2,3m+2),(B2,3m) : m\ge0\right\} \, ,
\end{aligned}\end{equation}
and all other coefficients vanish. Pressure increases the TB parameters, which drive both the orbital gap and the wavefunction coefficients. Hence, the wavefunctions are more complex at elevated pressures.

\begin{figure*}[ht]
    \centering
    \includegraphics[width=8cm]{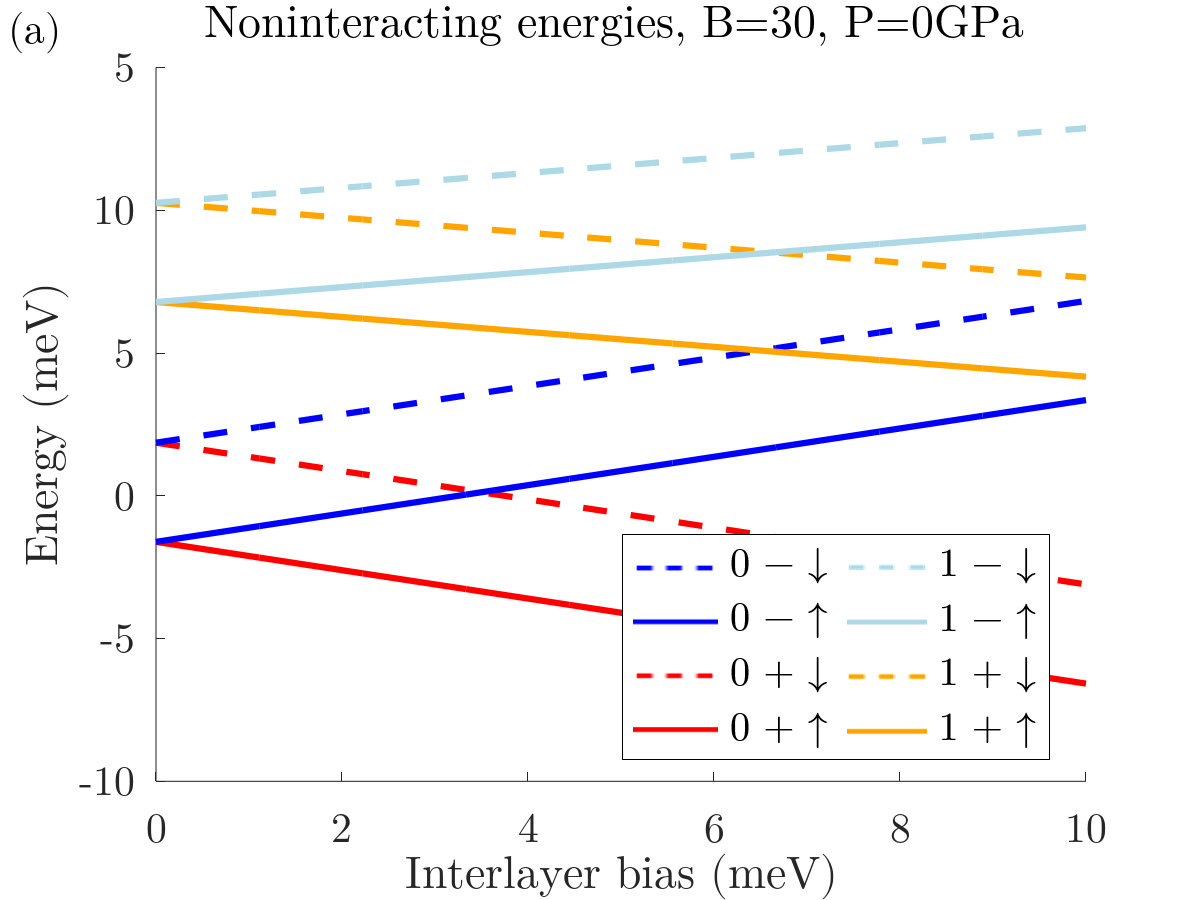}
    \includegraphics[width=8cm]{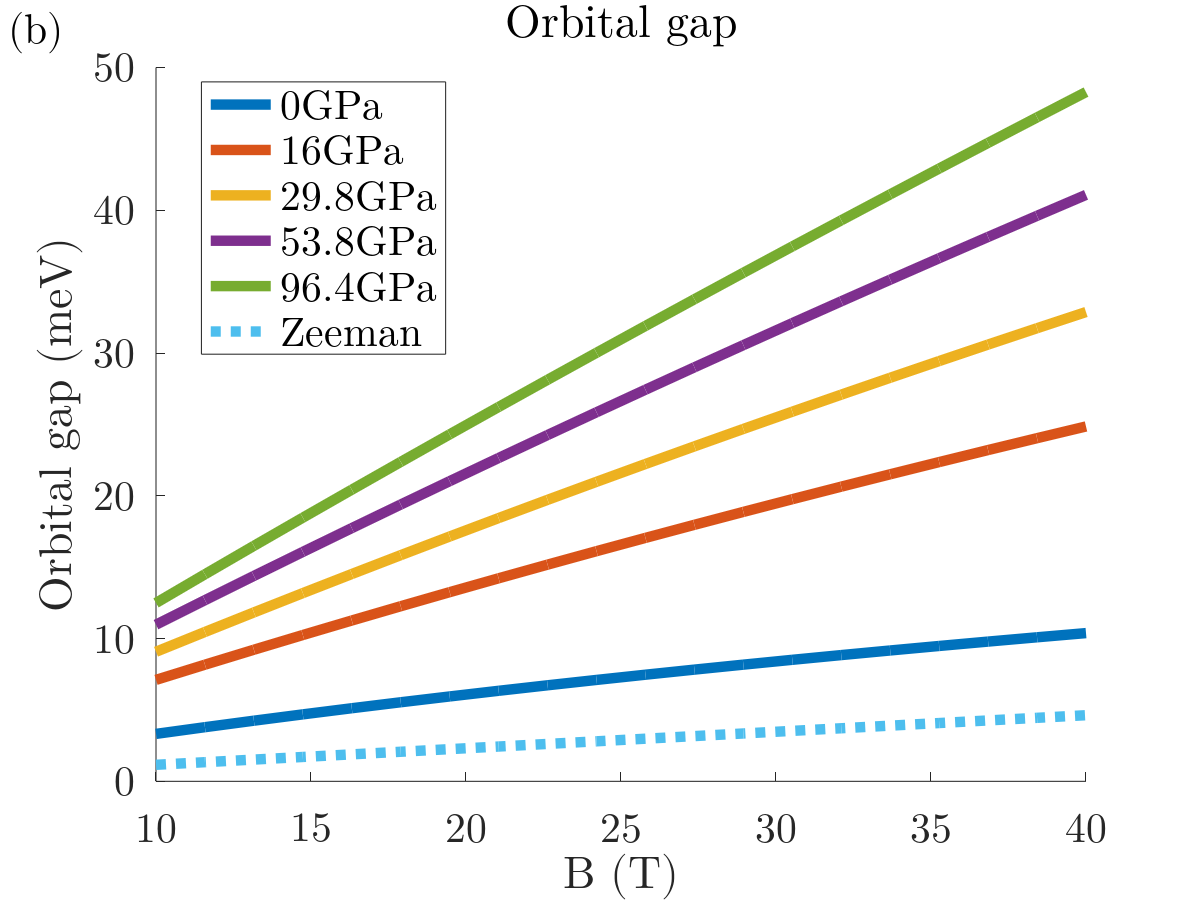}
    
    \vspace{0.5cm}
    
    \includegraphics[width=8cm]{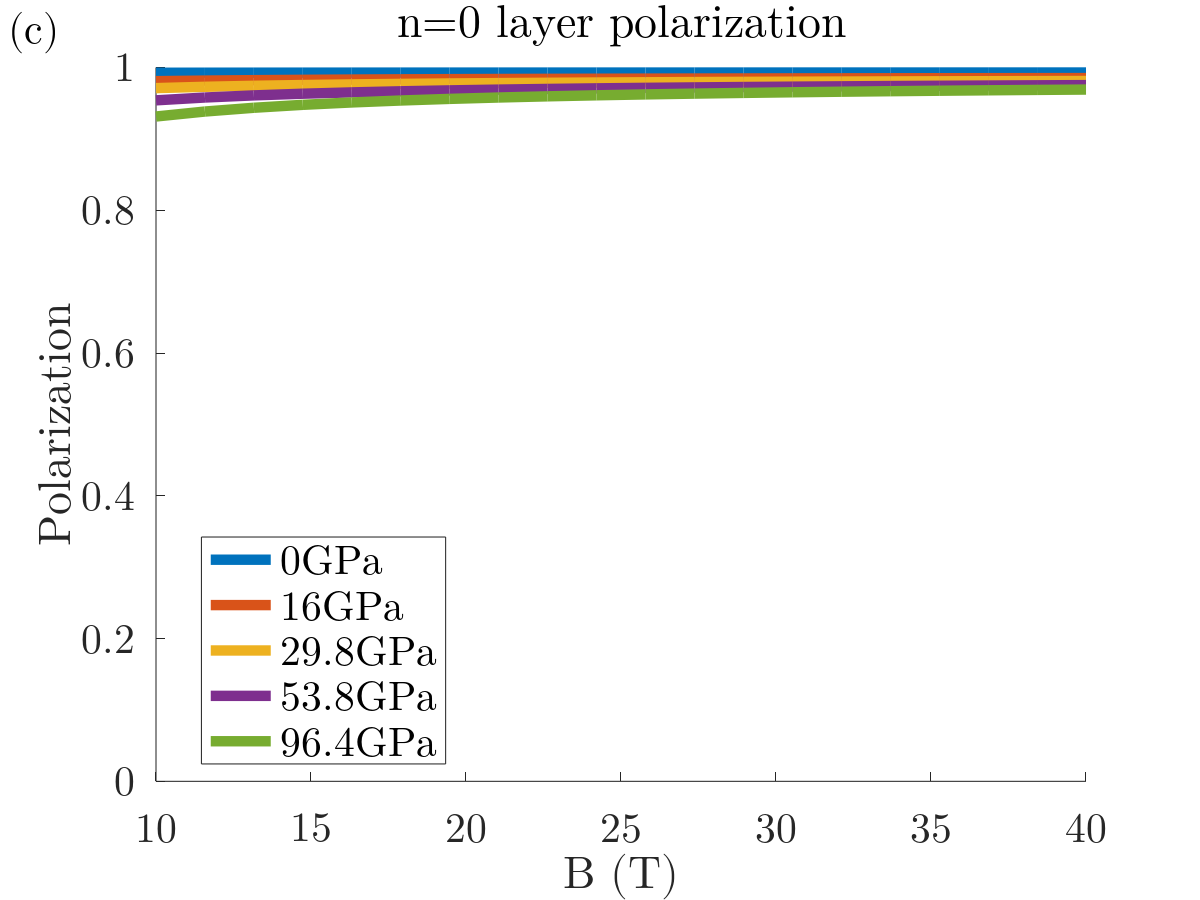}
    \includegraphics[width=8cm]{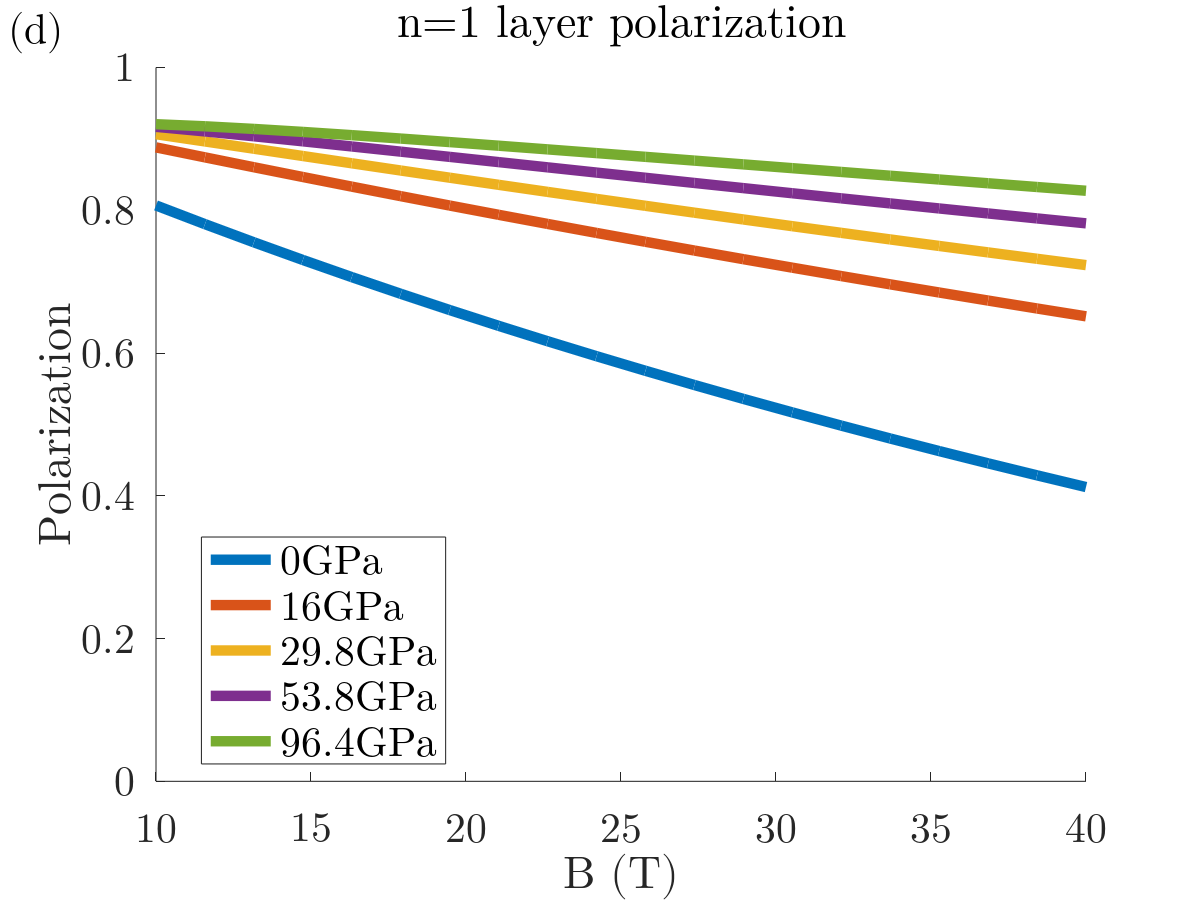}
    
    \refstepcounter{SIfig}\label{fig:spresults1}
    
    \caption{(a) Except under large bias, noninteracting dynamics favor orbitally polarized states. The $n=0$ state is more strongly affected by bias than is the $n=1$ state because the former is more layer-polarized. (b) The orbital gap $E_1-E_0$ increases with both magnetic field and pressure, and the Zeeman splitting (i.e., spin gap) is plotted alongside for comparison. (c) While the $n=0$ layer polarization is nearly constant, (d) the $n=1$ layer polarization decreases steeply with magnetic field for low pressure.}
\end{figure*}

\begin{figure*}
    \centering
    \includegraphics[width=8cm]{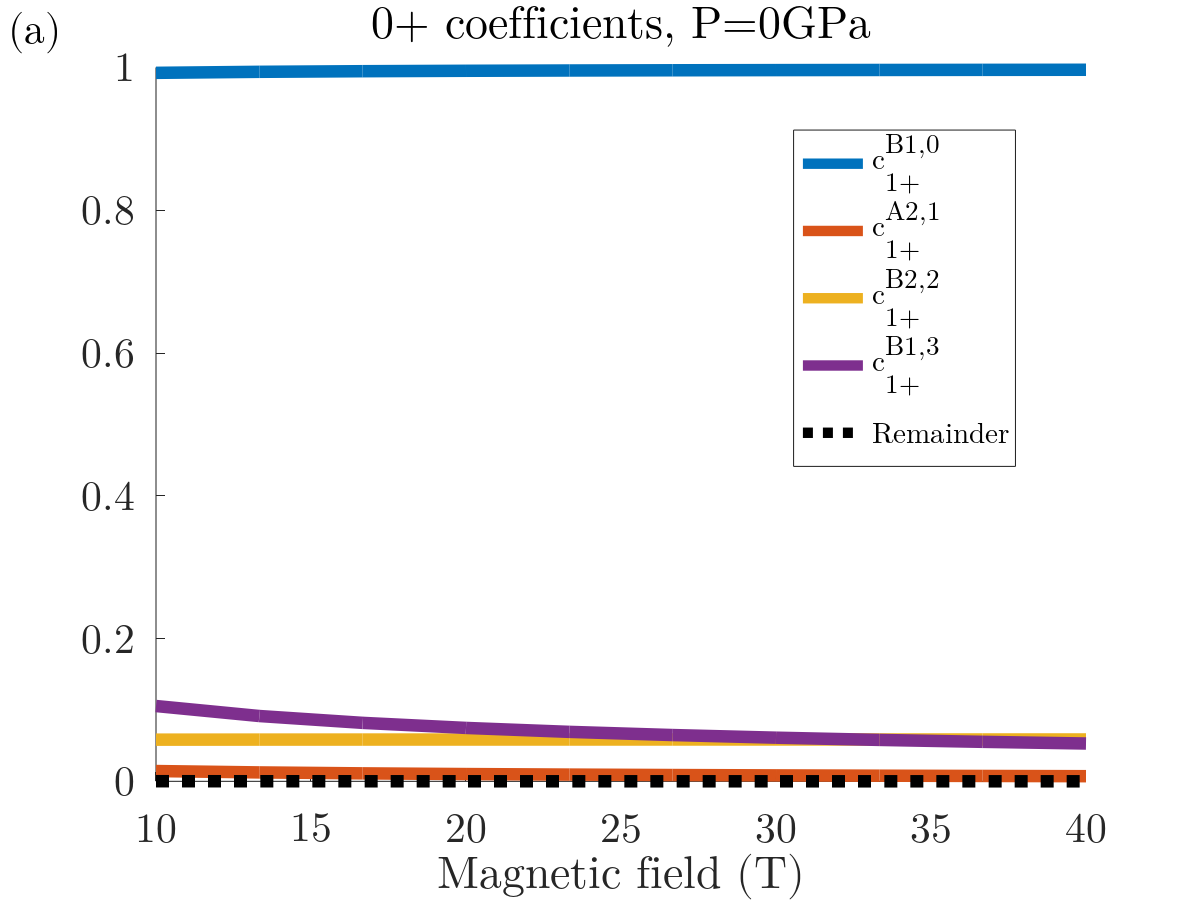}
    \includegraphics[width=8cm]{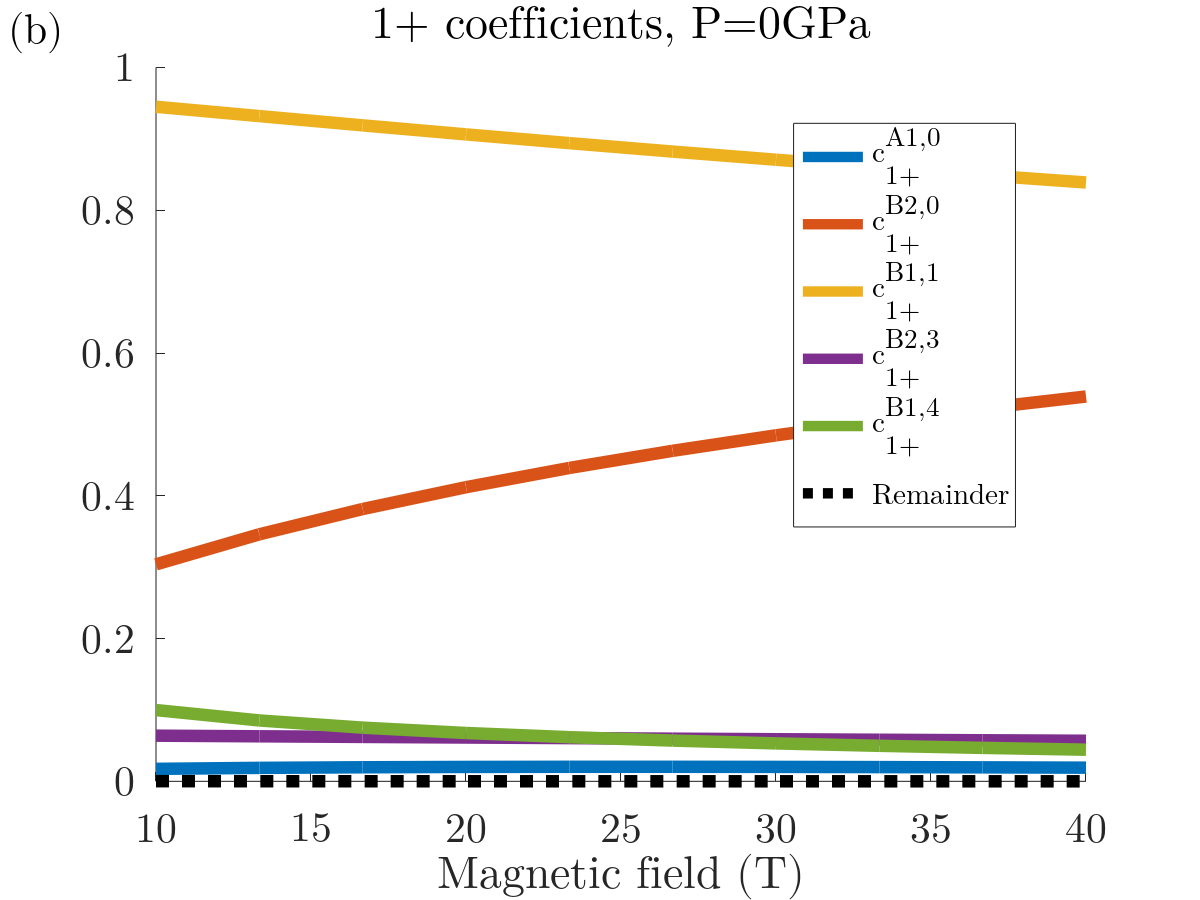}
    
    \vspace{0.5cm}
    
    \includegraphics[width=8cm]{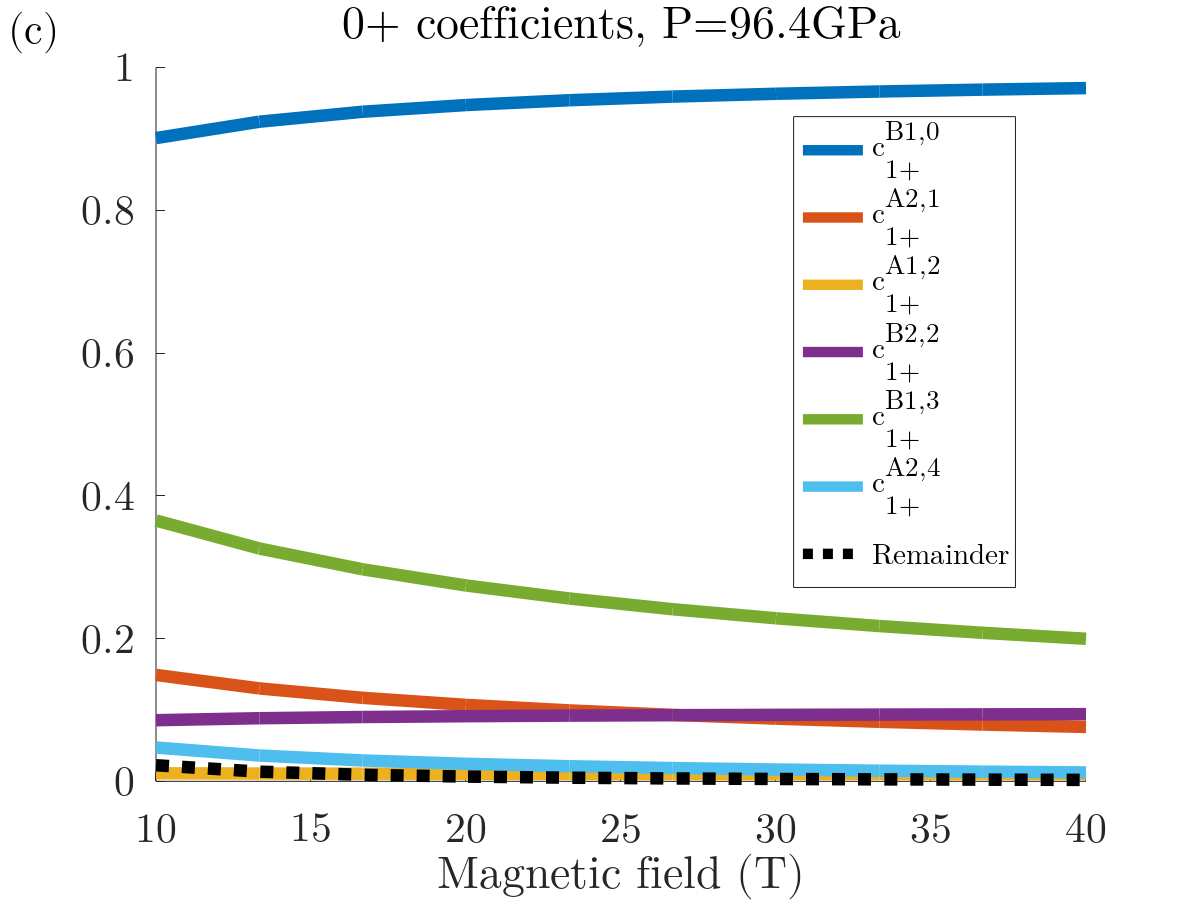}
    \includegraphics[width=8cm]{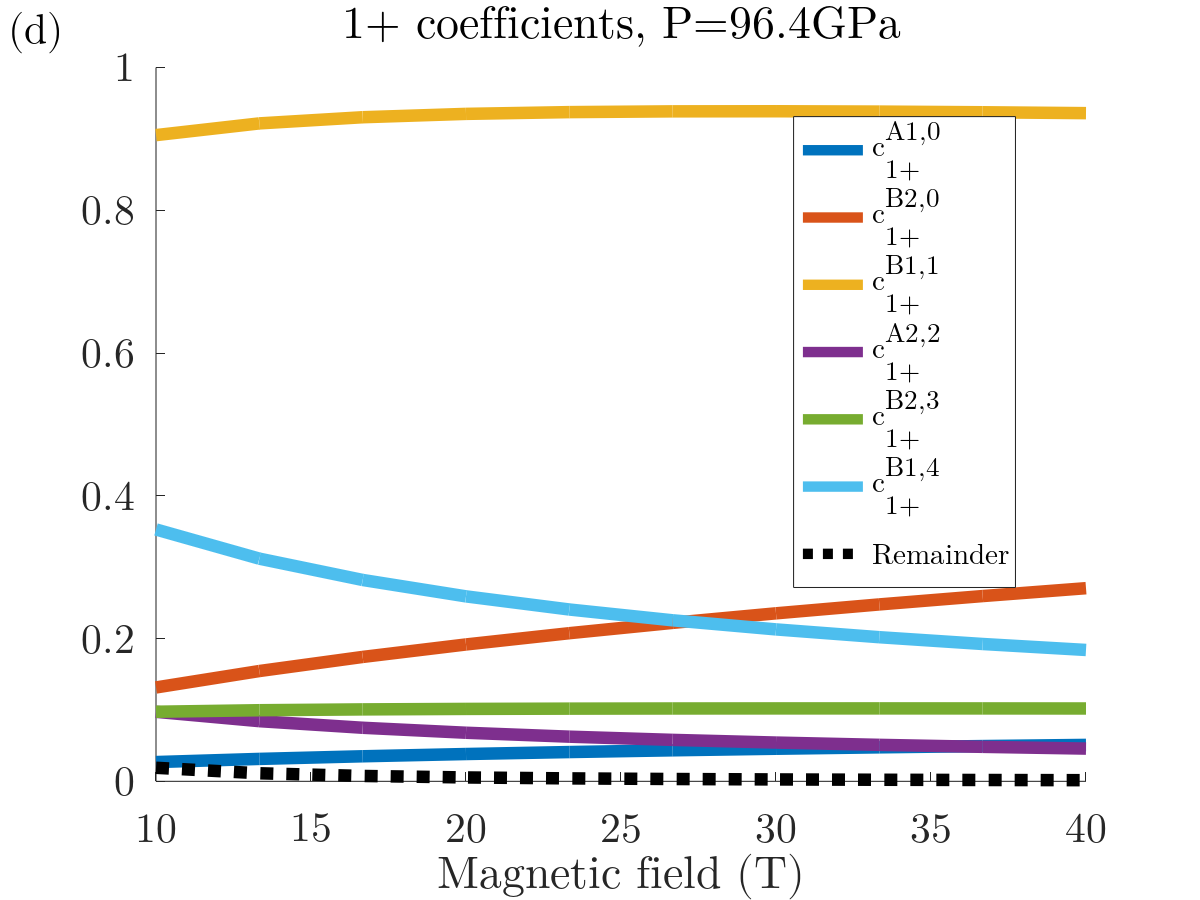}
    
    \refstepcounter{SIfig}\label{fig:spresults2}
    
    \caption{Here we plot $\left|c_{n\xi}^{Tj}\right|$ to demonstrate necessity of a larger basis to characterize the LLLs and the justification for truncating $j>4$ in calculating the exchange integrals. (See Appendix~\ref{appendix_exchangeintegrals}.) The remainder is defined by $1-\sum_{j=0}^{4} \sum_{T}\left|c_{n \xi}^{T j}\right|^{2}$. Results for other pressures interpolate between the two extremes shown. Only coefficients distinguishable from $0$ are shown.}
\end{figure*}

\section{Form factors and 3D Coulomb interaction} \refstepcounter{SIsect}\label{appendix_FFand3DCoul}

To derive the form factors, we begin by calculating
\begin{equation}\begin{aligned} \refstepcounter{SIeq}\label{firstFFintegral} & \int d^{2} \bm{r} e^{i \bm{q} \cdot \bm{r}} \phi_{n_{1} \xi \sigma X_1}^{*}(\bm{r}, z) \phi_{n_{4} \xi \sigma X_4}(\bm{r}, z) \\ & \quad = \sum_{T_{1} T_{4} \bm{R_{1} R_{4}}} \sum_{j_{1} j_{4}} {c_{n_{1} \xi}^{T_{1} j_{1}}}^{*} c_{n_{4} \xi}^{T_{4} j_{4}} h_{j_{1} X_{1}}^{*}\left(\bm{R_{1}}\right) h_{j_{4} X_4}\left(\bm{R_4}\right) \\ & \quad \quad \times \int d^{2} \bm{r} e^{i \bm{q} \cdot \bm{r}} \psi_{2 p_{z}}^{*}\left(\bm{r}-\bm{\tau^{2D}_{T_1}}-\bm{R_{1}}, z-\tau^z_{T_1}\right) \psi_{2 p_{z}}\left(\bm{r}-\bm{\tau^{2D}_{T_4}}-\bm{R_{4}}, z-\tau^z_{T_4}\right) \, . \end{aligned}\end{equation}
%\begin{align} \refstepcounter{SIeq}\label{firstFFintegral} & \int d^{2} \bm{r} e^{i \bm{q} \cdot \bm{r}} \phi_{n_{1} \xi \sigma X_1}^{*}(\bm{r}, z) \phi_{n_{4} \xi \sigma X_4}(\bm{r}, z) \\ & \quad = \!\!\!\! \sum_{T_{1} T_{4} \bm{R_{1} R_{4}}} \sum_{j_{1} j_{4}} {c_{n_{1} \xi}^{T_{1} j_{1}}}^{*} c_{n_{4} \xi}^{T_{4} j_{4}} h_{j_{1} X_{1}}^{*} \! \left(\bm{R_{1}}\right) h_{j_{4} X_4} \! \left(\bm{R_4}\right) \int d^{2} \bm{r} e^{i \bm{q} \cdot \bm{r}} \psi_{2 p_{z}}^{*} \! \left(\bm{r} \textrm{--} \bm{\tau^{2D}_{T_1}} \textrm{--} \bm{R_{1}}, z \textrm{--} \tau^z_{T_1}\right) \psi_{2 p_{z}} \! \left(\bm{r} \textrm{--} \bm{\tau^{2D}_{T_4}} \textrm{--} \bm{R_{4}}, z \textrm{--} \tau^z_{T_4}\right) . \nonumber \end{align}
The tight-binding orbitals will have negligible overlap unless $T_1=T_4$, $\bm{R_1}=\bm{R_4}$ so we drop their subscripts and sum over only a single pair $T,\bm{R}$. 

Now we note the relevant length scales. The atomic orbital scale $\sim0.1~\text{nm}$, basis vector length $a=0.142~\text{nm}$, and lattice vector spacing $a\sqrt{3}=0.246~\text{nm}$ are all much smaller than length scale of the Fourier transform wavevector given by the size of the LL envelopes, which is the magnetic length $l_B \approx \left(25.7~\text{nm}\cdot\text{T}^{-1/2}\right)\sqrt{B}$. We will use this fact to make several useful approximations. Proceeding with the integration in Eq.~(\ref{firstFFintegral}),
\begin{equation}\begin{aligned} \refstepcounter{SIeq}
& \int d^{2} \bm{r} e^{i \bm{q} \cdot \bm{r}} \psi_{2 p_{z}}^{*}\left(\bm{r}-\bm{\tau^{2D}}-\bm{R}, z-\tau^z_T\right) \psi_{2 p_{z}}\left(\bm{r}-\bm{\tau^{2D}}-\bm{R}, z-\tau^z_T\right) \\
& = e^{i \bm{q} \cdot(\bm{\tau^{2D}}+\bm{R})} \int d^{2} \bm{r} e^{i \bm{q} \cdot \bm{r}} \psi_{2 p_{z}}^{*}(\bm{r}, z-\tau^z_T) \psi_{2 p_{z}}(\bm{r}, z-\tau^z_T) \approx e^{i \bm{q} \cdot \bm{R}}P(z-\tau^z_T) \, .
\end{aligned}\end{equation}
Here we used the facts that $e^{i\bm{q}\cdot \bm{r}}\approx 1$ is essentially constant over the atomic orbitals, and likewise that $e^{i\bm{q}\cdot \bm{T^{2D}}}\approx 1$ because $\bm{T^{2D}}$ is small compared to $l_B$, and $q\sim 1/l_B$. $P(z)=\int d^{2} \bm{r} \left|\psi_{2 p_{z}}(\bm{r}, z)\right|^2$ is the probability density in the z-direction. In our calculations, we used the parametrization of $\psi_{2p_z}(\bm{r})$ and hence $P(z)$ given by Clementi and Raimondi \cite{clementiatomic1963}. Therefore
\begin{equation}\begin{aligned} \refstepcounter{SIeq}
& \int d^{2} \bm{r} e^{i \bm{q} \cdot \bm{r}} \phi_{n_{1} \xi \sigma X_1}^{*}(\bm{r}, z) \phi_{n_{4} \xi \sigma X_4}(\bm{r}, z) \\
& \quad = \sum_{T j_{1} j_{4}} {c_{n_{1} \xi}^{T j_{1}}}^{*} c_{n_{4} \xi}^{T j_{4}} P(z-\tau^z_T) \sum_{\bm{R}} e^{i\bm{q}\cdot \bm{R}} h_{j_{1} X_{1}}^{*}\left(\bm{R}\right) h_{j_{4} X_4}\left(\bm{R}\right) \, .
\end{aligned}\end{equation}
%\begin{equation} \refstepcounter{SIeq}
%\int d^{2} \bm{r} e^{i \bm{q} \cdot \bm{r}} \phi_{n_{1} \xi \sigma X_1}^{*}(\bm{r}, z) \phi_{n_{4} \xi \sigma X_4}(\bm{r}, z) = \sum_{T j_{1} j_{4}} {c_{n_{1} \xi}^{T j_{1}}}^{*} c_{n_{4} \xi}^{T j_{4}} P(z-\tau^z_T) \sum_{\bm{R}} e^{i\bm{q}\cdot \bm{R}} h_{j_{1} X_{1}}^{*}\left(\bm{R}\right) h_{j_{4} X_4}\left(\bm{R}\right) \, .
%\end{equation}

Since $h_{jX}(\bm{R})$ is already normalized as a continuous variable, no normalization factor is needed to take $\sum_{\bm{R}}\to\int d^2 \bm{R}$. From this we derive the expression that defines the elementary form factors,
\begin{equation}\begin{aligned} \refstepcounter{SIeq}
& \int d^2 \bm{R} e^{i \bm{q} \cdot \bm{R}} h_{j_{1} X_{1}}^{*}\left(\bm{R}\right) h_{j_{4} X_{4}}\left(\bm{R}\right) \\
& = \frac{1}{L_{y}} \int d y e^{i\left(q_{y} l_{B}^{2}+X_{4}-X_{1}\right) \frac{y}{l_{B}^{2}}} \int d x e^{i q_{x} x} Q_{j_{1}}\left(x-X_{1}\right) Q_{j_{4}}\left(x-X_{4}\right) \\
& = \delta_{X_{4},X_{1}-q_{y} l_{B}^{2}} e^{i \frac{q_{x}}{2}\left(X_{1}+X_{4}\right)} \int d x e^{i q_{x} x} Q_{j_{1}}\left(x-\frac{q_{y} l_{B}^{2}}{2}\right) Q_{j_{4}}\left(x+\frac{q_{y} l_{B}^{2}}{2}\right) \\
& = \delta_{X_{4},X_{1}-q_{y} l_{B}^{2}} e^{i \frac{q_{x}}{2}\left(X_{1}+X_{4}\right)} K_{j_{1} j_{4}}(\bm{q}) \, .
\end{aligned}\end{equation}
%\begin{equation}\begin{aligned} \refstepcounter{SIeq}
%& \int d^2 \bm{R} e^{i \bm{q} \cdot \bm{R}} h_{j_{1} X_{1}}^{*}\left(\bm{R}\right) h_{j_{4} X_{4}}\left(\bm{R}\right) = \frac{1}{L_{y}} \int d y e^{i\left(q_{y} l_{B}^{2}+X_{4}-X_{1}\right) \frac{y}{l_{B}^{2}}} \int d x e^{i q_{x} x} Q_{j_{1}}\left(x-X_{1}\right) Q_{j_{4}}\left(x-X_{4}\right) \\
%& \quad = \delta_{X_{4},X_{1}-q_{y} l_{B}^{2}} e^{i \frac{q_{x}}{2}\left(X_{1}+X_{4}\right)} \int d x e^{i q_{x} x} Q_{j_{1}}\left(x-\frac{q_{y} l_{B}^{2}}{2}\right) Q_{j_{4}}\left(x+\frac{q_{y} l_{B}^{2}}{2}\right) = \delta_{X_{4},X_{1}-q_{y} l_{B}^{2}} e^{i \frac{q_{x}}{2}\left(X_{1}+X_{4}\right)} K_{j_{1} j_{4}}(\bm{q}) \, .
%\end{aligned}\end{equation}
Thus we can write the Fourier transform of the wavefunction overlap
\begin{equation}\begin{aligned} \refstepcounter{SIeq} & \int d^{2} \bm{r} e^{i \bm{q} \cdot \bm{r}} \phi_{n_{1} \xi \sigma X_1}^{*}(\bm{r}, z) \phi_{n_{4} \xi \sigma X_4}(\bm{r}, z) \\ & \quad = \delta_{X_{4},\left(X_{1}-q_{y} l_{B}^{2}\right)} e^{i \frac{q_{x}}{2}\left(X_{1}+X_{4}\right)} \sum_{T j_{1} j_{4}} {c_{n_{1} \xi}^{T j_{1}}}^{*} c_{n_{4} \xi}^{T j_{4}} P \left(z-\tau^z_T\right) K_{j_{1} j_{4}}(\bm{q}) \\ & \quad = \delta_{X_{4},X_{1}-q_{y} l_{B}^{2}} e^{i \frac{q_{x}}{2}\left(X_{1}+X_{4}\right)} \sum_{T_{z}} P \! \left(z+(-1)^{T_z}\frac{d}{2}\right) J_{\substack{n_1 \xi \\ n_4 \xi}}^{T_z}(\bm{q}) \end{aligned}\end{equation}
%\begin{equation}\begin{aligned} \refstepcounter{SIeq} & \int d^{2} \bm{r} e^{i \bm{q} \cdot \bm{r}} \phi_{n_{1} \xi \sigma X_1}^{*}(\bm{r}, z) \phi_{n_{4} \xi \sigma X_4}(\bm{r}, z) = \delta_{X_{4},\left(X_{1}-q_{y} l_{B}^{2}\right)} e^{i \frac{q_{x}}{2}\left(X_{1}+X_{4}\right)} \sum_{T j_{1} j_{4}} {c_{n_{1} \xi}^{T j_{1}}}^{*} c_{n_{4} \xi}^{T j_{4}} P \left(z-\tau^z_T\right) K_{j_{1} j_{4}}(\bm{q}) \\ & \quad \quad = \delta_{X_{4},X_{1}-q_{y} l_{B}^{2}} e^{i \frac{q_{x}}{2}\left(X_{1}+X_{4}\right)} \sum_{T_{z}} P \! \left(z+(-1)^{T_z}\frac{d}{2}\right) J_{\substack{n_1 \xi \\ n_4 \xi}}^{T_z}(\bm{q}) \end{aligned}\end{equation}
in terms of the layer-projected form factors
\begin{equation} \refstepcounter{SIeq}
J_{\substack{n_1 \xi \\ n_4 \xi}}^{T_z}(\bm{q})=\sum_{j_{1} j_{4}} K_{j_{1} j_{4}}(\bm{q}) \sum_{T_{2 D}} {c_{n_{1} \xi}^{T_{2 D} T_{z} j_{1}}}^* c_{n_{4} \xi}^{T_{2 D} T_{z} j_{4}} \, .
\end{equation}
Finally, the z-dependence of the Coulomb interaction can be isolated by defining a layer-resolved Coulomb interaction;
\begin{equation}\begin{aligned} \refstepcounter{SIeq} \hat{V} = & \sum_{\substack{n_j \xi_j \sigma_j X_j \\ j=1,2,3,4}} \sum_{\bm{q}} \Bigl(\int d z \int d z^{\prime} V\left(q, z, z^{\prime}\right) \\ & \quad \quad \quad \quad \times \left( \int d^{2} \bm{r} e^{i \bm{q} \cdot \bm{r}} \phi_{n_{1} \xi_{1} \sigma_{1} X_1}^{*}\left(\bm{r}, z\right) \phi_{n_{4} \xi_{4} \sigma_{4} X_4}\left(\bm{r}, z\right)\right) \\ & \quad \quad \quad \quad \times \Bigl(\int d^{2} \bm{r^{\prime}} e^{-i \bm{q} \cdot \bm{r^{\prime}}} \phi_{n_{2} \xi_{2} \sigma_{2} X_2}^{*}\left(\bm{r^{\prime}}, z^{\prime}\right) \phi_{n_{3} \xi_{3} \sigma_{3} X_3}\left(\bm{r^{\prime}}, z^{\prime}\right)\Bigr) \Bigr) \\ & \quad \quad \quad \quad \times c^+_{n_1 \xi_1 \sigma_1 X_1}c^+_{n_2 \xi_2 \sigma_2 X_2}c_{n_3 \xi_3 \sigma_3 X_3}c_{n_4 \xi_4 \sigma_4 X_4} \\ = & \sum_{\substack{n_j \xi_j X_j \\ j=1,2,3,4}} \sum_{\sigma \sigma^{\prime}} \sum_{\bm{q}} \sum_{T_z T_{z^\prime}} V_{T_{z} T_{z}^{\prime}}(q) \\ & \quad \quad \quad \quad \times \left(\delta_{X_{4},X_{1}-q_{y} l_{B}^{2}} e^{i \frac{q_{x}}{2}\left(X_{1}+X_{4}\right)} J_{\substack{n_{1} \xi_{1} \\ n_{4} \xi_{4}}}^{T_{z}}(\bm{q})\right) \\ & \quad \quad \quad \quad \times \left(\delta_{X_{3},X_{2}+q_{y} l_{B}^{2}} e^{-i \frac{q_{x}}{2}\left(X_{2}+X_{3}\right)} J_{\substack{n_2 \xi_2 \\ n_3 \xi_3}}^{T_{z}}(-\bm{q})\right) \\ & \quad \quad \quad \quad \times c^+_{n_1 \xi_1 \sigma_1 X_1}c^+_{n_2 \xi_2 \sigma_2 X_2}c_{n_3 \xi_3 \sigma_3 X_3}c_{n_4 \xi_4 \sigma_4 X_4} \end{aligned}\end{equation}
%\begin{equation}\begin{aligned} \refstepcounter{SIeq} \hat{V} & = \frac{1}{2} \sum_{\substack{n_j \xi_j \sigma_j X_j \\ j=1,2,3,4}} \sum_{\bm{q}} c^+_{n_1 \xi_1 \sigma_1 X_1}c^+_{n_2 \xi_2 \sigma_2 X_2}c_{n_3 \xi_3 \sigma_3 X_3}c_{n_4 \xi_4 \sigma_4 X_4} \Bigg( \int d z \int d z^{\prime} V\left(q, z, z^{\prime}\right) \\ & \quad \quad \quad \times \left( \int d^{2} \bm{r} e^{i \bm{q} \cdot \bm{r}} \phi_{n_{1} \xi_{1} \sigma_{1} X_1}^{*}\left(\bm{r}, z\right) \phi_{n_{4} \xi_{4} \sigma_{4} X_4}\left(\bm{r}, z\right)\right) \left(\int d^{2} \bm{r^{\prime}} e^{-i \bm{q} \cdot \bm{r^{\prime}}} \phi_{n_{2} \xi_{2} \sigma_{2} X_2}^{*}\left(\bm{r^{\prime}}, z^{\prime}\right) \phi_{n_{3} \xi_{3} \sigma_{3} X_3}\left(\bm{r^{\prime}}, z^{\prime}\right)\right) \Bigg) \\ & = \sum_{\substack{n_j \xi_j X_j \\ j=1,2,3,4}} \sum_{\sigma \sigma^{\prime}} \sum_{\bm{q}} \sum_{T_z T_{z^\prime}} V_{T_{z} T_{z}^{\prime}}(q) \left(\delta_{X_{4},X_{1}-q_{y} l_{B}^{2}} e^{i \frac{q_{x}}{2}\left(X_{1}+X_{4}\right)} J_{\substack{n_{1} \xi_{1} \\ n_{4} \xi_{4}}}^{T_{z}}(\bm{q})\right) \left(\delta_{X_{3},X_{2}+q_{y} l_{B}^{2}} e^{-i \frac{q_{x}}{2}\left(X_{2}+X_{3}\right)} J_{\substack{n_2 \xi_2 \\ n_3 \xi_3}}^{T_{z}}(-\bm{q})\right) \\ & \quad \quad \quad \times c^+_{n_1 \xi_1 \sigma_1 X_1}c^+_{n_2 \xi_2 \sigma_2 X_2}c_{n_3 \xi_3 \sigma_3 X_3}c_{n_4 \xi_4 \sigma_4 X_4} \end{aligned}\end{equation}
where
\begin{equation} \refstepcounter{SIeq} \label{Vlayerprojection_appendix}
V_{T_{z} T_{z}^{\prime}}(q)=\int d z \int d z^{\prime} V\left(q, z, z^{\prime}\right) P\left(z+(-1)^{T_z}\frac{d}{2}\right) P\left(z+(-1)^{T_z^\prime}\frac{d}{2}\right) \, .
\end{equation}\end{widetext}

For calculating the exchange integrals, it is useful to find an analytic approximation to the result of the integral of Eq.~(\ref{Vlayerprojection_appendix}), which also depends on pressure through the layer separation when $T_z \ne T_z^\prime$. To construct such an approximation, note that if one neglects $d$ when compared to $D$, i.e. takes $D+d \approx D$, in the propagator $V(q,z,z^\prime)$, it reduces to
\begin{equation} \refstepcounter{SIeq}
V\left(q,+\frac{d}{2},-\frac{d}{2}\right) \approx \frac{1}{N_\Phi} \alpha \frac{1}{q l_{B}} \tanh (q D) e^{-q d} \, .
\end{equation}
Using this form but replacing the physical layer separation $d$ with an "effective layer separation" $d^{T_{z} T_{z}^{\prime}}_{eff}$ yields an excellent fit, and we take
\begin{equation} \refstepcounter{SIeq} \label{Vfit_appendix}
V_{T_{z} T_{z}^{\prime}}(q) = \frac{1}{N_\Phi} \alpha \frac{1}{ql_B}\tanh{(qD)}e^{-q d^{T_{z} T_{z}^{\prime}}_{eff}}
\end{equation}
in our calculations. Here $d^{T_{z} T_{z}^{\prime}}_{eff}$ is indexed by whether or not $T_{z} = T_{z}^{\prime}$ and by pressure.

The Coulomb blockade, as shown in Appendix~\ref{appendix_CoulombBlockade}, is obtained by the $q \to 0$ limit, ${\Delta V = N_\Phi \left(V_{11}(0)-V_{12}(0)\right)}$. Without the spatial extent of the p\textsubscript{z} orbitals, taking $D+d \approx D$, this difference would be
\begin{equation} \refstepcounter{SIeq}
N_\Phi\lim_{q\to0}\left(V\left(q,\frac{d}{2},\frac{d}{2}\right)- V\left(q,\frac{d}{2},-\frac{d}{2}\right)\right) \approx\alpha \frac{d}{l_B} .
\end{equation}
Hence, the Coulomb blockade strength $\Delta V$ can be written in terms of an effective layer separation through $\Delta V = \alpha \frac{d^{CB}_{eff}}{l_B}$. Because $d^{CB}_{eff}$ is a single-point calculation at $q=0$ and should not be constrained by results for $q \ne 0$, we calculate it independently of the $q$-dependent fit parameters $d^{T_{z} T_{z}^{\prime}}_{eff}$ using Eq.~(\ref{VCB_appendix}).

\begin{figure*}
    \centering
    \includegraphics[width=8cm]{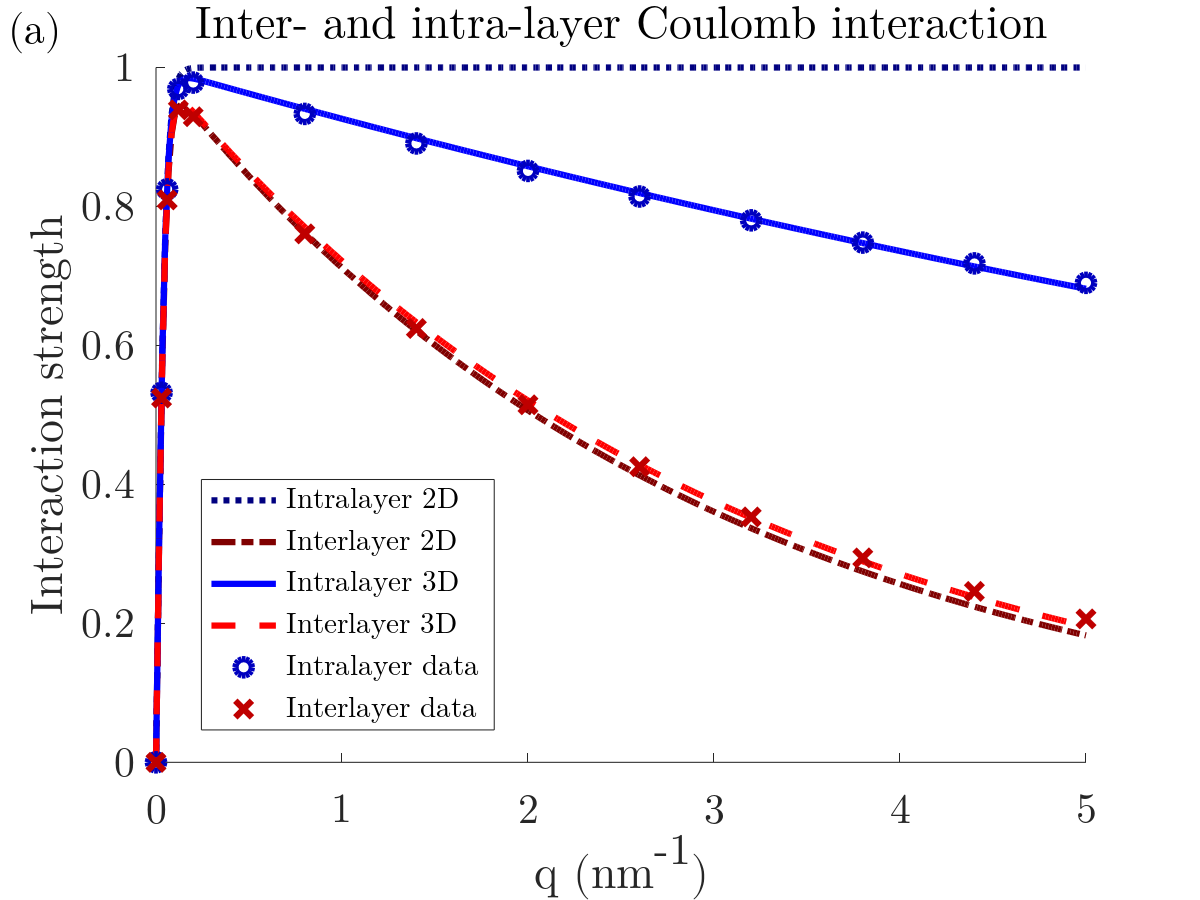}
    \includegraphics[width=8cm]{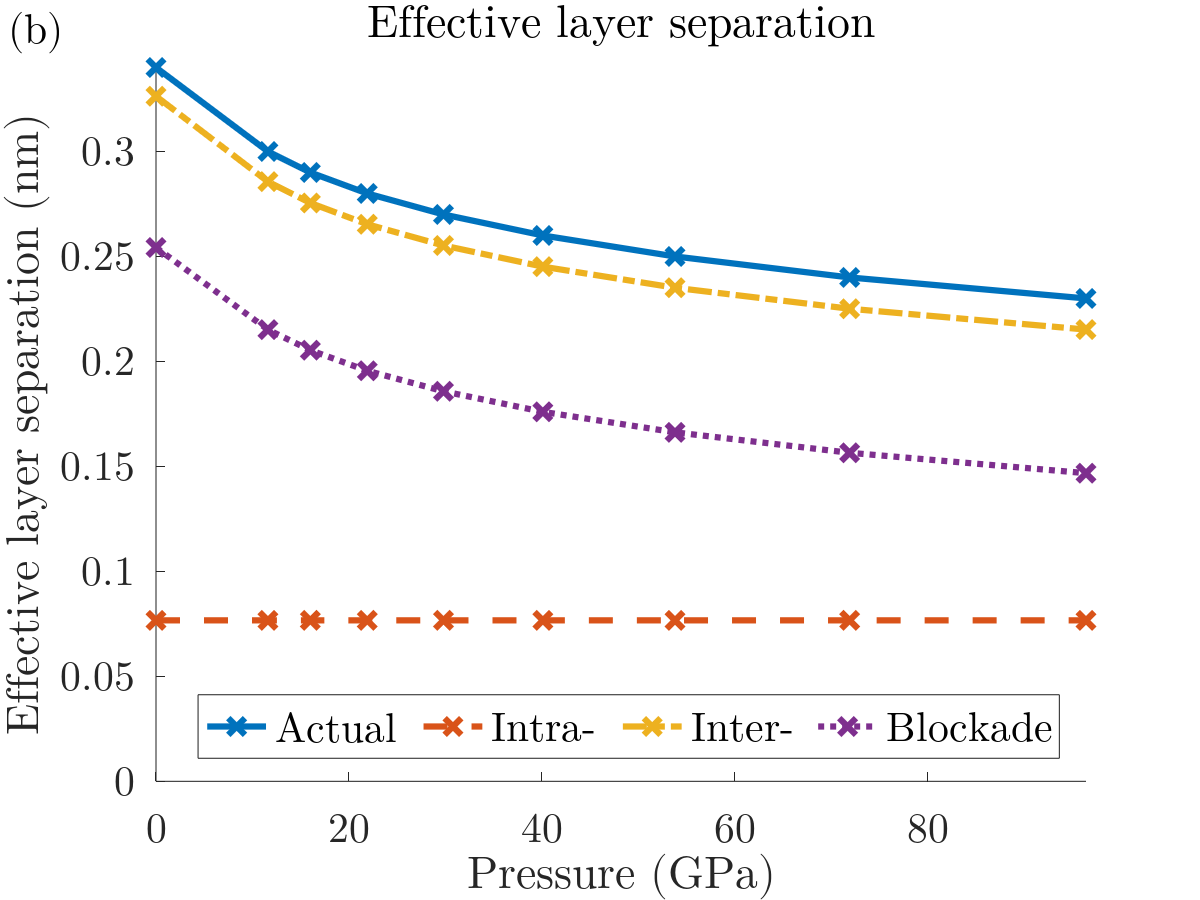}
    
    \refstepcounter{SIfig}\label{fig:Vfitplots}
    
    \caption{(a) Here we compare Coulomb interaction propagators corresponding to the Coulomb propagator calculated exactly or using a fit, with or without spatially extended 3D p\textsubscript{z} orbitals, in units of $\frac{1}{N_\Phi}\alpha\frac{1}{ql_B}$ at zero pressure. Plots at elevated pressures are similar. "2D", "3D", and "data" respectively refer to Eq.~(\ref{Vlayerprojection_appendix}) with $P(z)\to\delta(z)$, Eq.~(\ref{Vfit_appendix}), and Eq.~(\ref{Vlayerprojection_appendix}) calculated exactly. (b) We also give the variation of the effective layer separations $d^{T_{z} T_{z}^{\prime}}_{eff}$ and $d^{CB}_{eff}$ with pressure, compared to the actual layer separation $d$. In this plot, lines are guides to the eye.}
\end{figure*}

Thus we have the four pressure-varying layer separations - actual, effective intralayer, effective interlayer, and effective Coulomb blockade - plotted in Fig.~\ref{fig:Vfitplots}(b). Layer separation weakens the exchange interaction and strengthens the Coulomb blockade, and neglecting the spatial extent of the p\textsubscript{z} orbitals is equivalent to taking $d_{eff}^{11}=0$, $d_{eff}^{12}=d$, and $d_{eff}^{CB}=d$. Hence, the spatial extended of the p\textsubscript{z} orbitals weakens intralayer exchange, marginally strengthens interlayer exchange, and weakens the Coulomb blockade.

\begin{widetext}
\section{Coulomb blockade} \refstepcounter{SIsect}\label{appendix_CoulombBlockade}

Restricting our attention to the LLLs and $q=0$, and writing the interaction in terms of the intra-/interlayer interaction difference ${\Delta V = N_\Phi \left(V_{11}(0)-V_{12}(0)\right)}$, we have
\begin{equation} \refstepcounter{SIeq}
\hat{V}_{D} = N_\Phi \sum_{n n^{\prime} \xi \xi^{\prime}} \sum_{T_{z} T_{z}^{\prime}}\left( N_\Phi V_{T_{z} T_{z}}(0)-\Delta V \delta_{T_z \left(-T_z^\prime\right)}\right)
J_{\substack{n \xi \\ n \xi}}^{T_{z}}(0) J_{\substack{n^\prime \xi^\prime \\ n^\prime \xi^\prime}}^{T_{z}^\prime}(0)
\sum_{\sigma \sigma^{\prime}}\left\langle\rho_{n n}^{\xi \xi \sigma \sigma}\right\rangle \rho_{n^{\prime} n^{\prime}}^{\xi^{\prime} \xi^{\prime} \sigma^{\prime} \sigma^{\prime}} \, .
\end{equation}

For the form factors, we have $K_{jj^\prime}(0)=\delta_{jj^\prime}$, so
\begin{equation} \refstepcounter{SIeq}
J_{\substack{n_{1} \xi \\ n_{4} \xi}}^{T_{z}}(0)=\sum_{j=0}^{\infty} \sum_{T_{2 D}} \left(c_{n_{1} \xi}^{T_{2 D} T_{z} j}\right)^* c_{n_{4} \xi}^{T_{2 D} T_{z} j} = \delta_{n_1 n_4} \sum_{j=0}^{\infty} \sum_{T_{2 D}} \left\lvert c_{n_{4} \xi}^{T_{2 D} T_{z} j}\right\rvert^2 = \delta_{n_1 n_4}\frac{1-(-1)^{T_z}\xi \Pi_{n_1}}{2} \, ,
\end{equation}
since we never simultaneously have $c_{0\xi}^{T_{2D}T_{z} j} \, , \, c_{1\xi}^{T_{2D}T_{z} j}\ne0$, as can be seen from Eq.~(\ref{nonzerocoeffs_appendix}). Hence, $J_{\substack{n \xi \\ n \xi}}^{T_{z}}(0)$ is the density in the layer $T_z$ due to the LLL with orbital $n$ and valley $\xi$.

Since $\sum_{T_z}J_{\substack{n \xi \\ n \xi}}^{T_{z}}(0)=1$ by normalization, the contribution of the $V_{T_{z} T_{z}}(0)$ term is a constant diagonal shift $N_\Phi^2 V_{T_{z} T_{z}}(0) \nu \sum_{n^{\prime} \xi^{\prime} \sigma^{\prime}} \rho_{n^{\prime} n^{\prime}}^{\xi^{\prime} \xi^{\prime} \sigma^{\prime} \sigma^{\prime}}$ which may be discarded. Redefining the interaction to exclude this constant shift,
\begin{equation} \refstepcounter{SIeq} \label{VDeqn_in_appendix}
\hat{V}_{D} = - N_\Phi \Delta V \sum_{T_{z} T_{z}^{\prime}} \delta_{T_z \left(-T_z^\prime\right)} \left(\sum_{n \xi \sigma}\left\langle\rho_{n n}^{\xi \xi \sigma \sigma}\right\rangle J_{\substack{n \xi \\ n \xi}}^{T_{z}}(0) \right)\left(\sum_{n^{\prime} \xi^{\prime} \sigma^{\prime}} \rho_{n^{\prime} n^{\prime}}^{\xi^{\prime} \xi^{\prime} \sigma^{\prime} \sigma^{\prime}} J_{\substack{n^\prime \xi^\prime \\ n^\prime \xi^\prime}}^{T_{z}^\prime}(0)\right) \, ,
\end{equation}
\end{widetext}
which may be simplified in terms of the layer occupations
\begin{equation}\begin{aligned} \refstepcounter{SIeq} \label{layeroccupationdef_app}
\nu_{T_{Z}} & = \sum_{n \xi \sigma} \left\langle\rho_{n n}^{\xi \xi \sigma \sigma}\right\rangle J_{\substack{n \xi \\ n \xi}}^{T_{z}}(0) \\ & = \frac{1}{2} \Big( \tilde{\nu} - (-1)^{T_z} \sum_{n \xi \sigma} \left\langle\rho_{n n}^{\xi \xi \sigma \sigma}\right\rangle \xi \Pi_{n} \Big) .
\end{aligned}\end{equation}

The interaction is therefore
\begin{equation} \refstepcounter{SIeq}
\hat{V}_{D} = - N_\Phi \Delta V \sum_{n \xi \sigma}\left(\nu_{2} \frac{1+\xi \Pi_{n}}{2}+\nu_{1} \frac{1-\xi \Pi_{n}}{2}\right) \rho_{n n}^{\xi \xi \sigma \sigma} \, .
\end{equation}

For the system energy, using Eqs.~(\ref{VDeqn_in_appendix}) and (\ref{layeroccupationdef_app}) we find
\begin{equation} \refstepcounter{SIeq}
\frac{1}{2}\left\langle\hat{V}_{D}\right\rangle = - N_\Phi \Delta V \nu_1 \nu_2 \, .
\end{equation}

Note that this agrees with the capacitive correction derived by Refs.~\cite{coteorbital2010,lambertquantum2013,lambertferro-aimants2013} and subsequently used in other work \cite{knothephase2016,huntdirect2017} of the form $N_\Phi\frac{1}{4}\Delta V\left(\nu_1-\nu_2\right)^{2}$. Since $\frac{1}{4}\left(\nu_1-\nu_2\right)^{2}=\frac{\nu^{2}}{4}-\nu_1 \nu_2$, the only difference is a shift $\frac{\nu^{2}}{4}$ dependent only on total filling factor, and hence irrelevant to LLL filling order at fixed filling factor. The shift originates in interactions with the positive background \cite{coteorbital2010,lambertquantum2013,lambertferro-aimants2013}.

Finally, we calculate $\Delta V$. For $q=0$ the Coulomb interaction propagator is
\begin{equation} \refstepcounter{SIeq}
V\left(0, z, z^{\prime}\right) = \frac{1}{N_\Phi} \alpha\left(\frac{D}{l_{B}}-\frac{\left|z^{\prime}-z\right|}{l_{B}}-\frac{z z^{\prime}}{D l_{B}}\right) \, ,
\end{equation}
so that the layer-resolved interaction difference is
\begin{align} \refstepcounter{SIeq} \label{VCB_appendix} \Delta V & = \frac{\alpha}{l_{B}} \int dz \int \! dz^{\prime} P(z) P\left(z^{\prime}\right) \\ & \quad \quad \times \left( \left|z^{\prime} - z + d\right| - \left|z^{\prime} - z\right| - \frac{d(2z + d)}{2 D}\right) . \nonumber \end{align}
This may be conveniently written in terms of an effective layer separation as $\Delta V = \alpha \frac{d_{eff}^{CB}}{l_B}$.

\section{Exchange integrals} \refstepcounter{SIsect}\label{appendix_exchangeintegrals}

The exchange interaction in the $n=0,1$ space, $X_{n_1 n_2 n_3 n_4}^{\xi \xi^{\prime}}(0)$, is calculated as a linear combination,
\begin{align} \refstepcounter{SIeq}
& X_{n_{1} n_{2} n_{3} n_{4}}^{\xi \xi^{\prime}}(0) = \alpha \sum_{T_{z} T_{z}^{\prime}} \sum_{j_{1} j_{2} j_{3} j_{4}=0}^{\infty} F_{j_{1} j_{2} j_{3} j_{4}}^{T_{Z} T_{Z}^{\prime}}(0) \\ & \quad \times \! \left(\sum_{T_{2 D}} {c_{n_{1} \xi}^{T_{2 D} T_{z} j_{1}}}^{*} c_{n_{2} \xi}^{T_{2 D} T_{z} j_{2}}\right) \! \left(\sum_{T_{2 D}^\prime} {c_{n_{3} \xi^\prime}^{T_{2 D}^\prime T_{z}^\prime j_{3}}}^{*} c_{n_{4} \xi^\prime}^{T_{2 D}^\prime T_{z}^\prime j_{4}}\right) , \nonumber
\end{align}
of the elementary exchange integrals $F_{n_1 n_2 n_3 n_4}^{\xi \xi^{\prime}}(0)$ which give the exchange interaction between elementary form factors,
\begin{equation}\begin{aligned} \refstepcounter{SIeq}
F_{j_1 j_2 j_3 j_4}^{T_{z} T_{z}^{\prime}}(\bm{q}) & = \int \! d^{2} \bm{p} V_{T_z T_z^\prime} \left(\bm{p}\right) \\ & \quad \quad \times K_{j_1 j_2}(\bm{p}) K_{j_3 j_4}(-\bm{p}) e^{i\bm{q}l_{B}\times\bm{p}l_{B}} .
\end{aligned}\end{equation}

These integrals are readily evaluated in polar coordinates using
\begin{widetext}
\begin{equation} \refstepcounter{SIeq} \label{formfactor_generalformula}
K_{j_{1} j_{4}}(\bm{p})=e^{-\left(\frac{p l_B}{2}\right)^{2}} \frac{1}{\sqrt{2^{j_1+j_4}j_1!j_4!}} \left( \sum_{k=0}^{j_{1}} \sum_{m=0}^{j_{4}} 2^{k} k! \binom{j_1}{k}\binom{j_4}{m} \left(i p l_{B} e^{-i \theta_{p}}\right)^{j_{1}-k} \left(i p l_{B} e^{i \theta_{p}}\right)^{j_{4}-m} \delta_{km} \right) \, .
\end{equation}
\end{widetext}

In our computations, we truncate $j>4$ and renormalize $c_{n \xi}^{T j} \to \frac{c_{n \xi}^{T j}}{\sqrt{\sum_{T^{\prime}} \sum_{j^{\prime}=0}^{4} \left|c_{n \xi}^{T^{\prime} j^{\prime}}\right|^2}}$. This is reasonable in light of the miniscule $j>4$ remainder from our exact diagonalization results given in Fig.~\ref{fig:spresults2}. Because the exchange integrals vary smoothly and slowly with magnetic field and are costly to calculate explicitly, we calculate the exchange integrals using the expressions we have derived and presented at $1~\text{T}$ intervals, and interpolate between them using cubic splines when higher resolution is needed.

Several symmetries reduce the number of independent exchange integrals. First, we have the valley symmetries
\begin{equation} \refstepcounter{SIeq}
X_{klmn}^{\xi \xi^{\prime}}(\bm{q})=X_{klmn}^{\xi^{\prime} \xi}(\bm{q}) \, , \quad X_{klmn}^{++}(\bm{q})=X_{klmn}^{--}(\bm{q}) \, .
\end{equation}
Two form factor symmetries also induce corresponding exchange integral symmetries;
\begin{equation}\begin{aligned} \refstepcounter{SIeq}
& J_{\substack{m \xi \\ n \xi}}^{T_{z}}(\bm{q}) = \left( J_{\substack{n \xi \\ \xi}}^{T_{z}}(-\bm{q}) \right)^{*} \\ & \quad \quad \Rightarrow \left(X_{k l m n}^{\xi \xi^{\prime}}(\bm{q})\right)^{*} = X_{l k n m}^{\xi \xi^{\prime}}(\bm{q})
\end{aligned}\end{equation}
and
\begin{equation}\begin{aligned} \refstepcounter{SIeq}
& J_{\substack{m \xi \\ n \xi}}^{T_{z}}(\bm{q})=(-1)^{m+n}J_{\substack{m \xi \\ n \xi}}^{T_{z}}(-\bm{q}) \\ & \quad \quad \Rightarrow X_{k l m n}^{\xi \xi^{\prime}}(\bm{q})=(-1)^{(m+n+k+l)} X_{m n k l}^{\xi \xi^{\prime}}(\bm{q}) \, .
\end{aligned}\end{equation}
One also has $F_{j_1 j_2 j_4 j_4}^{T_{z} T_{z}^{\prime}}(\bm{q})=0$ if $j_2+j_4 \ne j_1+j_3$.

\begin{widetext}
\section{State energies} \refstepcounter{SIsect}\label{appendix_stateenergies}

The energy of the FSP state is
\begin{equation}\begin{aligned} \refstepcounter{SIeq} \tilde{\nu} \frac{E_{H F}}{N_{e}} & = E_{0+\uparrow}+\left(E_{1+\uparrow}\!+\!\Delta_{Lamb}\right)+E_{0-\uparrow}+\left(E_{1-\uparrow}\!+\!\Delta_{Lamb}\right) \\ & \quad -4 \Delta V-\alpha\left(X_{0000}^{++}+X_{1111}^{++}+2 X_{0110}^{++}\right) \, . \end{aligned}\end{equation}
%\begin{equation} \refstepcounter{SIeq} \tilde{\nu} \frac{E_{H F}}{N_{e}} = E_{0+\uparrow}+\left(E_{1+\uparrow}\!+\!\Delta_{Lamb}\right)+E_{0-\uparrow}+\left(E_{1-\uparrow}\!+\!\Delta_{Lamb}\right) -4 \Delta V-\alpha\left(X_{0000}^{++}+X_{1111}^{++}+2 X_{0110}^{++}\right) \, . \end{equation}

The energy of the FVP state is
\begin{equation}\begin{aligned} \refstepcounter{SIeq} \tilde{\nu} \frac{E_{H F}}{N_{e}} & = E_{0+\uparrow}+\left(E_{1+\uparrow}\!+\!\Delta_{Lamb}\right)+E_{0+\downarrow}+\left(E_{1+\downarrow}\!+\!\Delta_{Lamb}\right) \\ & \quad -\Delta V\left(4-\left(\Pi_{0}+\Pi_{1}\right)^{2}\right) -\alpha\left(X_{0000}^{++}+X_{1111}^{++}+2 X_{0110}^{++}\right) \, . \end{aligned}\end{equation}
%\begin{equation} \tilde{\nu} \frac{E_{H F}}{N_{e}} = E_{0+\uparrow}+\left(E_{1+\uparrow}\!+\!\Delta_{Lamb}\right)+E_{0+\downarrow}+\left(E_{1+\downarrow}\!+\!\Delta_{Lamb}\right) -\Delta V\left(4-\left(\Pi_{0}+\Pi_{1}\right)^{2}\right) -\alpha\left(X_{0000}^{++}+X_{1111}^{++}+2 X_{0110}^{++}\right) \, . \end{equation}

The energy of the FOP state is
\begin{equation} \refstepcounter{SIeq}
\tilde{\nu} \frac{E_{H F}}{N_{e}} = E_{0+\uparrow}+E_{0-\uparrow}+E_{0+\downarrow}+E_{0-\downarrow}-4 \Delta V-2 \alpha X_{0000}^{++} \, .
\end{equation}

The energy of the POP state is
\begin{equation}\begin{aligned} \refstepcounter{SIeq} \tilde{\nu} \frac{E_{H F}}{N_{e}} & = E_{0+\uparrow}+E_{0-\uparrow}+E_{0+\downarrow}+\left(E_{1+\uparrow}\!+\!\Delta_{Lamb}\right) \\ & \quad -\Delta V\left(4-\frac{1}{4}\left(\Pi_{0}+\Pi_{1}\right)^{2}\right) - \frac{1}{2} \alpha\left(3 X_{0000}^{++}+X_{1111}^{++}+2 X_{0110}^{++}\right) \, . \end{aligned}\end{equation}
%\begin{equation} \refstepcounter{SIeq} \tilde{\nu} \frac{E_{H F}}{N_{e}} = E_{0+\uparrow}+E_{0-\uparrow}+E_{0+\downarrow}+\left(E_{1+\uparrow}\!+\!\Delta_{Lamb}\right) -\Delta V\left(4-\frac{1}{4}\left(\Pi_{0}+\Pi_{1}\right)^{2}\right) - \frac{1}{2} \alpha\left(3 X_{0000}^{++}+X_{1111}^{++}+2 X_{0110}^{++}\right) \, . \end{equation}

The energy of the OSP state is
\begin{equation}\begin{aligned} \refstepcounter{SIeq} \tilde{\nu} \frac{E_{H F}}{N_{e}} & = E_{0+\uparrow}+E_{0-\uparrow}+E_{0+\downarrow}+\left(E_{1-\uparrow}\!+\!\Delta_{Lamb}\right) \\ & \quad -\Delta V\left(4-\frac{1}{4}\left(\Pi_{0}-\Pi_{1}\right)^{2}\right) - \frac{1}{2} \alpha\left(3 X_{0000}^{++}+X_{1111}^{++}+2 X_{0110}^{++}\right) \, . \end{aligned}\end{equation}
%\begin{equation} \refstepcounter{SIeq} \tilde{\nu} \frac{E_{H F}}{N_{e}} = E_{0+\uparrow}+E_{0-\uparrow}+E_{0+\downarrow}+\left(E_{1-\uparrow}\!+\!\Delta_{Lamb}\right) -\Delta V\left(4-\frac{1}{4}\left(\Pi_{0}-\Pi_{1}\right)^{2}\right) - \frac{1}{2} \alpha\left(3 X_{0000}^{++}+X_{1111}^{++}+2 X_{0110}^{++}\right) \, . \end{equation}

The energy of the FSP-FVP LLC state is
\begin{equation}\begin{aligned} \refstepcounter{SIeq} \label{FSPFVPenergy} \tilde{\nu} \frac{E_{H F}}{N_{e}} & = E_{0+\uparrow}+\left(E_{1+\uparrow}\!+\!\Delta_{Lamb}\right)+E_{0-\uparrow} \cos ^{2} \theta_{0}+E_{0+\downarrow} \sin ^{2} \theta_{0} \\ & \quad +\left(E_{1-1}\!+\!\Delta_{Lamb}\right) \cos ^{2} \theta_{1}+\left(E_{1+\downarrow}\!+\!\Delta_{Lamb}\right) \sin ^{2} \theta_{1} \\ & \quad -\Delta V\left(4-\left(\Pi_{0} \sin ^{2} \theta_{0}+\Pi_{1} \sin ^{2} \theta_{1}\right)^{2}\right) \\ & \quad - \frac{1}{4}\alpha \Big( X_{0000}^{++}\left(4-\sin ^{2} 2 \theta_{0}\right)+X_{1111}^{++}\left(4-2 \sin ^{2} \theta_{1}\right) \\ & \quad \quad \quad \quad +4 X_{0110}^{++}\left(\sin ^{2} \theta_{0}+\sin ^{2} \theta_{1}+2 \cos ^{2} \theta_{0} \cos ^{2} \theta_{1}\right) \\ & \quad \quad \quad \quad +\left(X_{0000}^{+-} \sin ^{2} 2 \theta_{0}+2 X_{0110}^{+-} \sin 2 \theta_{0} \sin 2 \theta_{1}+X_{1111}^{+-} \sin ^{2} 2 \theta_{1}\right) \Big) \, . \end{aligned}\end{equation}
%\begin{align} \refstepcounter{SIeq} \label{FSPFVPenergy} \tilde{\nu} \frac{E_{H F}}{N_{e}} & = E_{0+\uparrow}+\left(E_{1+\uparrow}\!+\!\Delta_{Lamb}\right)+E_{0-\uparrow} \cos ^{2} \theta_{0}+E_{0+\downarrow} \sin ^{2} \theta_{0} +\left(E_{1-1}\!+\!\Delta_{Lamb}\right) \cos ^{2} \theta_{1}+\left(E_{1+\downarrow}\!+\!\Delta_{Lamb}\right) \sin ^{2} \theta_{1} \nonumber \\ & \quad -\Delta V\left(4-\left(\Pi_{0} \sin ^{2} \theta_{0}+\Pi_{1} \sin ^{2} \theta_{1}\right)^{2}\right) - \frac{1}{4}\alpha \Big( X_{0000}^{++}\left(4-\sin ^{2} 2 \theta_{0}\right)+X_{1111}^{++}\left(4-2 \sin ^{2} \theta_{1}\right) \\ & \quad +4 X_{0110}^{++}\left(\sin ^{2} \theta_{0}+\sin ^{2} \theta_{1}+2 \cos ^{2} \theta_{0} \cos ^{2} \theta_{1}\right) +\left(X_{0000}^{+-} \sin ^{2} 2 \theta_{0}+2 X_{0110}^{+-} \sin 2 \theta_{0} \sin 2 \theta_{1}+X_{1111}^{+-} \sin ^{2} 2 \theta_{1}\right) \Big) \, . \nonumber \end{align}
Under the approximation $\theta_0=\theta_1\equiv\theta$ the optimal parameter can be found analytically and is
\begin{equation} \refstepcounter{SIeq} \label{FSPFVPopteqn}
\cos ^{2} \theta = \frac{1}{2} - \frac{ E_{0-\uparrow}-E_{0+\downarrow}+E_{1-\uparrow}-E_{1+\downarrow} - \Delta V \left(\Pi_0+\Pi_1\right)^2 }{ 2 \Delta V\left(\Pi_0+\Pi_1\right)^2 + 2 \alpha\left(\left(X_{0000}^{+-}+2 X_{0110}^{+-}+X_{1111}^{+-}\right)-\left(X^{++}_{0000}+2 X^{++}_{0110}+X^{++}_{1111}\right)\right) } \, .
\end{equation}
This is a good approximation and serves well as an ansatz to find $\theta_0,\,\theta_1$ numerically. The energy concavity can also be calculated analytically with respect to this parameter and is
\begin{equation}\begin{aligned} \refstepcounter{SIeq} \left(\frac{d}{d\cos^{2} \theta}\right)^{2}\nu\frac{E_{HF}}{N_e} & = 2\Delta V\left(\Pi_{0}+\Pi_{1}\right)^{2} \\ & \quad + 2\alpha\left( \left(X_{0000}^{+-}+2 X_{0110}^{+-}+X_{1111}^{+-}\right) - \left(X_{0000}^{++}+2 X_{0110}^{++}+X_{1111}^{++}\right) \right) \, . \end{aligned}\end{equation}
%\begin{equation} \refstepcounter{SIeq} \left(\frac{d}{d\cos^{2} \theta}\right)^{2}\nu\frac{E_{HF}}{N_e} = 2\Delta V\left(\Pi_{0}+\Pi_{1}\right)^{2} + 2\alpha\left( \left(X_{0000}^{+-}+2 X_{0110}^{+-}+X_{1111}^{+-}\right) - \left(X_{0000}^{++}+2 X_{0110}^{++}+X_{1111}^{++}\right) \right) \, . \end{equation}

The energy of the OSP-POP LLC state is
\begin{equation}\begin{aligned} \refstepcounter{SIeq} \tilde{\nu} \frac{E_{H F}}{N_{e}} & = E_{0+\uparrow}+E_{0-\uparrow}+E_{0+\downarrow}+\left(E_{1+\uparrow}\!+\!\Delta_{Lamb}\right) \cos ^{2} \theta+\left(E_{1-\uparrow}\!+\!\Delta_{Lamb}\right) \sin ^{2} \theta \\ & \quad -\Delta V\left(4-\frac{1}{4}\left(\Pi_{0}+\Pi_{1} \cos 2 \theta\right)^{2}\right) \\ & \quad -\frac{1}{4} \alpha\left(6 X_{0000}^{++}+X_{1111}^{++}\left(2-\sin ^{2} 2 \theta\right)+4 X_{0110}^{++}+X_{1111}^{+-} \sin ^{2} 2 \theta\right) \, . \end{aligned}\end{equation}
%\begin{equation}\begin{aligned} \refstepcounter{SIeq} \tilde{\nu} \frac{E_{H F}}{N_{e}} & = E_{0+\uparrow}+E_{0-\uparrow}+E_{0+\downarrow}+\left(E_{1+\uparrow}\!+\!\Delta_{Lamb}\right) \cos ^{2} \theta+\left(E_{1-\uparrow}\!+\!\Delta_{Lamb}\right) \sin ^{2} \theta \\ & \quad -\Delta V\left(4-\frac{1}{4}\left(\Pi_{0}+\Pi_{1} \cos 2 \theta\right)^{2}\right) -\frac{1}{4} \alpha\left(6 X_{0000}^{++}+X_{1111}^{++}\left(2-\sin ^{2} 2 \theta\right)+4 X_{0110}^{++}+X_{1111}^{+-} \sin ^{2} 2 \theta\right) \, . \end{aligned}\end{equation}
The optimal parameter $\theta$ is exactly
\begin{equation} \refstepcounter{SIeq}
\cos ^{2} \theta = \frac{E_{1+\uparrow}-E_{1-\uparrow} + \Delta V \Pi_1\left(\Pi_0-\Pi_1\right) + \alpha\left(X^{++}_{1111}-X_{1111}^{+-}\right)}{2 \alpha\left(X^{++}_{1111}-X_{1111}^{+-}\right)-2 \Delta V\Pi_1^{2}} \, .
\end{equation}
The energy concavity is
\begin{equation} \refstepcounter{SIeq}
\left(\frac{d}{d \cos ^{2} \theta}\right)^{2}\tilde{\nu}\frac{E_{H F}}{N_{e}}=2 \Delta V \Pi_{1}^{2}-2 \alpha\left(X_{1111}^{++}-X_{1111}^{+-}\right) \, .
\end{equation}

Because LLC states appear when two LLSD states are close in energy, it is useful to calculate the energy concavities for hypothetical LLC states mixing each pair of LLSD states that share a boundary. The energy concavity for the FSP-FOP state is
\begin{equation} \refstepcounter{SIeq}
\left(\frac{d}{d \cos ^{2} \theta}\right)^{2}\tilde{\nu}\frac{E_{H F}}{N_{e}}=2 \alpha\left(2 X_{0011}^{++}-X_{0000}^{++}-X_{1111}^{++}\right) \, .
\end{equation}

The energy concavity for the FOP-POP, FSP-POP and FVP-POP states is
\begin{equation} \refstepcounter{SIeq}
\left(\frac{d}{d \cos ^{2} \theta}\right)^{2}\tilde{\nu}\frac{E_{H F}}{N_{e}}=\frac{1}{2} \Delta V\left(\Pi_{0}+\Pi_{1}\right)^{2}+\alpha\left(2 X_{0011}^{+-}-X_{0000}^{++}-X_{1111}^{++}\right) \, .
\end{equation}

The energy concavity for the FSP-OSP and FOP-OSP states is
\begin{equation} \refstepcounter{SIeq}
\left(\frac{d}{d \cos ^{2} \theta}\right)^{2}\tilde{\nu}\frac{E_{H F}}{N_{e}}=\frac{1}{2} \Delta V\left(\Pi_{0}-\Pi_{1}\right)^{2}+\alpha\left(2 X_{0011}^{++}-X_{0000}^{++}-X_{1111}^{++}\right) \, .
\end{equation}
\end{widetext}

\section{Stabilizing LLC states with interactions} \refstepcounter{SIsect}\label{appendix_EntanglementInteractionAndExchange}

The noninteracting contribution to the energy of an LLC state, say the $\Psi-\Psi^\prime$ state, is always between that of the $\Psi$ and $\Psi^\prime$ LLSD states which it mixes - in particular, either the $\Psi$ or the $\Psi^\prime$ state has lower noninteracting energy that the $\Psi-\Psi^\prime$ state. Therefore, the only way that the $\Psi-\Psi^\prime$ state could be the ground state in a fully interacting model is if its superposition lowers the interaction energy. Physically, superpositions lower energy by delocalizing electrons. (This explains why, when we explored the effects of 3D p\textsubscript{z} orbitals, layer separation, gating, and form factors, the only LLC states that appeared involved intervalley superpositions, which spread electron density across the two layers.) Delocalization reduces the repulsive Coulomb interaction, which lowers both the Coulomb blockade and the exchange interaction. Since the Coulomb blockade raises energy while the exchange interaction lowers it, a superposition must decrease the Coulomb blockade more than it decreases the exchange interaction. If this is not the case, the superposition will not be favorable.

As a demonstration, suppose the $\Psi-\Psi^\prime$ state involves only one superposition, between the LLLs $n\xi\sigma$ and $n^\prime\xi^\prime\sigma^\prime$. The density matrix elements corresponding to these LLLs are
\begin{gather} \refstepcounter{SIeq}
\left\langle\rho_{nn}^{\xi\xi\sigma\sigma}\right\rangle = \cos^2\theta \, , \quad \left\langle\rho_{n^\prime n^\prime}^{\xi^\prime\xi^\prime\sigma^\prime\sigma^\prime}\right\rangle = \sin^2\theta \, , \\ \left\langle\rho_{nn^\prime}^{\xi\xi^\prime\sigma\sigma^\prime}\right\rangle = \left\langle\rho_{n^\prime n}^{\xi^\prime\xi\sigma^\prime\sigma}\right\rangle = \frac{1}{1}\sin 2\theta
\end{gather}
and the matrix elements corresponding to LLLs not involved in the superposition are fixed to be $0$ or $1$, so the state's energy concavity is
\begin{widetext}
\begin{equation} \refstepcounter{SIeq} \label{generalsinglesuperposconcav}
\left(\frac{d}{d \cos ^{2} \theta}\right)^{2}\tilde{\nu}\frac{E_{H F}}{N_{e}} = \frac{1}{2} \Delta V\left(\xi \Pi_{n}-\xi^{\prime} \Pi_{n^{\prime}}\right)^{2}+\alpha\left(2 X_{n n n^{\prime} n^{\prime}}^{\xi \xi^{\prime}}-X_{n n n n}^{\xi \xi}-X_{n^{\prime} n^{\prime} n^{\prime} n^{\prime}}^{\xi^{\prime} \xi^{\prime}}\right) \, .
\end{equation}
\end{widetext}
The first term is nonnegative and represents weakening the Coulomb blockade, and the second term is nonpositive and represents weakening the exchange interaction. The latter is nonpositive because the "off-diagonal" exchange integral $X_{n n n^{\prime} n^{\prime}}^{\xi \xi^{\prime}}$, arising from the overlap of the superimposed LLLs, is less than either "diagonal" integral, $X_{n n n n}^{\xi \xi}$, representing the overlap of an LLL with itself; any LLL has greater overlap with itself than with another LLL.

\bibliographystyle{apsrev4-2}

\end{document}